\documentclass[fleqn,usenatbib]{config/mnras}

\usepackage[T1]{fontenc}

\DeclareRobustCommand{\VAN}[3]{#2}
\let\VANthebibliography\thebibliography
\def\thebibliography{\DeclareRobustCommand{\VAN}[3]{##3}\VANthebibliography}

\usepackage{amsmath}	
\usepackage{graphicx}
\usepackage{booktabs}
\usepackage{multirow}	
\usepackage{paralist}
\usepackage{textgreek}
\usepackage{xargs}
\usepackage{xcolor}
\usepackage{xspace}
\usepackage{xfrac}
\usepackage{txfonts}
\usepackage{hyperref}
\hypersetup{
    colorlinks=true,
    linkcolor=blue,
    citecolor=blue,
    filecolor=magenta,      
    urlcolor=cyan,
    }
\usepackage{subcaption}
\usepackage{adjustbox}  

\graphicspath{{./}{figs/}}
\usepackage{etoolbox}
\makeatletter
\newcommand\sendemail[3]{
\edef\@tempa{mailto:#1?subject=#2 }%
\edef\@tempb{\expandafter\html@spaces\@tempa\@empty}%
\href{\@tempb}{#3}}

\catcode\%=11
\def\html@spaces#1 #2{#1
\catcode\%=14
\makeatother

\newcommand{\todo}[1]{\textcolor{magenta}{[#1]}}
\newcommand{\orcid}[2]{\href{http://orcid.org/#2}{#1}}
\newcommand{\orcidsymb}[2]{\href{http://orcid.org/#2}{#1\adjustbox{trim={-.15\width} {0\height} {-.15\width} {0\height},clip}{\includegraphics[height=10pt]{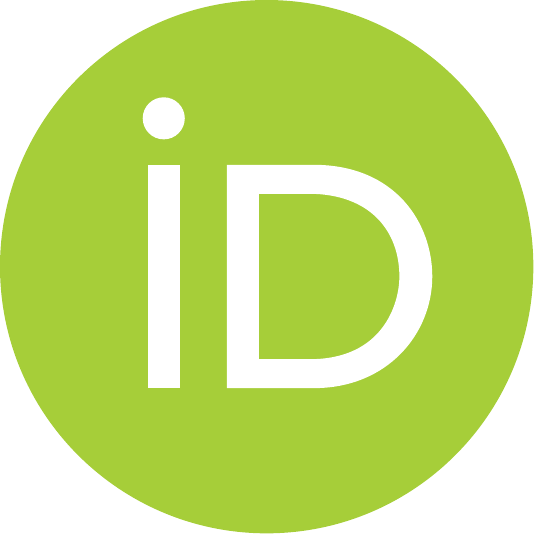}}}}

\newcommand{\citationneeded}{\textcolor{ForestGreen}{$^{\rm citation\;needed}$}}
\let\oldtextsigma\textsigma
\renewcommand{\textsigma}{\oldtextsigma\xspace}
\let\oldAA\AA
\renewcommand{\AA}{\text{\oldAA}\xspace}
\let\oldtextdegree\textdegree
\renewcommand{\textdegree}{\oldtextdegree\xspace}

\newcommand{\kms}{\ensuremath{\mathrm{km\,s^{-1}}}\xspace}
\newcommand{\Msun}{\ensuremath{{\rm M}_\odot}\xspace}
\newcommand{\Zsun}{\ensuremath{{\rm Z}_\odot}\xspace}
\newcommand{\yr}{\ensuremath{{\rm yr}}\xspace}
\newcommand{\Myr}{\ensuremath{{\rm Myr}}\xspace}
\newcommand{\Gyr}{\ensuremath{{\rm Gyr}}\xspace}
\newcommand{\peryr}{\ensuremath{{\rm yr^{-1}}}\xspace}
\newcommand{\Lsun}{\hbox{\,${\rm L}_\odot$}}
\newcommand{\mum}{\text{\textmu m}\xspace}
\newcommand{\kpc}{\text{kpc}\xspace}
\newcommand{\Gpc}{\text{Gpc}\xspace}

\newcommand{\ZH}{\text{[Z/H]}\xspace}

\newcommandx{\lambdar}[2][1=R,2=]{\ensuremath{\lambda_{\rm {#1}}{#2}}\xspace}
\newcommand{\eps}{\ensuremath{\epsilon}\xspace}
\newcommand{\msurv}{\ensuremath{M_{\star\mathrm{,surv}}}\xspace}
\newcommand{\mstar}{\ensuremath{M_\star}\xspace}
\newcommand{\mdyn}{\ensuremath{M_\mathrm{dyn}}\xspace}
\newcommand{\re}{\ensuremath{R_\mathrm{e}}\xspace}
\newcommand{\vstar}{\ensuremath{v_\star}\xspace}
\newcommand{\vnai}{\ensuremath{v_{\NaI}}\xspace}
\newcommand{\sigmastar}{\ensuremath{\sigma_\star}\xspace}
\newcommand{\sigmaestar}{\ensuremath{\sigma_{\star,\mathrm{e}}}\xspace}
\newcommand{\vperc}[1]{\ensuremath{v_{#1}}\xspace}

\newcommand{\vesc}{\ensuremath{v_\mathrm{esc}}\xspace}
\newcommand{\nelec}{\ensuremath{n_\mathrm{e}}\xspace}
\newcommand{\Rout}{\ensuremath{R_\mathrm{out}}\xspace}
\newcommand{\vout}{\ensuremath{v_\mathrm{out}}\xspace}
\newcommandx{\Mout}[2][1=,2=]{\ensuremath{M_{\mathrm{out}{#2}}^{#1}}\xspace}
\newcommandx{\Mdotout}[2][1=,2=]{\ensuremath{\dot{M}_{\mathrm{out}{#2}}^{#1}}\xspace}

\newcommandx{\fluxdcgs}[1][1=-20]{$\times 10^{[#1]}$~erg~s$^{-1}$~cm$^{-2}$~\AA$^{-1}$\xspace}
\newcommandx{\fluxcgs}[2][1=-20,2=\ensuremath{\times}]{${#2}10^{#1}$~erg~s$^{-1}$~cm$^{-2}$\xspace}
\newcommandx{\powercgs}[1][1=44]{$\times 10^{#1}$~erg~s$^{-1}$\xspace}
\newcommand{\Av}{\ensuremath{A_V}\xspace}

\newcommand{\fHbetash}{$f(\mathrm{H}\text{\textbeta})_\mathrm{sh}$\xspace}
\newcommandx{\target}[1][1=]{\text{ulema{#1}}\xspace}

\newcommand{\jwst}{\textit{JWST}\xspace}
\newcommand{\hst}{\textit{HST}\xspace}
\newcommand{\ppxf}{{\sc ppxf}\xspace}
\newcommand{\prospector}{{\sc prospector}\xspace}
\newcommand{\emcee}{{\sc emcee}\xspace}
\newcommand{\cloudy}{{\sc cloudy}\xspace}
\newcommandx{\mappings}[1][1=]{{\sc mappings{#1}}\xspace}
\newcommand{\shinbnzv}{{\sc shinbnzv}\xspace}
\newcommand{\fsps}{{\sc fsps}\xspace}
\newcommand{\flamingo}{{\sc flamingo}\xspace}
\newcommand{\lcdm}{\ensuremath{\mathrm{\Lambda}}CDM\xspace}

\newcommand{\Mdynvalue}{$\Mdyn = 2.0\pm0.5 \times 10^{11}$~\MSun}

\defcitealias{Labb__2023}{L23}
\defcitealias{Glazebrook_2024}{GB24}

\newcommand{\Lyalpha}{\text{Ly\textalpha}\xspace}
\newcommand{\Halpha}{\text{H\textalpha}\xspace}
\newcommand{\Hbeta}{\text{H\textbeta}\xspace}
\newcommand{\Hgamma}{\text{H\textgamma}\xspace}
\newcommand{\Hdelta}{\text{H\textdelta}\xspace}
\newcommand{\Paalpha}{\text{Pa\textalpha}\xspace}
\newcommand{\Pabeta}{\text{Pa\textbeta}\xspace}
\newcommand{\Hepsilon}{\text{H\textepsilon}\xspace}

\newcommandx{\permittedEL}[6][1=O,2=III,3=,4=,5=,6=]{\text{{#1}\,{\sc {#2}}{#3}{#4}{#5}{#6}}\xspace}
\newcommandx{\semiforbiddenEL}[6][1=O,2=III,3=,4=,5=,6=]{\text{{#1}\,{\sc{#2}}]{#3}{#4}{#5}{#6}}\xspace}
\newcommandx{\forbiddenEL}[6][1=O,2=III,3=,4=,5=,6=]{\text{[{#1}\,{\sc{#2}}]{#3}{#4}{#5}{#6}}\xspace}

\newcommand{\EW}[1]{\text{EW(#1)}\xspace}

\newcommand{\HI}{\permittedEL[H][i]}
\newcommand{\HII}{\permittedEL[H][ii]}

\newcommand{\NV}{\permittedEL[N][v]}
\newcommandx{\NVL}[1][1=1243]{\permittedEL[N][v][\textlambda][#1]}
\newcommandx{\NVall}{\permittedEL[N][v][\textlambda][\textlambda][1239,][1243]}

\newcommandx{\CIIL}[1][1=232x]{\semiforbiddenEL[C][ii][\textlambda][#1]}
\newcommandx{\CIIall}{\semiforbiddenEL[C][ii][\textlambda][\textlambda][2323.5--][2328.1]}

\newcommand{\NIV}{\semiforbiddenEL[N][iv]}
\newcommandx{\NIVL}[1][1=1486]{\semiforbiddenEL[N][iv][\textlambda][#1]}

\newcommand{\CIV}{\permittedEL[C][iv]}
\newcommandx{\CIVL}[1][1=1550]{\permittedEL[C][iv][\textlambda][#1]}
\newcommand{\CIVall}{\permittedEL[C][iv][\textlambda][\textlambda][1549,][1551]}

\newcommand{\HeII}{\permittedEL[He][ii]}
\newcommandx{\HeIIL}[1][1=1640]{\permittedEL[He][ii][\textlambda][#1]}

\newcommand{\semiOIII}{\semiforbiddenEL[O][iii]}
\newcommandx{\semiOIIIL}[1][1=1666]{\semiforbiddenEL[O][iii][\textlambda][#1]}
\newcommand{\semiOIIIall}{\semiforbiddenEL[O][iii][\textlambda][\textlambda][1661,][1666]}

\newcommand{\NIII}{\semiforbiddenEL[N][iii]}
\newcommandx{\NIIIL}[1][1=1750]{\semiforbiddenEL[N][iii][\textlambda][#1]}
\newcommand{\NIIIall}{\semiforbiddenEL[N][iii][\textlambda][\textlambda][1747--][1754]}

\newcommandx{\CIII}{\semiforbiddenEL[C][iii]}
\newcommandx{\CIIIL}[1][1=1909]{\semiforbiddenEL[C][iii][\textlambda][#1]}
\newcommand{\CIIIall}{\semiforbiddenEL[C][iii][\textlambda][\textlambda][1907,][1909]}

\newcommand{\NeIV}{\forbiddenEL[Ne][iv]}
\newcommandx{\NeIVL}[1][1=2424]{\forbiddenEL[Ne][iv][\textlambda][#1]}
\newcommand{\NeIVall}{\forbiddenEL[Ne][iv][\textlambda][\textlambda][2422,][2424]}

\newcommand{\MgII}{\permittedEL[Mg][ii]}
\newcommandx{\MgIIL}[1][1=2803]{\permittedEL[Mg][ii][\textlambda][#1]}
\newcommand{\MgIIall}{\permittedEL[Mg][ii][\textlambda][\textlambda][2796,][2803]}

\newcommand{\NeV}{\forbiddenEL[Ne][v]}
\newcommandx{\NeVL}[1][1=3426]{\forbiddenEL[Ne][v][\textlambda][#1]}
\newcommand{\NeVall}{\forbiddenEL[Ne][v][\textlambda][\textlambda][3346,][3426]}

\newcommand{\OII}{\forbiddenEL[O][ii]}
\newcommandx{\OIIL}[1][1=3727]{\forbiddenEL[O][ii][\textlambda][#1]}
\newcommand{\OIIall}{\forbiddenEL[O][ii][\textlambda][\textlambda][3726,][3729]}

\newcommand{\NeIII}{\forbiddenEL[Ne][iii]}
\newcommandx{\NeIIIL}[1][1=3869]{\forbiddenEL[Ne][iii][\textlambda][#1]}
\newcommand{\NeIIIall}{\forbiddenEL[Ne][iii][\textlambda][\textlambda][3869,][3967]}

\newcommand{\OIII}{\forbiddenEL[O][iii]}
\newcommandx{\OIIIL}[1][1=5007]{\forbiddenEL[O][iii][\textlambda][#1]}
\newcommand{\OIIIall}{\forbiddenEL[O][iii][\textlambda][\textlambda][4959,][5007]}

\newcommandx{\NIL}[1][1=5200]{\forbiddenEL[N][i][\textlambda][#1]}
\newcommand{\NIall}{\forbiddenEL[N][i][\textlambda][\textlambda][5198,][5200]}

\newcommand{\OI}{\forbiddenEL[O][i]}
\newcommandx{\OIL}[1][1=6300]{\forbiddenEL[O][i][\textlambda][#1]}
\newcommand{\OIall}{\forbiddenEL[O][i][\textlambda][\textlambda][6300,][6363]}

\newcommand{\HeI}{\permittedEL[He][i]}

\newcommand{\NII}{\forbiddenEL[N][ii]}
\newcommandx{\NIIL}[1][1=6584]{\forbiddenEL[N][ii][\textlambda][#1]}
\newcommand{\NIIall}{\forbiddenEL[N][ii][\textlambda][\textlambda][6549,][6584]}

\newcommand{\SII}{\forbiddenEL[S][ii]}
\newcommand{\SIIL}[1][1=6717]{\forbiddenEL[S][ii][\textlambda][#1]}
\newcommand{\SIIall}{\forbiddenEL[S][ii][\textlambda][\textlambda][6717,][6730]}

\newcommandx{\OIIAuL}[1][1=7325]{\forbiddenEL[O][ii][\textlambda][#1]}
\newcommand{\OIIAuall}{\forbiddenEL[O][ii][\textlambda][\textlambda][7319--][7332]}

\newcommandx{\CIIFIRL}{\forbiddenEL[C][ii][\textlambda][158\,\mum]}

\newcommand{\hda}{\ensuremath{\mathrm{H\text{\textdelta}_A}}\xspace}
\newcommand{\hga}{\ensuremath{\mathrm{H\text{\textgamma}_A}}\xspace}

\title[Age dating early galaxies]{Age-dating early quiescent galaxies: high star-formation efficiency, but consistent with direct, higher-redshift observations}

\author[Turner, Tacchella, D'Eugenio et al.]{\parbox{\textwidth}{
\orcidsymb{Crispin Turner}{0009-0002-0520-1189}$^{\hyperlink{aff1}{1},\hyperlink{aff2}{2}}$\thanks{E-mail: cyhft2@cantab.ac.uk},
\orcidsymb{Sandro Tacchella}{0000-0002-8224-4505}$^{\hyperlink{aff1}{1},\hyperlink{aff2}{2}}$\thanks{E-mail: st578@cam.ac.uk},
\orcidsymb{Francesco D'Eugenio}{0000-0003-2388-8172}$^{\hyperlink{aff1}{1},\hyperlink{aff2}{2}}$\thanks{E-mail: fd391@cam.ac.uk},
\orcidsymb{Stefano Carniani}{0000-0002-6719-380X}$^{\hyperlink{aff3}{3}}$,
\orcidsymb{Mirko Curti}{0000-0002-2678-2560}$^{\hyperlink{aff4}{4}}$,
\orcidsymb{Karl Glazebrook}{0000-0002-3254-9044}$^{\hyperlink{aff5}{5}}$,
\orcidsymb{Benjamin D. Johnson}{0000-0002-9280-7594}$^{\hyperlink{aff6}{6}}$,
\orcidsymb{Seunghwan Lim}{0000-0002-3642-2446}$^{\hyperlink{aff1}{1},\hyperlink{aff2}{2}}$,
\orcidsymb{Tobias Looser}{0000-0002-3642-2446}$^{\hyperlink{aff1}{1},\hyperlink{aff2}{2}}$,
\orcidsymb{Roberto Maiolino}{0000-0002-4985-3819}$^{\hyperlink{aff1}{1},\hyperlink{aff2}{2}}$,
\orcidsymb{Themiya Nanayakkara}{0000-0003-2804-0648}$^{\hyperlink{aff5}{5}}$,
\orcidsymb{Jenny Wan}{0000-0001-8872-4991}$^{\hyperlink{aff7}{7}, \hyperlink{aff8}{8}}$
}\vspace{0.4cm}
\\
\parbox{\textwidth}{
\hypertarget{aff1}{$^{1}$}Kavli Institute for Cosmology, University of Cambridge, Madingley Road, Cambridge, CB3 0HA, United Kingdom\\
\hypertarget{aff2}{$^{2}$}Cavendish Laboratory - Astrophysics Group, University of Cambridge, 19 JJ Thomson Avenue, Cambridge, CB3 0HE, United Kingdom\\
\hypertarget{aff3}{$^{3}$}Scuola Normale Superiore, Piazza dei Cavalieri 7, 56126 Pisa, Italy\\
\hypertarget{aff4}{$^{4}$}European Southern Observatory, Karl-Schwarzschild-Strasse 2, 85748 Garching, Germany\\
\hypertarget{aff5}{$^{5}$}Centre for Astrophysics and Supercomputing, Swinburne University of Technology, P.O. Box 218, Hawthorn, 3122, VIC, Australia\\
\hypertarget{aff6}{$^{6}$}Center for Astrophysics $|$ Harvard \& Smithsonian, 60 Garden Street, Cambridge, MA 02138, USA\\\hypertarget{aff7}{$^{7}$}Department of Physics, Stanford University, 382 Via Pueblo Mall, Stanford, CA 94305, USA\\
\hypertarget{aff8}{$^{8}$}Kavli Institute for Particle Astrophysics and Cosmology, P.O. Box 2450, Stanford University, Stanford, CA 94305, USA\\
}
}

\date{Accepted XXX. Received YYY; in original form ZZZ}

\pubyear{2024}

\begin{document}
\label{firstpage}
\pagerange{\pageref{firstpage}--\pageref{lastpage}}
\maketitle

\begin{abstract}
We present a detailed analysis of JWST/NIRSpec and NIRCam observations of ZF-UDS-7329, a massive, quiescent galaxy at redshift $z=3.2$, which has been put forward to challenge cosmology and galaxy formation physics. We study on the impact of different star formation history (SFH) priors, stellar libraries, metallicity, and initial mass function assumptions. Our results show that ZF-UDS-7329, with a formed stellar mass of $M_{\star} \approx 10^{11.4}~\Msun$ (surviving mass $M_{\star\mathrm{,surv}} \approx 10^{11.2}~\Msun$) and a specific star-formation rate of $\mathrm{sSFR} \approx 0.03~\Gyr^{-1}$, formed efficiently in the first billion years of the Universe. In agreement with previous work, we find that the spectrum is consistent with mass-weighted stellar ages of $1.3-1.8$ Gyr, depending on the SFH prior used. A physically motivated rising SFH prior makes the formation history of ZF-UDS-7329 compatible with stellar mass and star-formation rate estimates of $z>6$ galaxies. Using NIRCam imaging, we identify a colour gradient indicative of an old, quiescent bulge and a younger disc component, as expected from a complex formation history. The inferred SFH is consistent a high stellar fraction of $f_{\star}=M_{\star}/(f_b \cdot M_{\rm h}) \approx 100\%$ at $z=7-12$, implying an extremely high integrated star-formation efficiency. However, when considering cosmic variance and possible mergers as expected in over-dense environments --  as traced by ZF-UDS-7329 -- the stellar fractions could be reduced to $f_{\star} \approx 50\%$, which is more consistent with galaxy formation models and the stellar-to-halo mass relation at lower redshifts. We conclude that ZF-UDS-7329 forms extremely efficient in the early universe, but does not necessitate unseen galaxies at higher redshifts since the inferred SFR of ancestors are consistent with those seen in $z>6$ galaxies.
\end{abstract}


\section{Introduction}

The sensitivity and near-infrared wavelength coverage of JWST enables us to trace the galaxy population in the first 500 Myr (redshift $z>10$). After two years of observation, numerous high-redshift dropouts have been discovered \citep{castellano22, naidu22_highz, finkelstein23_ceers, hainline24, adams24}, with seven spectroscopically confirmed at $z\gtrsim12$ \citep{curtis-lake23, robertson23, deugenio24_C, fujimoto23_zspec, wang23, castellano24, carniani24_z14}. Results show that the bright-end of the ultra-violet (UV) luminosity function only weakly evolves with redshift \citep{finkelstein24, harikane23_uvlf, donnan24, robertson24}, thereby challenging theoretical models -- such as numerical simulations \citep{dave19, vogelsberger20, shen20, kannan23, wilkins23_frontier}, semi-analytical galaxy formation \citep{dayal14, cowley18, yung19_uvlf}, or empirical models \citep{tacchella13, tacchella18, mason15, sun16, behroozi20}. 

The surprisingly high number density of galaxies at $z>10$, including the extreme UV brightness of individual systems, has sparked a range of theoretical ideas and developments. One way to overcome this tension between observations and theory is to modify cosmology, i.e., not assume $\Lambda$CDM, to boost the number density of dark matter haloes. This can be achieved by enhancing the matter power spectrum \citep{sabti24} or increasing the expansion of the Universe at early times with ``Early Dark Energy'' \citep{shen24_ede}. In contrast, other solutions focus on modifying the baryonic physics of galaxy formation, which has mostly been calibrated with lower-redshift observations, and hence may need some refinement for the low-metallicity and dense environment of cosmic dawn. Such ideas include a higher star-formation efficiency \citep{dekel23, li24_fbb}, a top-heavy stellar initial mass (IMF) function \citep{inayoshi22, cueto24, trinca24, ventura24, yung24}, and a UV contribution from accreting black holes \citep{inayoshi22, trinca24, hegde24}. In addition to boosting the UV luminosity of the whole galaxy population, scatter of the UV luminosity at fixed halo mass can have a drastic effect on the bright end of the UV luminosity function because of the steepness of the dark matter halo mass function at these early times \citep{mason23, mirocha23, shen23, sun23_bursty, gelli24, kravtsov24}. 

These different ideas and scenarios can be tested. Current JWST observations of $z>10$ galaxies reveal a great diversity of stellar populations and morphologies. Galaxies such as GN-z11 probably host an active galactic nucleus (AGN; \citealt{maiolino24_gnz11}) and a significant fraction ($>50\%$) of the UV output stems from the central point source \citep{tacchella23}. GN-z11, after removing the point source, has a stellar mass of $M_{\star}\approx10^{9}~\Msun$ and a star-formation rate (SFR) of $\mathrm{SFR}\approx10~\Msun~\mathrm{yr}^{-1}$. However, not all bright, $z>10$ galaxies are AGN: JADES-GS-z14-0, the most distant spectroscopically confirmed galaxy at $z=14.3$, is unexpectedly luminous, is spatially resolved with a half-light radius of 260 pc, and consistent with being dominated by stellar continuum emission \citep{carniani24_z14, helton24}. This object, with a stellar mass of $M_{\star}\approx10^{8.7}~M_{\odot}$, has a $\mathrm{SFR}\approx22~M_{\odot}~\mathrm{yr}^{-1}$, which is an order of magnitude higher than the dark matter accretion rate of typical haloes at this redshift \citep{tacchella18}. The less luminous galaxies at $z>10$ are typically doubling their stellar mass every few tens of Myr and are morphologically compact, with sizes of the order of 100 pc, implying high stellar mass and SFR surface densities \citep{robertson23, morishita24, wang23}. While these observations directly probe the earliest galaxies known, they are probably biased towards the brightest systems and cover only the rest-frame UV emission, giving us a limited view on the stellar populations of those galaxies (but see \citealt{helton24} and \citealt{zavala24}).

A complementary approach to probe the first generation of galaxies is via archaeological lookback studies. JWST enables us to probe the crucial rest-frame optical wavelength range out to $z\approx9$. While star-forming galaxies are abundant, extracting lookback information such as the stellar mass, the stellar age and the star-formation history (SFH) is extremely challenging, because the bursty SFHs of galaxies outshine older stellar populations \citep{papovich23, tacchella23_metal, endsley24, witten24}. In the case where young stars dominate the emission and older stellar populations do not leave any spectral signatures, the priors related to the SFHs play a crucial role \citep{tacchella22_highz, whitler23_sfh}.  Furthermore, AGN introduce significant uncertainties in both stellar mass and SFH estimations because of unknown contribution of the AGN to continuum emission \citep{kocevski23, kocevski24, wang24_sp, williams24}.

Contrarily, quiescent galaxies are less affected by outshining and offer a detailed view of the early Universe. Observing quiescent galaxies at $z>3$ is challenging, as older stellar populations are both rarer and fainter, and a signal-to-noise ratio (S/N) of $\approx20$ per \AA in the spectral continuum is required to provide meaningful constraints on early star formation, metallicity and chemical abundance \citep{conroy13_rev, nanayakkara22}. JWST provided new constraints on quiescent galaxies at cosmic noon, $z\approx1-3$, highlighting the importance of supermassive black holes in rapidly suppressing star formation (``quenching'') in massive galaxies by efficiently ejecting gas \citep{belli24, deugenio24_psb, beverage24, park24_quench, scholtz24,davies24}. Quiescent galaxies have also been probed at earlier times, both at low stellar masses \citep{looser24, strait23} and high stellar masses \citep{carnall23, de-graaff24, glazebrook24, nanayakkara24, setton24}.

A particularly notable example of a massive quiescent galaxy at $z=3.2$ is ZF-UDS-7329. This galaxy was initially selected by \citet{schreiber18} as a quiescent candidate using the UVJ colour-colour diagram \citep[e.g.,][]{williams09}. Using JWST spectroscopy, \citet{glazebrook24} confirmed this galaxy to be quiescent and inferred a stellar mass $\mstar > 10^{11}~\Msun$ forming at $z \gtrsim 11$. Such an extreme SFH challenges our understanding of early stellar populations, galaxy formation and/or the nature of dark matter, and may point to the presence of undetected populations of early galaxies. \citet{carnall24} obtained three and analysed two medium resolution grating/filter combinations of ZF-UDS-7329, emphasising that this galaxy aligns with the most massive galaxies expected under the assumption of 100 per cent conversion of baryons to stars. Their results suggest extreme galaxy formation physics during the first billion years, but no conflict with $\Lambda$CDM cosmology.

In this paper, we re-reduce and analyse the JWST data of ZF-UDS-7329 (Section~\ref{datac}). Specifically, we extend the analysis of \citet{glazebrook24} by studying the impact of different SFH priors, stellar libraries, metallicity and initial mass function (IMF) assumptions (Section \ref{prospectorm}). Using NIRCam imaging data, we show clear indication for a colour gradient in ZF-UDS-7329, which is consistent with an old, quiescent bulge component and a younger disc component (Section \ref{morphology}). Our results highlight efficient star formation in the first billion years, though uncertainties remain regarding self-consistent modelling of $\alpha$-enhanced stellar population models, mergers diluting the archaeological signal and cosmic variance (Section \ref{discussion}). We conclude in Section \ref{conclusion}. 

Throughout this work, we assume a Planck18 flat $\Lambda$CDM cosmology, with $\Omega_m = 0.315$, $H_0 = 67.4$~\kms, and a baryon fraction of $f_b=\Omega_b/\Omega_m=0.156$ \citep{planck-collaboration20}. We assume a Kroupa IMF \citep{kroupa01} and a solar metallicity $\mathrm{Z}_\odot = 0.014$ \citep{asplund09}.

\section{Observations and data}\label{datac}

\begin{figure}
\includegraphics[width=\columnwidth]{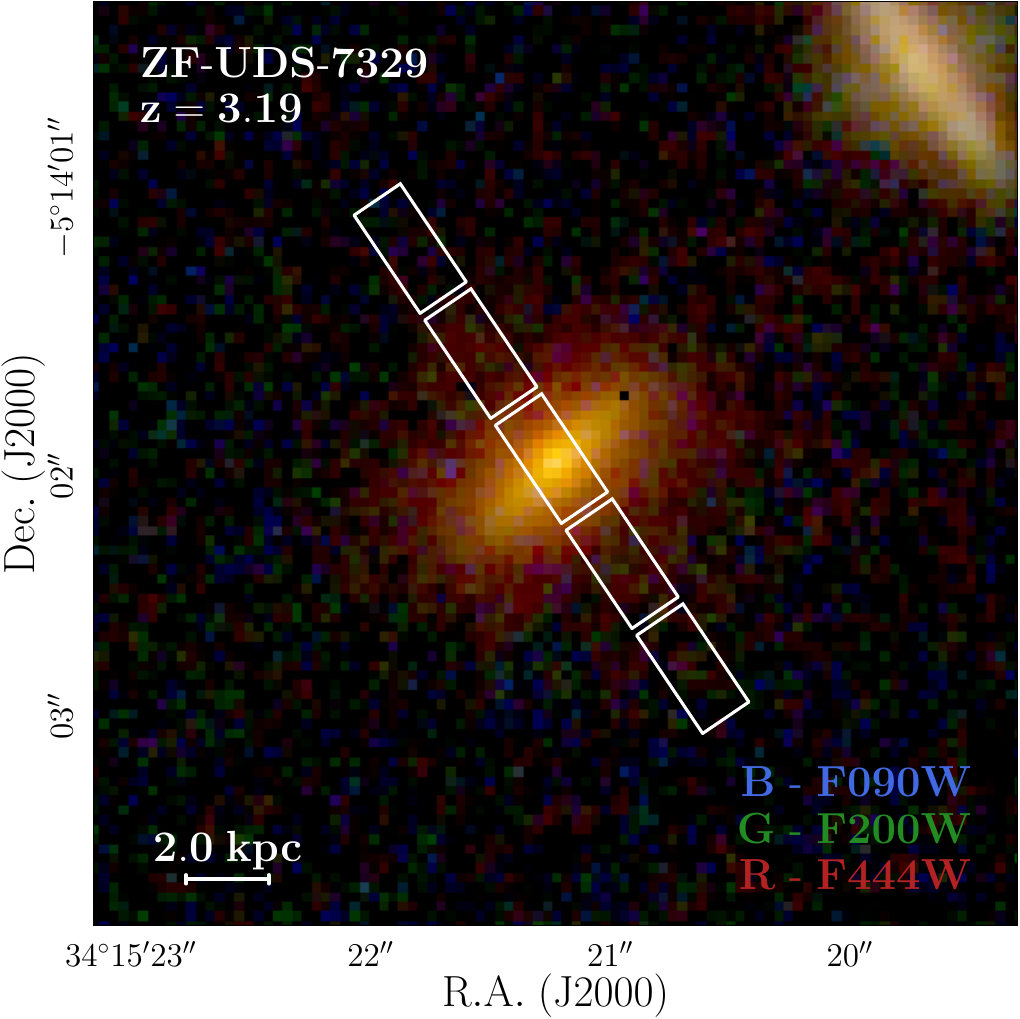}
       \caption{Colour composite image of ZF-UDS-7329 from NIRCam. The galaxy appears flattened, suggesting a nearly edge-on disc morphology (Section~\ref{morphology}). The red/green/blue colours are F444W/F200W/F090W. The white rectangles mark the shutters of the NIRSpec/PRISM observations.}\label{fig.RGB}
\end{figure}

\subsection{NIRCam and NIRSpec observations}

ZF-UDS-7329 was observed with the JWST NIRSpec instrument \citep{jakobsen22} in multi-object mode using the MSA \citep[Micro-Shutter Assembly;][]{ferruit22}, as part of Program \#2565 as described in \citet{glazebrook24}. The MSA target ID is 12629 \citep[which is also the 3D-HST ID;][]{brammer12,skelton14,momcheva16}. The MSA was configured with 5-shutter slitlets and three nodded exposures. Each exposure consisted of a single integration of 45 groups in NRSIRS2RAPID mode \citep{rauscher07}, giving 667~s per exposure and 2001~s total observing time. The disperser used was the prism, which produced spectral data covering wavelengths $0.6\text{--}5.2$ \mum at a resolving power of $\sim 50\text{--}350$ \citep{jakobsen22}.

In addition to spectroscopy, we use JWST NIRCam data from the public PRIMER survey\footnote{\href{https://primer-jwst.github.io/}{https://primer-jwst.github.io/}} (Program \#1837; PI J. Dunlop). These data consist of imaging in 8 NIRCam bands across $1.0\text{--}5.0$ \mum (F090W, F115W, F150W, F200W, F277W, F356W, F410M and F444W). A composite image using three of these is displayed in Fig.~\ref{fig.RGB}. The flux measurements in the NIRCam bands are from \citet{glazebrook24}. In addition, we use cutouts obtained from the DAWN JWST Archive\footnote{\href{https://dawn-cph.github.io/dja/index.html}{https://dawn-cph.github.io/dja/index.html}.} \citep[DJA;][]{valentino23}.

\subsection{NIRSpec reduction}

To assess the impact of possible systematics in the data reduction on the measurements, we compared the data reduction from two different pipelines. The spectrum from \citet[][$f^{\mathrm{K}}_{\lambda}$ in Fig.~\ref{fig.compare}]{glazebrook24} was reduced with the publicly available \textsc{jwst} pipeline v1.12.5 and the Calibration Reference Data System (CRDS) context 1149 \citep[see][]{nanayakkara24}. As an alternative, we use the GTO pipeline (Carniani in~prep.; $f^{\mathrm{C}}_{\lambda}$ in Fig.~\ref{fig.compare}), described in \citet{curti24_Z}. The reduced spectra are compared in Fig.~ \ref{fig.compare} (see Appendix \ref{a.red}), alongside the NIRCam data highlighted with red circles. \citet{nanayakkara24} calibrated the $f^{\mathrm{K}}_{\lambda}$ spectrum to match the shape and normalisation of the photometry by fitting a 2\textsuperscript{nd}-order polynomial. In the case of $f^{\mathrm{C}}_{\lambda}$, we applied a uniform scaling to match the photometry. The difference in the spectral shape of the rescaling is chiefly responsible for $f^{\mathrm{K}}_{\lambda}$ appearing slightly bluer, on the order of 0.05 mags in the F200W$-$F356W colour, which is discernible from the plot of $\chi_{\mathrm{eff}} := |f^{\mathrm{K}}_{\lambda} - f^{\mathrm{C}}_{\lambda}| / \sqrt{\sigma_{\mathrm{C}}^2 + \sigma_{\mathrm{K}}^2}$. It should be noted that since these are not actually independent measurements, $\chi_{\mathrm{eff}}$ merely quantifies the discrepancy between the reductions relative to the estimates of statistical uncertainty, $\sigma_{\mathrm{C}}$ and $\sigma_{\mathrm{K}}$.

Even without any spectral calibration, the scaled $f^{\mathrm{C}}_{\lambda}$ is reasonably consistent with the photometry. Both spectra, however, contain artefacts larger than the estimated noise, $\sigma$, such as the prominent bumps at $\sim 6800$ \AA rest-frame, that do not correspond to any known emission features. In all subsequent fitting, we used the reduction from \citet{curti24_Z}, with the spectrum and photometry S/N capped to 20. Since ZF-UDS-7329 is neither a point source, nor covers the slit angle, the line-spread function (LSF) of \citet{glazebrook24} was chosen. This was derived using the NIRCam images and NIRSpec/MSA instrument model to estimate the wavelength-dependent LSF, parameterised by its Gaussian full-width half maximum \citep[FWHM;][]{de-graaff24_kin}. The wavelength-dependent spectral resolution ranges from 4,000 km s$^{-1}$ (around 1~\mum) to 400 km s$^{-1}$ (5~\mum). We note that the overall normalisation and large-scale wavelength dependence of the spectrum is not important in our analysis because we use a polynomial to match the spectrum to the photometry, see Section~\ref{prospectormodel}.

\subsection{Preliminary comments}

The spectrum (see Fig.~ \ref{fig.MAPs} and \ref{fig.compare}) displays a prominent 4000-\AA break that is typical for old, $> 1 $ Gyr stellar populations, for which ionised metal absorption lines and an absence of hot, blue stars suppress the strong Balmer break seen in A-type stellar spectra. This is in contrast to so-called ``post-starburst'' quiescent galaxies, which have recently quenched their star formation (within $\sim500 \, \Myr$ prior to observation) following a strong burst \citetext{e.g., \citealp{deugenio23_cii, looser23_pop}}. An old age is also consistent with the strength of the Mg\,\textsc{i}b 5175\AA triplet and TiO absorption bands at $\sim$ 7150 \AA and 9400 \AA, which may indicate some degree of $\alpha$-element enhancement. Particularly notable is the strong Na D 5890 \AA doublet absorption line, hypothesised to be either a result of cold interstellar medium (ISM) or an enhanced Na/Fe abundance \citep{jeong13}\footnote{The high rest-frame equivalent width (EW) of Na~D suggests an ISM rather than a stellar origin (10~\AA; Nanayakkara T., priv. comm.)}. No prominent emission lines are visible, as expected for a galaxy with little ongoing star formation.


\begin{table*}
    \begin{center}
    \caption{The parameters of the fiducial \prospector model with associated priors and posteriors.
    }\label{t.prosp}
    \setlength{\tabcolsep}{4pt}
    \begin{tabular}{llcllcc}
  \toprule
   & Parameter & Free & Description & Prior & Posterior (MILES) & Posterior (C3K)\\
   & (1)       & (2)  & (3)         & (4)   & (5) & (6)\\
   \midrule
   \multirow{5}{*}{\rotatebox[origin=c]{90}{Stellar}}
   & $z_\mathrm{obs}$ & Y & Redshift & $\mathcal{U}(3.0,3.4)$ & $3.200^{+0.001}_{-0.001}$ & $3.192^{+0.001}_{-0.001}$ \\
   & $\log_{10}(\mstar/\Msun)$ & Y & Total stellar mass formed & $\mathcal{U}(7, 12)$ & $11.34^{+0.01}_{-0.02}$ & $11.37^{+0.02}_{-0.02}$\\
   & $\log_{10}(Z/\Zsun)$ & Y & Stellar metallicity & $\mathcal{U}(-2, 0.19)$ & $0.03^{+0.10}_{-0.10}$ & $0.11^{+0.05}_{-0.06}$ \\
   & $\log_{10}(\mathrm{SFR\,ratios})$ & Y & Ratios of the SFRs of adjacent bins of the SFH & $\mathcal{T}(0, 0.3, 2)$ & --- & --- \\
   & $t_{50} [\Gyr]$ & N & Median mass-weighted age & --- & $1.63^{+0.02}_{-0.03}$ & $1.61^{+0.04}_{-0.04}$ \\

   \multirow{4}{*}{\rotatebox[origin=c]{90}{Dust}}
   & $\tau_{2,V}$ & Y & Optical depth of the diffuse dust & $\mathcal{G}(0.3,1;0,2)$ & $0.15^{+0.07}_{-0.06}$ & $0.27^{+0.04}_{-0.04}$ \\
   & $\tau_{1,V}/\tau_{2,V}$ & Y & Ratio of the birth cloud and diffuse optical depths & $\mathcal{G}(1.0,0.3,0,2.0)$ & $1.16^{+0.13}_{-0.11}$ & $1.09^{+0.20}_{-0.22}$ \\
   & $n$ & Y & Slope index of the \citet{kriek13} dust attenuation curve & $\mathcal{U}(-1.0,0.4)$ & $-0.21^{+0.17}_{-0.18}$ & $-0.24^{+0.16}_{-0.17}$\\
   & $t_{\mathrm{esc}} [\Myr]$ & N & Age below which stars are attenuated by $\tau_{1,V}$ & $10 $ & --- & ---\\
   \multirow{3}{*}{\rotatebox[origin=c]{90}{Nebulae}}
   & $\sigma_\mathrm{gas} \; [\kms]$ & Y & Gas intrinsic velocity dispersion & $\mathcal{U}(30,550)$ & $155^{+76}_{-69}$ & $139^{+86}_{-63}$\\
   & $\log_{10}(Z_\mathrm{gas}/\Zsun)$ & Y & Nebular metallicity & $\mathcal{U}(-2, 0.19)$ & $-1.5^{+0.5}_{-0.3}$ &$-0.3^{+0.5}_{-0.6}$\\
   & $\log_{10}U$ & Y & Gas ionisation parameter & $\mathcal{U}(-4, -1)$ & $-3.4^{+0.3}_{-0.3}$ &$-2.8^{+0.4}_{-0.4}$\\
   \midrule
   \multirow{3}{*}{\rotatebox[origin=c]{90}{Other}}
   & $j_\mathrm{spec}$ & Y & Multiplicative noise inflation factor for spectral data & $\mathcal{U}(0.5,5.0)$ & $2.4^{+0.7}_{-0.7}$ &$3.4^{+0.9}_{-0.9}$\\
   & $f_\mathrm{outlier}$ & Y & Fraction of outlier data points & $\mathcal{U}(10^{-5},10^{-2})$ & $0.003^{+0.001}_{-0.001}$ & $0.004^{+0.002}_{-0.002}$\\
   & $f_\mathrm{norm}$ & Y & Normalisation factor of the whole spectrum & $\mathcal{N}(1.0,0.1)$ & $1.03^{+0.07}_{-0.07}$ & $0.92^{+0.06}_{-0.07}$\\
  \bottomrule
  \end{tabular}
  \end{center}
(1) Parameter name and units (where applicable). (2) Parameters marked with `Y' are optimised by \prospector; parameters marked with `N' are fixed in the model (with value in Column~4), or are calculated after the fit from the posterior distributions (in this case, Column~4 is empty). (3) Parameter description. (4) Parameter prior probability distribution.
$\mathcal{N}(\mu, \sigma)$ is a Gaussian distribution with mean $\mu$ and standard deviation $\sigma$; $\mathcal{U}(a, b)$ is a uniform distribution between $a$ and $b$; $\mathcal{T}(\mu, \sigma, \nu)$ is a Student's $t$ distribution with mean $\mu$, dispersion $\sigma$ and $\nu$ degrees of freedom.
$\mathcal{G}(\mu, \sigma, a, b)$ is a clipped Gaussian distribution with mean $\mu$ and dispersion $\sigma$, between $a$ and $b$.
(5) Median and the 16\textsuperscript{th}--84\textsuperscript{th} percentiles of the marginalised posterior distribution with the MILES stellar library. (6) Same as previous but using the C3K stellar library.  For brevity we do not show the posteriors for the SFR ratios.
\end{table*}


\section{Stellar population analysis}\label{prospectorm}

\subsection{Prospector setup}\label{prospectormodel}

The spectrum and photometry are simultaneously fit with the Bayesian code \prospector \citep{johnson19, johnson21}. \prospector uses a Python implementation \citep{speagle20, koposov22} of the Dynamic Nested Sampling algorithm \citep{higson18} in which the posterior probability distribution is estimated by Monte Carlo sampling over nested shells bounded by iso-likelihood contours. The stellar population synthesis model is based on \textsc{fsps} code \citep[Felixible Stellar Population Synthesis ;][]{conroy09a, conroy10_code}, which allows for a great deal of flexibility in the inputted spectral templates, isochrones, dust laws, and emission models. We explore both the empirical MILES \citep{sanchez-blazquez06} or the synthetic C3K \citep{conroy19} stellar libraries, in conjunction with the MIST isochrones \citep{choi16, dotter16}. In the former case, the limited resolution of the templates means that only the $0.8\text{--}3.0$ \mum observed range ($0.2\text{--}0.7$ \mum in rest-frame) of the SED is fitted. 

Within the \prospector framework, we fit a model with 20 free parameters that describe the stellar populations of ZF-UDS-7329 in conjunction with dust and nebular effects. Table \ref{t.prosp} summarises all free parameters of the fiducial model, alongside their priors and resulting posterior probabilities.

We use a flexible (sometimes referred to as non-parametric) ``continuity'' SFH prior, which penalises drastic changes in SFR by fitting for the ratios of the SFRs between adjacent time bins \citep{leja19_nonparm}. Specifically, while no parametric form for the SFH as a function of time is assumed, time bins and a prior still need to be specified. As done in previous works \citep{leja19_nonparm, tacchella22_quench, wan24}, we assume the continuity prior, which assumes a Student-t distribution for the log ratio of the SFR in adjacent time bins, with $\nu=2$ degrees of freedom and a scale of $\sigma=0.3$. This choice was originally motivated by the SFH in the Illustris simulations \citep{leja19_nonparm}. The scale $\sigma$, which we do not vary in this work, controls the amount of variable star formation, and has been increased for high-$z$ galaxies in some works to account for possible bursty SFHs \citep{tacchella22_highz}. Importantly, this prior enforces a constant expectation value for the SFH, something we will address in Section~\ref{sfh}. Furthermore, in this fiducial model, no star formation is allowed before $z=20$, leaving nine bins that are approximately logarithmically-spaced except between lookback times $1\text{--}1.75$ Gyr, to properly characterise the formation period while capturing any recent star formation. 

Dust attenuation is implemented using the two-component model proposed in \citet{charlot00} and \citet{noll09}, in which the attenuation due to birth-clouds and ambient ISM are approximated as two separate screens. The diffuse component follows an attenuation curve with a variable slope index, $n$, and a UV dust absorption bump,
\begin{equation}
    \tau_{\mathrm{2}}(\lambda) = \frac{\tau_{\mathrm{2,V}}}{4.05}(k(\lambda)+D(\lambda)) \left( \frac{\lambda}{5500 \, \AA} \right) ^ n,
\end{equation}
where $\tau_{\mathrm{2,V}}$ is the effective V-band optical depth, and $k(\lambda)$ is the Calzetti curve \citep{calzetti00}, which has total-to-selective ratio $A_{\mathrm{V}} / E(B-V) = 4.05$. $D(\lambda)$ is a Lorentzian-like Drude profile approximating the UV bump, for which we tie the strength to the best-fit value of $n$ according to the results of \citet{kriek13}. We adopt uniform, and clipped-Gaussian priors on $n$, and $\tau_{\mathrm{2,V}}$, respectively. The birth-cloud component, $\tau_{\mathrm{1}}(\lambda)$, only attenuates nebular and stellar emission from stars that are younger than $10~\Myr$, and follows a simple powerlaw:
\begin{equation}\label{dust1}
        \tau_{\mathrm{1}}(\lambda) = \tau_{\mathrm{1,V}} \left( \frac{\lambda}{5500 \, \AA} \right) ^ {-1}.
\end{equation}
We fit for the ratio $\tau_{\mathrm{1,V}}/\tau_{\mathrm{2,V}}$ with a clipped Gaussian prior. Dust emission is typically negligible at rest-frame $\lambda \lesssim 10 \, \mum$, hence it is omitted from all the models.

Nebular emission is included using the default grids of \fsps \citep{byler19}, which are based on the simulation code \cloudy \citep{ferland13}. This is controlled by three parameters, the gas ionisation parameter, $U$; the gas metallicity, $Z_{\mathrm{gas}}$; and the gas velocity dispersion $\sigma_{\mathrm{gas}}$ (which for us is a nuisance parameter). Additional flexibility, by the means of directly fitting for, and marginalising over, emission line fluxes, was not considered given the lack of any observable emission in the data. As our templates cannot reproduce the strong Na D absorption, this line is masked from the fitting procedure. 

An outlier model and noise jitter term are also included to mitigate bad pixels, possible data-model mismatches, and underestimated noise as follows. We fit for a fraction $f_{\mathrm{out}}$ of the data points treated as outliers, which have errors inflated by a factor of 50. The outlier locations are marginalised over as described in Appendix D of \citet{johnson21}. A multiplicative inflation of all the errors by factor $j_\mathrm{spec}$ is fitted for by including kernels for uncorrelated noise in the likelihood calculations.

Since the spectrum, $f^{\mathrm{C}}_{\lambda}$, is not flux-calibrated, we fit a Chebyshev polynomial to match the model continuum to the data at each likelihood call. Specifically, at each likelihood call, we match the model spectrum to the normalization of the spectroscopic data by fitting a polynomial in wavelength to their ratio. The net effect of our approach is that the large-scale continuum shape and normalization of the model are set by the photometry. By fitting a moderate-order polynomial to the ratio between the observed and physical model spectrum at each likelihood call, the spectroscopic calibration model basically removes all information content from the continuum shape of the spectroscopic data. This means that the continuum shape of the observed spectrum does not inform any of the galaxy's physical parameters. Instead, information about physical parameters that affect the continuum shape is derived from photometry, which does not include any multiplicative calibration model. Therefore, there is no degeneracy between the spectroscopic calibration model and the galaxy's parameters, such as the dust content or the star-formation history. Overall, the fitting of this polynomial is found to have only a small effect on the inferred population parameters (typically less than $\sim 5\%$). The order of the polynomial was chosen as five with MILES, or seven with C3K templates, such that it is only sensitive to spectral variations across $\gtrsim 500 \, \AA$.

\begin{figure*}
   \includegraphics[width=\textwidth]{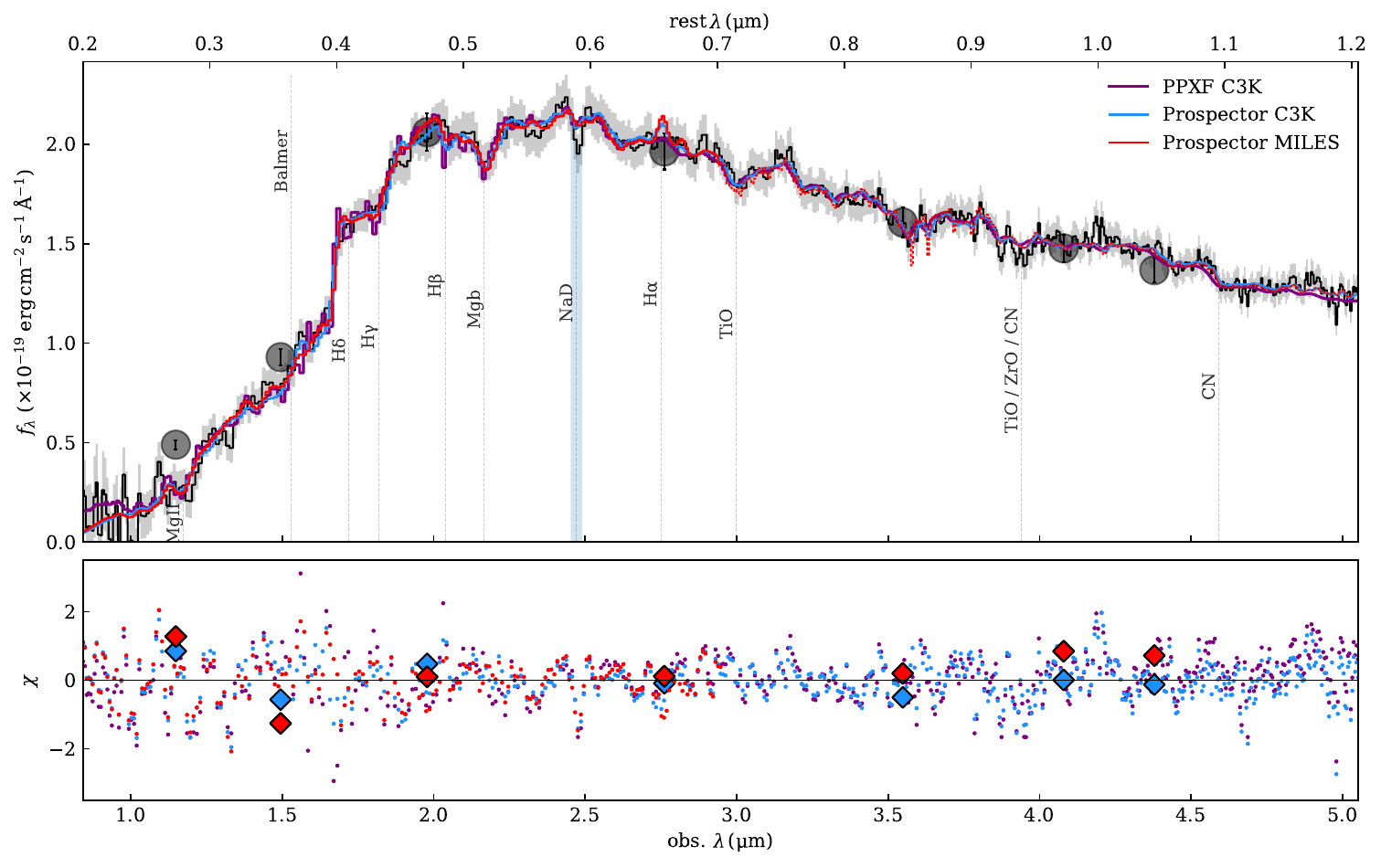}
   \caption{(Top) NIRSpec data for ZF-UDS-7329 in black with statistical errors shaded in grey. NIRCam photometry is included as error bars highlighted with grey circles. Best-fit spectra from \ppxf (purple), and \prospector with the empirical MILES (red) or synthetic C3K (blue) templates. The MILES library resolution is too poor to be fitted beyond $3 \, \mum$ so the model is extrapolated with a dotted red line. Our model fits a polynomial to the model continuum at each likelihood call, such that only the photometric information constrains the overall shape. Pixels in the blue shaded band (the Na D doublet) were masked from the fit. (Bottom) Residuals from each model spectrum as circles, colour-coded accordingly. Photometric residuals are displayed as diamonds.
   }\label{fig.MAPs}
\end{figure*}

\begin{figure*}
   \includegraphics[width=1.0\textwidth]{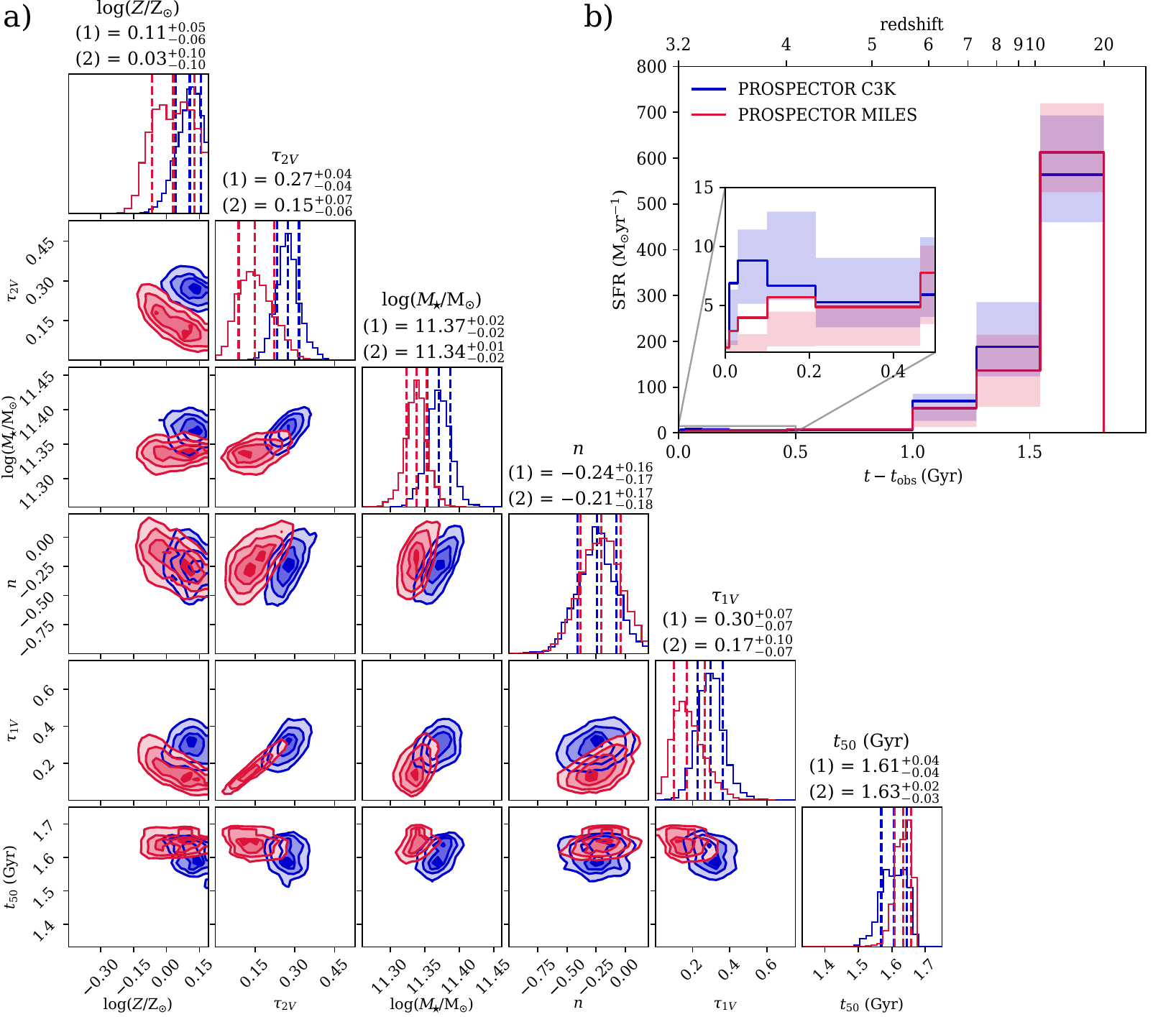}
   \caption{(a) Marginalised posterior distributions for six key parameters (metallicity, diffuse dust optical depth, total formed stellar mass, dust attenuation slope, birth cloud optical depth, and median age) of the \prospector model with MILES (red) or C3K (blue) stellar libraries. Their median values with MILES (1) and C3K (2) are also displayed. (b)  MAP SFHs (SFR versus look-back time) inferred by our \prospector models, colour-coded likewise. No star formation is allowed past $z=20$ in the model. Shaded regions denote the $16\textsuperscript{th}\text{--}84\textsuperscript{th}$ percentiles of the mass formed in each time bin. Both stellar libraries lead to a similar SFH, where most of the mass formed in the earliest SFH bin ($z\approx10-20$), leading to a high stellar age of $t_{\rm 50}\approx1.6$ Gyr. We find that the C3K library prefers a slightly higher metallicity and higher dust attenuation than the MILES library.
   }\label{fig.posteriors}
\end{figure*}

\subsection{Fiducial results}\label{prospectorfiducial}

The maximum a posteriori (MAP) spectra of the fiducial C3K and MILES models are presented in Fig.~ \ref{fig.MAPs} and the posterior distribution of the \prospector model parameters are listed in Table \ref{t.prosp}. Setting $j_\mathrm{spec}=1$, we find reduced $\chi^2$ values of 0.37 and 0.44 for C3K and MILES, respectively, which show good agreement given S/N of $\sim20$ across $\sim200\text{--}500$ spectral pixels with 20 free parameters. We find $j_{\rm spec}=2.5-3.5$, which is possibly caused by data-model mismatches (in addition to what $f_{\rm outlier}$ covers), such as $\alpha$-enhancement or TP-AGB phase contributions (Section~\ref{subsec:alphenc} and \ref{subsec:tpagb}. The extrapolation, without convolution with the LSF, of the MILES best-fit past 3 \mum, is also displayed with a dotted red line (it should be noted that the NIR photometry is still included in the fitting). A corner plot of the posterior distributions of some of the key free parameters is shown in Fig.~ \ref{fig.posteriors}, alongside the inferred SFHs. There are markedly stronger degeneracies between the dust attenuation parameters and stellar metallicity in the MILES posterior, as well as a larger uncertainty in these values overall. We speculate that this is primarily due to the exclusion of the rest-frame NIR spectral data from the fit. The derived dust optical depths, $\tau_{\mathrm{2,V}} \sim 0.2$, are relatively low, with a dust law steeper than Calzetti, $n \approx -0.2$, which is consistent with the galaxy having little recent star formation.

We find a stellar age of $t_{50} \approx 1.6 \, \Gyr$ (corresponding to $z_{50}\approx11$), defined as the look-back time at which half of the stellar mass has formed. The degeneracy of the other parameters with the age is relatively weak. As expected, all the nebular emission parameters are broadly unconstrained. The models preferentially add some weak emission for a small reduction in the overall $\chi^2$, however, the average star-formation rate over the last 100 Myr remains $\lesssim 10 \, \Msun~\mathrm{yr}^{-1}$. This translates to a specific SFR (sSFR) of $\sim 0.03 \, \Gyr^{-1}$ and mass doubling timescale of $\sim 30 \, \Gyr$, which is 15 times larger than the age of the Universe at that epoch ($t_{\rm H} \approx 2 \, \Gyr$), implying that this galaxy is well quiescent. We note that the SFH in Fig.~ \ref{fig.posteriors} shows a downward trend to $\sim2~M_{\odot}/\mathrm{yr}$ in the past 10 Myr, which is caused by the absence of emission line in the spectrum. The SFR in the time period of $10~\mathrm{Myr}< t - t_{\rm obs} < 1~\mathrm{Gyr}$ is generally low ($<10~M_{\odot}/\mathrm{yr}$) and driven by the low UV flux, but less constrained than the SFR in the past 10 Myr due to the degeneracy with the dust attenuation law and stellar metallicity.

The fits also give high formed stellar masses of $\log_{10}{(\mstar/\Msun)} = 11.34^{+0.01}_{-0.02}$, and $11.37^{+0.02}_{-0.02}$, or surviving stellar masses of $\log_{10}{(\msurv/\Msun)} = 11.17^{+0.01}_{-0.02}$, and $11.20^{+0.02}_{-0.02}$, for the MILES and C3K fits, respectively. This assembles in a strong burst with $\sim 70 \%$ of the mass forming between $z=10\text{--}20$ with a $\mathrm{SFR}\approx500-700~\Msun~\mathrm{yr}^{-1}$. We note that the strong constraint on the SFR in the oldest time bin is likely influenced by our decision to disallow any star formation before $z = 20$, which we explore in detail in Section \ref{sfh}.

Both stellar libraries lead to solutions with super-solar metallicities, [$Z$/H] $= 0.03 \pm 0.10$ (MILES), and $0.11 \pm 0.05$ (C3K), which is surprising for a galaxy formed at $z \approx 10-20$. The MILES results are all consistent with the findings of \citet{glazebrook24}. However, the fitting of spectral data beyond 3 \mum seems to push the best fit to adopt higher $Z$ due to the inclusion of metallicity-sensitive absorption bands. This effect is explored in the Section \ref{metallicitymethod}, and strongly implies a degree of $\mathrm{\alpha}$-element enhancement, which we discuss in Section \ref{subsec:alphenc}. \citet{carnall24} find an even higher value of [$Z$/H] $= 0.35^{+0.07}_{-0.08}$, which is beyond the range of our templates, while all other population parameters are relatively consistent with our findings. We also note that the inconsistency in the derived redshifts between MILES and C3K likely arises from a small wavelength calibration issue across the full spectral range \citep[e.g., ][their Section~9.3]{deugenio24}.

Lastly, we corroborate these results with those from the Penalised Pixel-Fitting method \citep[\ppxf;][]{cappellari23}, which proceeds via a $\chi^2$-minimisation process, in lieu of \prospector's Bayesian framework. In this model, following the approach of \citet{looser23_pop}, we use simple stellar population (SSP) templates that are a combination of the synthetic C3K model atmospheres \citep{conroy19} and MIST isochrones \citep{choi16}. Similarly to the Bayesian approach, we find a super-solar metallicity, high stellar-mass and age > 1.5 \Gyr. The full methodology and results are presented in Appendix \ref{ppxfm}, and the MAP spectrum in Fig.~ \ref{fig.MAPs}, alongside the \prospector best fits.

\begin{figure*}          
    \includegraphics[width=1.0\textwidth]{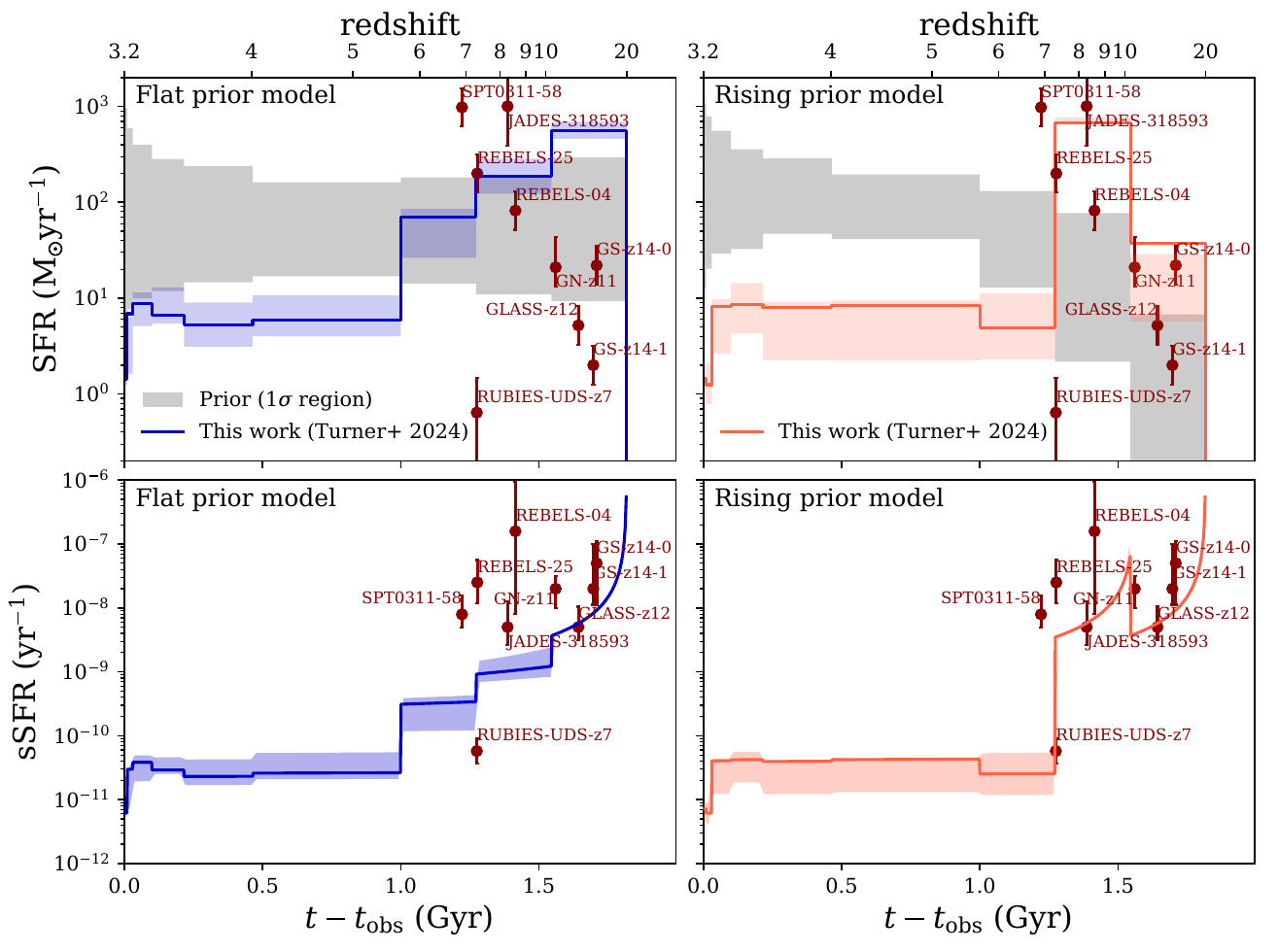}
       \caption{MAP and posterior SFHs for \prospector models with a flat continuity prior (left column) or a rising continuity prior (right column).  (top row) SFRs from our models, with shaded regions denoting the $16\textsuperscript{th}\text{--}84\textsuperscript{th}$ percentiles of the mass formed in each time bin from the fit (blue/orange) or normalised prior (grey). SFRs from several spectroscopically-confirmed galaxies, GN-z11 \citep[$\mstar \approx 10^9~\Msun$,][]{tacchella23}, GLASS-z12 \citep[$\mstar \approx 10^9~\Msun$,][]{castellano24}, GS-z14-0, GS-z14-1 \citep[$\mstar \approx 10^8~\Msun$,][]{carniani24_z14}, REBELS-04 \citep[$\mstar \approx 10^9~\Msun$,][]{bouwens22}, REBELS-25 \citep[$\mstar \approx 10^{10}~\Msun$,][]{bouwens22}, JADES-318593 \citep[$\mstar \approx 10^{11}~\Msun$,][]{simmonds24}, RUBIES-UDS-z7 \citep[$\mstar \approx 10^{10}~\Msun$,][]{weibel24},
       and proto-cluster SPT0311-58 \citep[$\mstar \sim 10^{11} \Msun$,][]{arribas24} are overlaid with red error bars.  On these we assume minimum uncertainties of 0.2 dex \citep[see e.g.][]{conroy13_rev}, and use quantities averaged over 10 \Myr prior to observation where possible.  The rising model is more in-line with such measurements.  (bottom row) The corresponding sSFRs for our models, with the $16\textsuperscript{th}\text{--}84\textsuperscript{th}$ percentiles shaded likewise, and the observations in red.  In both cases, our posterior gives very small uncertainties across the oldest time bin as an artefact of using the non-parametric SFH.  Once again, the rising prior is more consistent with high redshift observations, however our approach likely underestimates the uncertainty across the oldest and second-oldest bins.  
   }\label{fig.priorSFHs}
\end{figure*}

\begin{figure*}
   \includegraphics[width=1.0\textwidth]{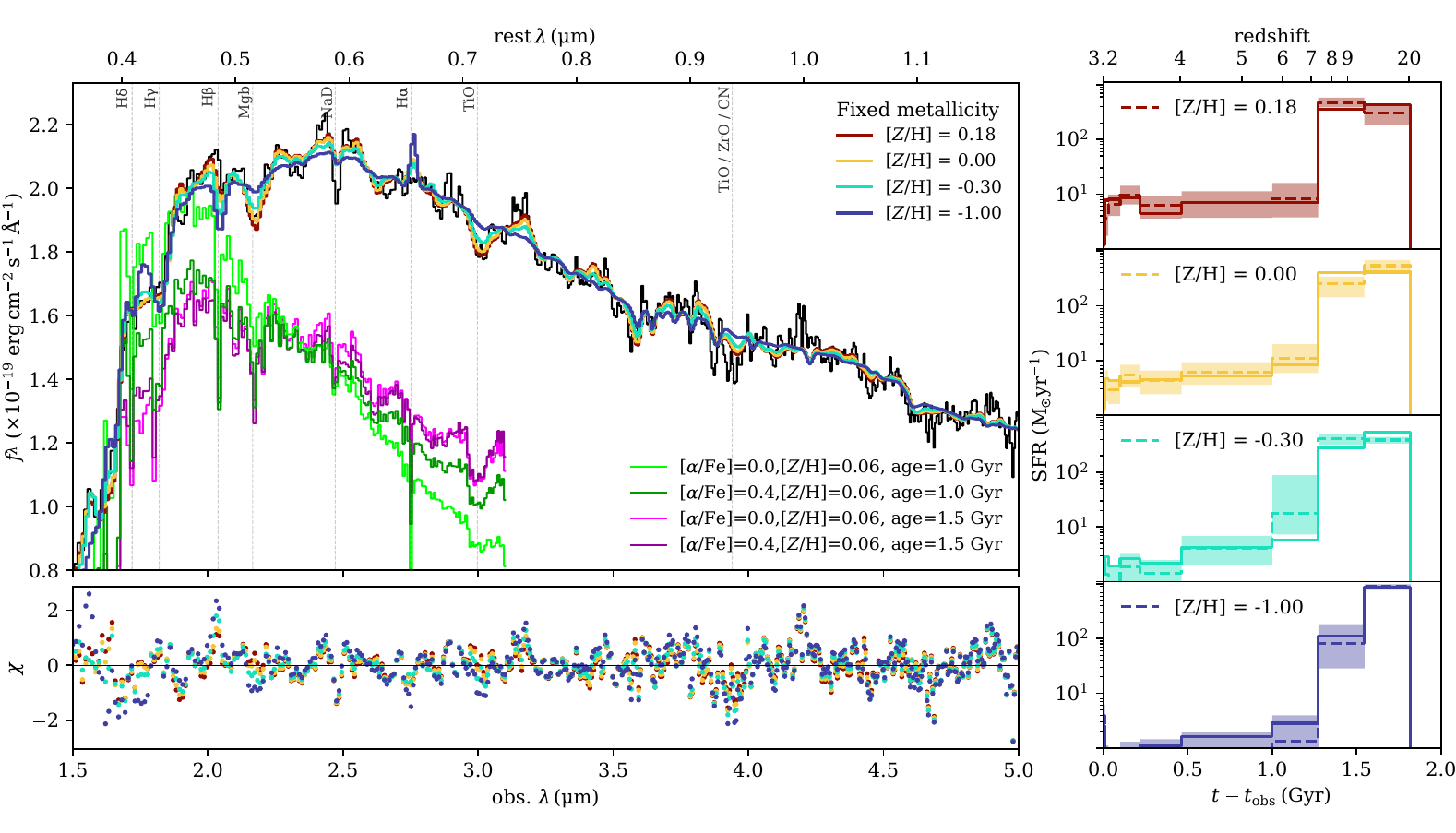}
   \caption{(Top-left) Best-fit spectra for \prospector (C3K) models with fixed metallicities from sub- (blue) to super-solar (red). We overlay without fitting a selection of SSP spectra from the $\alpha$-enhanced MILES \citet{vazdekis15}; these model spectra (vertically offset for clarity) show how $\alpha$-enhancement changes the SED shape at fixed age and total metallicity, in a manner that could increase the apparent age. Green lines indicate SSPs of age $1 \, \Gyr$, and magenta $1.5\, \Gyr$, with darker shades denoting a super-solar [$\alpha$/Fe]. (Bottom-left) Residuals from each of the models. The sub-solar models fail to reproduce the strong Mg or TiO absorption lines, as well as the ``bump'' around Na D. (Right) Inferred SFHs for each of the models, with the MAP (solid line) and median (dashed line) SFRs and 16\textsuperscript{th}--84\textsuperscript{th} percentiles as shaded regions. There is a general trend whereby the lower metallicity models favour older ages, but also more recent star formation (within the last $10 \, \Myr$) to account for the UV flux.
   }\label{fig.metallicity}
\end{figure*}

\subsection{Rising SFH prior and stellar age constraints}\label{sfh}

The flat continuity prior utilised in the fiducial model typically favours a median stellar age half that of the universe \citep{leja19}.  In our case, however, we argue it biases towards the oldest possible solution for the data by penalising substantial deviations from a constant SFR.  We explore this effect in Appendix~\ref{app.ssp} by fitting a mock SSP with age $= 1.5~\Gyr$.  Our results demonstrate that the fiducial model prior overestimates the age by $\sim 250~\Myr$ due to the strong degeneracy between solutions older than $\sim 1.3~\Gyr$.

In contrast, both observations and theory point toward increasing SFHs with cosmic time at $z>3$. Observations show that SFHs are increasing on average, with a high fraction of bursty galaxies with high-EW emission lines \citep{looser23_pop, tacchella23_metal, endsley24, simmonds24_bursty}. 

To account for the expectation for rising SFHs of high-$z$ galaxies, we now motivate a rising SFH prior, which is still flexible enough to include quenching SFHs. 
We motivate this prior based on dark matter accretion, which goes beyond recent considerations by \citet[][considered a fixed ratio of increase between adjacent time bins]{de-graaff24} and \citet[][considered an increasing prior based on the cosmic SFR density]{wang23_prospectorbeta}. We consider the $\Lambda$CDM cosmology in the Einstein-deSitter regime, which is a useful approximation at $z>1$ and becomes more and more accurate at higher redshifts. In such a regime, following \citet{dekel13}, the average accretion rate of mass into haloes of mass $M_{\rm h}$ at $z$ can be approximated by an expression of the form

\begin{equation}
    \dot{M}_{\rm h} \propto M_{\rm h,0}~e^{-\alpha(z-z_{0})}~(1+z)^{\mu},
    \label{eq:dm}
\end{equation}
where the power-law index of $\mu=5/2$ and $\alpha=4/5$ can be understood from the Press-Schechter (PS) and Extended-PS (EPS) approximations of gravitational structure formation in cosmology \citep{press74, bond91}. This functional form was indeed found to be a good fit to the halo growth in cosmological $N$-body simulations \citep{wechsler02, neistein08, dekel13}. 

To implement this rising SFH prior within \prospector, we shift the mean of the t-distribution for each SFR ratio such that the star formation in each bin obeys
\begin{equation}
    \mathrm{SFR}(z) \propto e^{-4/5\cdot(z - z_{\mathrm{obs}})} \cdot (1+z)^{5/2},
\end{equation}
where $z$ refers to the mean redshift of each time bin, motivated directly by Eq.~\ref{eq:dm}. For example, for $z_{\rm obs}=3.2$, the expectation is that the SFR at $z=10$ is reduced by a factor of $\approx20$. The 1$\sigma$ regions of both priors are shown in the left panel (flat continuity prior) and right panel (rising continuity prior) of Fig.~\ref{fig.priorSFHs}. Both priors have been normalised to a total stellar mass of $M_{\star}=10^{11.37}~\Msun$. Repeating the 1.5 \Gyr mock SSP test with the rising prior, we now find the best fit underestimates the age by $\sim 100~\Myr$ (see Appendix~\ref{app.ssp}). The test indicates the importance of the choice of prior when measuring stellar ages on an accuracy of $\sim10\%$ of quiescent galaxies.

In addition to the 1$\sigma$ prior region, Fig.~\ref{fig.priorSFHs} shows the resulting best-fit SFHs for the flat continuity SFH prior and the rising SFH prior in the left and right columns, respectively. We assume for both fits the C3K stellar library. Both fits reproduce the photometry and the spectrum equally well, with the flat continuity SFH prior and the rising SFH prior giving a reduced $\chi^2 = 0.37$ and 0.42 (calculated with $j_\mathrm{spec}=1$), respectively. Counter-intuitively, the rising prior has a lower noise inflation factor $j_\mathrm{spec}\sim2$ compared to the fiducial $j_\mathrm{spec}\sim3$, but the values are consistent within $1\sigma$.  The posterior distributions for the key population parameters are shown in Fig.~\ref{fig.priorposteriors} in Appendix~\ref{c.sfhprior}.  

The rising prior instigates a best fit that has the vast majority of the stellar mass in the second-oldest bin. This leads to a (very constrained) median age $\sim 200$ \Myr younger than the fiducial result. Repeating the fit with more time bins did not substantially widen the posterior on the age, as the model always favours a short burst of star-formation at $z \sim 8$. It is also worth noting that our use of a non-parametric SFH artificially gives a very constrained posterior for the specific SFR in the first time bin (see Fig.~\ref{fig.priorSFHs}). While the other parameters remain generally consistent (approximately within $1\sigma$), the rising prior favours overall a younger, more dusty solution. Importantly, the SFH obtained with the rising prior is more consistent with high-redshift observations of star-forming galaxies, which are overlaid in Fig.~\ref{fig.priorSFHs}. It also significantly reduces the tension with the inferred star-formation efficiency at $z>8$ and theoretical expectations, which we discuss in Section \ref{subsec:agediscussion}.

In addition to changing the base of the SFH prior, we now modify when the SFH starts. In our \prospector models described above, we assume a negligible SFR at $z>20$. We run two additional models, one assuming the SFH starts at $z=10$, the other starting at $z=\infty$, corresponding to the Big Bang. We run these two additional models for both the C3K and the MILES stellar libraries. We find a stellar age of $t_{50}=1.4$ Gyr ($t_{50}=1.3$ Gyr) and $t_{50}=1.7$ Gyr ($t_{50}=1.8$ Gyr) for the $z=10$ and $z=\infty$ model, assuming C3K (MILES). 

The MAP spectra for these tests are displayed in Appendix Fig.~\ref{a.sfh}. These demonstrate that forcing a $\sim 300 \, \Myr$ younger fit does not significantly disagree with the data, with most values remaining within the $5 \%$ error margin. From oldest to youngest age, we find reduced $\chi^2$ values (setting $j_\mathrm{spec}=1$ for comparison) of 0.43, 0.44, 0.49 for the MILES models and 0.37, 0.37, 0.38 for the C3K models, which cover a larger spectral range. We also note that the inferred noise inflation factor is relatively unchanged in these tests, with all values remaining consistent with $j_{\mathrm{spec}}\sim2.5$.

\subsection{Metallicity investigation} \label{metallicitymethod}

To quantify the poorness of the sub-solar models in fitting the data, the C3K fits were re-run with \textit{fixed} metallicities of [$Z$/H] $ = -1.0, -0.3, 0.0, 0.18$, giving reduced $\chi^2$ values of $0.58, 0.40, 0.37, 0.37$, respectively (calculated assuming $j_{\mathrm{spec}}=1$). The resulting MAP spectra are presented with lines coloured blue to red in Fig.~\ref{fig.metallicity}, which is zoomed into the rest-frame optical-to-NIR region for clarity. The figure demonstrates that the sub-solar models fail to reproduce many of the absorption features of the SED, particularly the Mg\,\textsc{i}b triplet or TiO band at $\sim$ 7150 \AA. 

The best-fit SFHs are also plotted for each case. Unsurprisingly, there is a rough trend in which the lower metallicity models compensate by having a greater proportion of the stellar mass form in the oldest time bin. Conversely, the sub-solar fits also have a higher very recent SFR $\sim 3 \, \Msun \mathrm{yr}^{-1}$ in the last 10 Myr, partially due to the model erroneously fitting the continuum with broadened emission. Overall, the median ages remain largely unchanged from the fiducial fits.  The possibility of $\mathrm{\alpha}$-element enhancement is discussed in Section \ref{subsec:alphenc}.

\subsection{IMF investigation}\label{imfmethod}

Finally, the effect of IMF variation is explored by re-running the fits after modifying the piece-wise power-law slope of the \citet{kroupa01} IMF,
\begin{equation}
    \xi(M) \propto M^{-\alpha_i}
\end{equation}
where
\begin{multline} \label{eq:kroupa}
    \alpha_1 = 1.3, \;\;\; 0.08 \leq M/\Msun \leq 0.50 \\
    \alpha_2 = 2.3, \;\;\; 0.50 \leq M/\Msun \leq 1.00 \\
    \alpha_3 = 2.3, \;\;\; 1.00 \leq M/\Msun \leq 100, \\
\end{multline}
which was used in our fiducial \prospector models. Here the C3K templates are used to ensure that any features sensitive to IMF variation in the mid-IR region are captured by the fit.

Several studies have found evidence that the IMF becomes more bottom-heavy in present-epoch, $z \sim 0$, ETGs than the Milky Way IMF \citep{conroy12, cappellari13d, la-barbera13, la-barbera17, spiniello15}. An analysis of optical-to-NIR absorption lines by \citet{conroy12} found the deviation is stronger at larger $\alpha$-enhancements and velocity dispersions. Given that we have derived a low level of dust attenuation, a bottom-heavy IMF (as opposed to cold ISM) would help to reproduce the strength of the Na D line if combined with some degree of Na-enhancement \citep{jeong13}. While this would increase our already unexpectedly-high \mstar, an increased prevalence of M-dwarf features in the NIR region might be slightly degenerate with the population's age at our limited resolution \citep{conroy12}. We test these possibilities by raising $\alpha_1$ to an extreme value of 3.0, above even the Salpeter slope of 2.35, to inflate the proportion of low-mass stars. This leads to a best-fit surviving stellar mass enlarged by $\sim 0.3$ dex, $\log_{10}{(\msurv/\Msun)} = 11.60^{+0.02}_{-0.02}$, with all other parameters largely unchanged. This is because faint, low-mass stars only contribute on the order of $1\%$ to the bolometric luminosity.

In contrast, observations at $z \sim 2$, such as an analysis of $^{13}\mathrm{C}/^{18}\mathrm{O}$ transitions by \citet{zhang18}, have suggested starburst galaxies emerge with top-heavy IMFs \citep{narayanan13}. A top-heavy IMF is an attractive possibility for our galaxy because an over-abundance of more luminous, high-mass stars could lower the $\mstar/L$ ratio. However, in our case lowering the high-mass slope to $\alpha_2 = \alpha_3 = 2.1$ (to increase the relative abundance of stars of $M \gtrsim 4 \Msun$) gives only a $\sim 0.1$ dex reduction in the surviving stellar mass, $\log_{10}{(\msurv/\Msun)} = 11.13^{+0.02}_{-0.02}$, with $\mstar$ largely unchanged. This is because the light from older, $> 1 \, \Gyr$, stellar populations is dominated by stars closer to $\sim 1 \, \Msun$ \citep{greggio11_book,esdaile21}. In fact, a much top-heavier IMF would only give a \textit{higher} dynamical mass-to-light ratio due to the surplus of stellar remnants \citep{dabringhausen09, narayanan13}. Overall, the effects of IMF variation are unlikely to significantly change the inferred SFH of this galaxy. In Appendix \ref{b.imf} we show the different IMF models give posteriors that are all in agreement, with the exception of the stellar mass.

\section{Morphology}\label{morphology}

\begin{figure*}
\includegraphics[width=\textwidth]{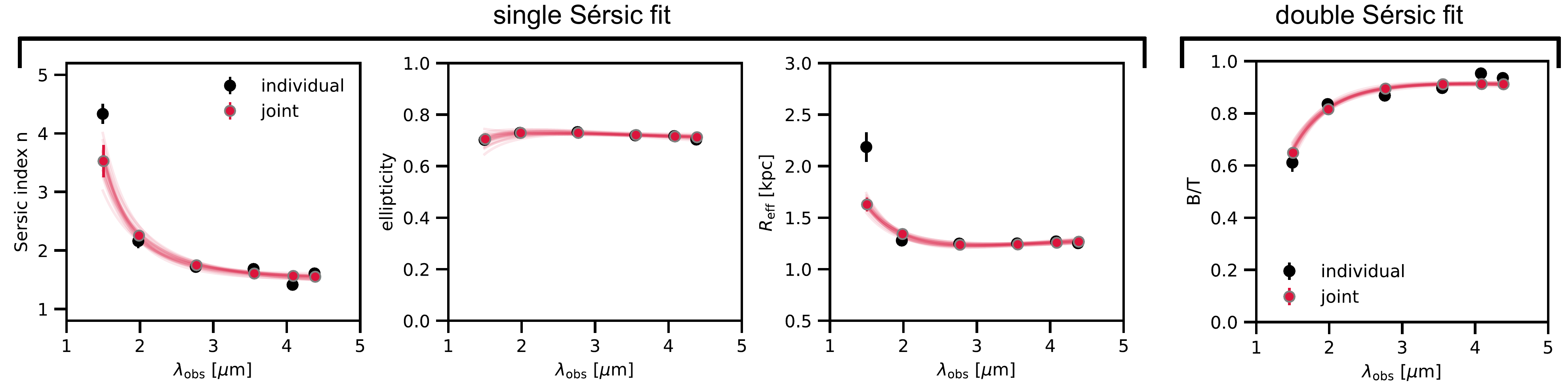}
       \caption{Results of the single S\'ersic model (first three panels from the left) and double, bulge+disc S\'ersic model (right panel). For each model fit, we fit the NIRCam filters (F150W, F200W, F277W, F356W, F410M and F444W) individually (black) and jointly (red). The red lines indicate the posterior distribution of the 2-order polynomials that are used to connect the morphological parameters across wavelength. The S\'ersic index and half-light size decrease with wavelength, while the ellipticity is nearly constant. From the bulge+disc fit, we see that the bulge-to-total ratio (B/T) increases from 0.65 to 0.91 in F150W ($\lambda_{\rm rest}\approx0.356~\micron$) to F444W ($\lambda_{\rm rest}\approx1.043~\micron$). In summary, ZF-UDS-7329 is consistent with an edge-on disc with a prominent bulge-like component in its core, which dominates the rest-frame 1-\micron~ emission.
   }\label{fig.morph}
\end{figure*}

Fig.~\ref{fig.RGB} shows that ZF-UDS-7329 has an overall regular and smooth morphology, with a central concentration (similar to a bulge) and an extended, edge-on disc-like structure. The NIRSpec slit traces the central region of the galaxy. In this section, we quantify the morphology of ZF-UDS-7329 in different NIRCam bands in order to further shed light onto the physical nature of this galaxy and put the stellar population analysis into context.

Both \citet{glazebrook24} and \citet{carnall24} have fit a single S\'ersic profile to ZF-UDS-7329. \citet{glazebrook24} used \textsc{Galight} \citep{ding22_code} to fit the F444W image, finding a half-light semi-major axis $R_{\rm e}=1.15_{-0.08}^{+0.08}$ kpc, $n=2.41_{-0.29}^{+0.24}$ and an axis ratio of $(b/a)=0.33_{-0.01}^{+0.01}$. \citet{carnall24} fitted the F277W band with the \textsc{PetroFit} code \citep{geda22}, finding $R_{\rm e}=0.91\pm0.01$ kpc and $n=2.5\pm0.1$. Both works commented on the consistency with the visual appearance of an edge-on disc. We expand here upon these studies by analysing the dependence of the morphology on wavelength in order to constrain radial colour gradients and thereby shedding light onto the formation of ZF-UDS-7329.

\subsection{Single S\'ersic fit}

We perform our morphological analysis with \textsc{pysersic} \citep{pasha23}. \textsc{Pysersic} is a code for fitting S\'ersic profiles to galaxy images using Bayesian inference while accounting for the effect of the point-spread function (PSF). It is written in python using \textsc{jax} with inference performed using \textsc{numpyro} \citep{phan19, bingham19}. In our analysis, we use model PSFs (mPSFs) which we derive by mosaicing  WebbPSF models \citep{webbpsf}, see \citet{ji24} for a description. We fit the F150W, F200W, F277W, F356W, F410M and F444W with both a single S\'ersic model and a double, bulge+disc S\'ersic model. In the latter model, we fix one of the S\'ersic models to have $n=1$ (disc component) and assume that both S\'ersic components have the same centre. We do not fit the F090W and F115W bands because of their low S/N.

For both the one-component and two-component models we perform a fit to the individual bands independently of each other as well as jointly. For the joint single S\'ersic fit, we assume that the S\'ersic index $n$, the ellipticity and half-light radius are linked across wavelength by a second-order polynomial in $1/\lambda$. The position angle and centre are the same for all bands. We find that the joint and individual fits are overall consistent, with S\'ersic index $n$ and half-light size $R_{\rm eff}$ decreasing with increasing wavelength (Fig.~\ref{fig.morph}). The largest, most significant difference is found in the bluest band (F150W), where the joint fit gives $n=3.53\pm0.28$ and $R_{\rm eff}=1.63\pm0.07$ kpc, while the fit to F150W individually gives $n=4.33\pm0.17$ and $R_{\rm eff}=2.18\pm0.14$ kpc. Beyond 2 \micron, both S\'ersic index and size do not vary significantly, and we find $n=1.55\pm0.03$ and $R_{\rm eff}=1.27\pm0.02$ kpc for F444W (rest-frame $\lambda_{\rm rest}\approx1~\micron$). For the ellipticity, which shows only a marginal wavelength dependence, we find $\varepsilon=0.71\pm0.01$ for F444W, which translates to a semi-to-major axis ratio of $0.29\pm0.01$. In summary, the single S\'ersic fit is consistent with an edge-on disc-like galaxy at $\lambda_{\rm rest}\approx1~\micron$, but with evidence for a larger size and S\'ersic index for F150W ($\lambda_{\rm rest}\approx0.356~\micron$). We discuss the implication of the galaxy size for the dark matter halo mass in Section~\ref{subsec:halomass_size}.

\subsection{Bulge-disc decomposition}

This clear indication for a wavelength-dependent morphology, implying a colour gradient, motivates us to perform a bulge-disc decomposition. For the bulge-disc decomposition, we assume that the fluxes of both bulge and disc can vary as a function of wavelength, while the structural parameters are constant with wavelength. Furthermore, we assume that the bulge has a smaller half-light radius than the disc, and the disc's S\'ersic index is fixed to 1. 

We find that the bulge-to-total ratio (B/T) increases with wavelength, $0.65\pm0.02$ to $0.91\pm0.01$ in F150W ($\lambda_{\rm rest}\approx0.356~\micron$) to F444W ($\lambda_{\rm rest}\approx1.043~\micron$). This is consistent with the results from above, where the bluer bands are better described by a more extended profile. Furthermore, the S\'ersic index of the bulge is $n=2.00\pm0.01$, which indicates that we are running up against the lower-bound of the prior. This means that this central structure is not a ``classical'', $n=4$ bulge. We find the size of the bulge and disc to be $1.04\pm0.01$ kpc and $3.45\pm0.10$ kpc, respectively.

\subsection{Colour gradient}

\begin{figure}
\includegraphics[width=\columnwidth]{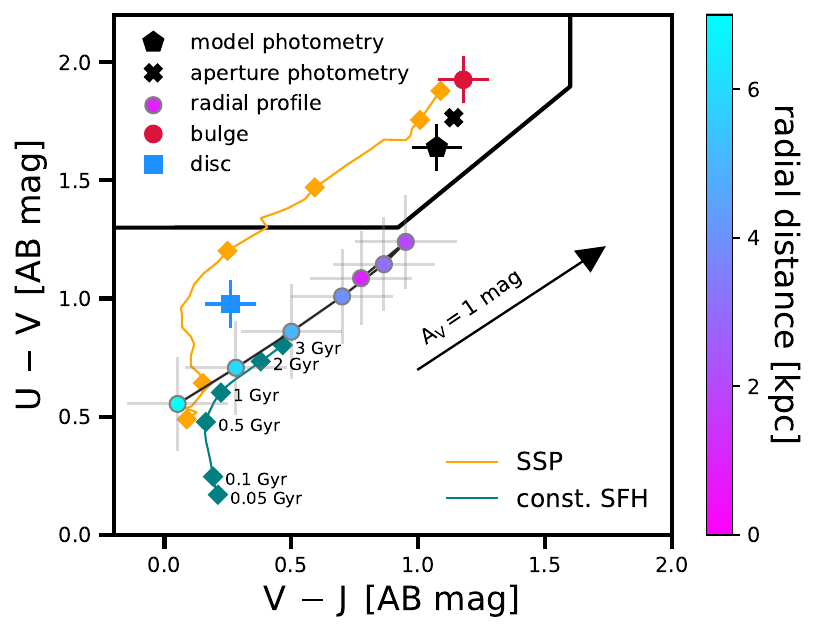}
       \caption{Rest-frame U$-$V versus V$-$J colour-colour (UVJ) diagram. The implied radial colour gradient profile from the single S\'ersic fit is shown by the coloured, connected circles, probing $1-7$ kpc (see colour bar on the right). The colour of the bulge and the disc as inferred from the bulge-disc decomposition are indicated with the red circle and blue square. The black pentagon and cross mark the total model colour (from the bulge-disc model) and the aperture photometry, respectively. For reference, we plot an SSP (orange line) and a stellar population with a constant SFH, with diamonds marking ages of 0.05, 0.1, 0.5, 1, 2 and 3 Gyr. A dust attenuation vector indicates 1 mag of attenuation in A$_{\rm V}$. We find that the disc and colour in the outskirt is consistent with a younger, less dust attenuated stellar population, while the central bulge is consistent with being older, within the quiescent box.
   }\label{fig.uvj}
\end{figure}

The single S\'ersic fit and the bulge-disc decomposition both imply a red centre and blue outskirt. While the NIRSpec spectrum mostly probes the central region of ZF-UDS-7329 (Fig.~\ref{fig.RGB}) and is consistent with an old, quiescent stellar population, we now address how much bluer the outskirt is relative to this central region. 

We base our analysis on the rest-frame U$-$V versus V$-$J colour-colour (UVJ) diagram \citep{williams09}, which is a powerful diagram to differentiate between reddening due to older stellar populations and due to dust attenuation (though this can be complicated by the fact of varying attenuation law; \citealt{leja19_uvj}). We take the NIRCam filters F150W ($\lambda_{\rm rest}\approx0.356~\micron$), F200W ($\lambda_{\rm rest}\approx0.471~\micron$), and F444W ($\lambda_{\rm rest}\approx1.043~\micron$) to trace the rest-frame U, V and J band, respectively. Since these filter curves differ from each other, we derive colour correction terms from the best-fit \prospector model. These corrections terms are 0.35 mag and $-0.17$ mag for the (U$-$V) and (V$-$J) colour, respectively, assuming the best-fit \prospector model.

Fig.~\ref{fig.uvj} shows the UVJ diagram. The black line marks the UVJ-quiescent box, i.e. colours in this region are consistent with being quiescent (i.e. old and low sSFR). The total, integrated photometry using apertures (black cross) and the bulge-disc decomposition give a red colour, within the quiescent region. Interestingly, the bulge-disc decomposition clearly shows that the bulge is significantly redder than the disc, with the bulge being consistent with a quiescent, old stellar population (SSP age of $1-3$ Gyr, with possibly some dust attenuation), while the disc lies in the star-forming region of the UVJ diagram wiht an SSP age of 0.1-0.5 Gyr. This is overall confirmed by the colour-colour profile that is implied from the single S\'ersic fit, which is shown by the connected circles in Fig.~\ref{fig.uvj}.

In summary, we find that the central region of ZF-UDS-7329, which is probed by the NIRSpec spectrum, is consistent with being quiescent, while the outer disc component is consistent with being significantly younger, consistent with a rapidly and recently quenched galaxy \citep[e.g.,][]{belli19} or with star formation. This is overall in agreement with the SFH shown in Fig.~\ref{fig.priorSFHs}, which shows an extended SFH with a non-negligible SFR over the past 1 Gyr. As we discuss below, such a colour gradient is consistent with inside-out quenching \citep{tacchella15, tacchella18_dust}. The high central stellar density can be formed through a gas compaction event \citep{zolotov15, tacchella16_MS}, which is triggered by mergers or counter-rotating streams \citep{dekel14_nugget, lapiner23}. This nuclear starburst will consume and expel most of the gas, leading to a quenching episode, soon after which the re-accreted and newly accreted gas with higher angular momentum will form mostly stars in an extended disc configuration \citep{tacchella16_profile}.

\section{Discussion}
\label{discussion}

\begin{figure*}
   \includegraphics[width=1.0\textwidth]{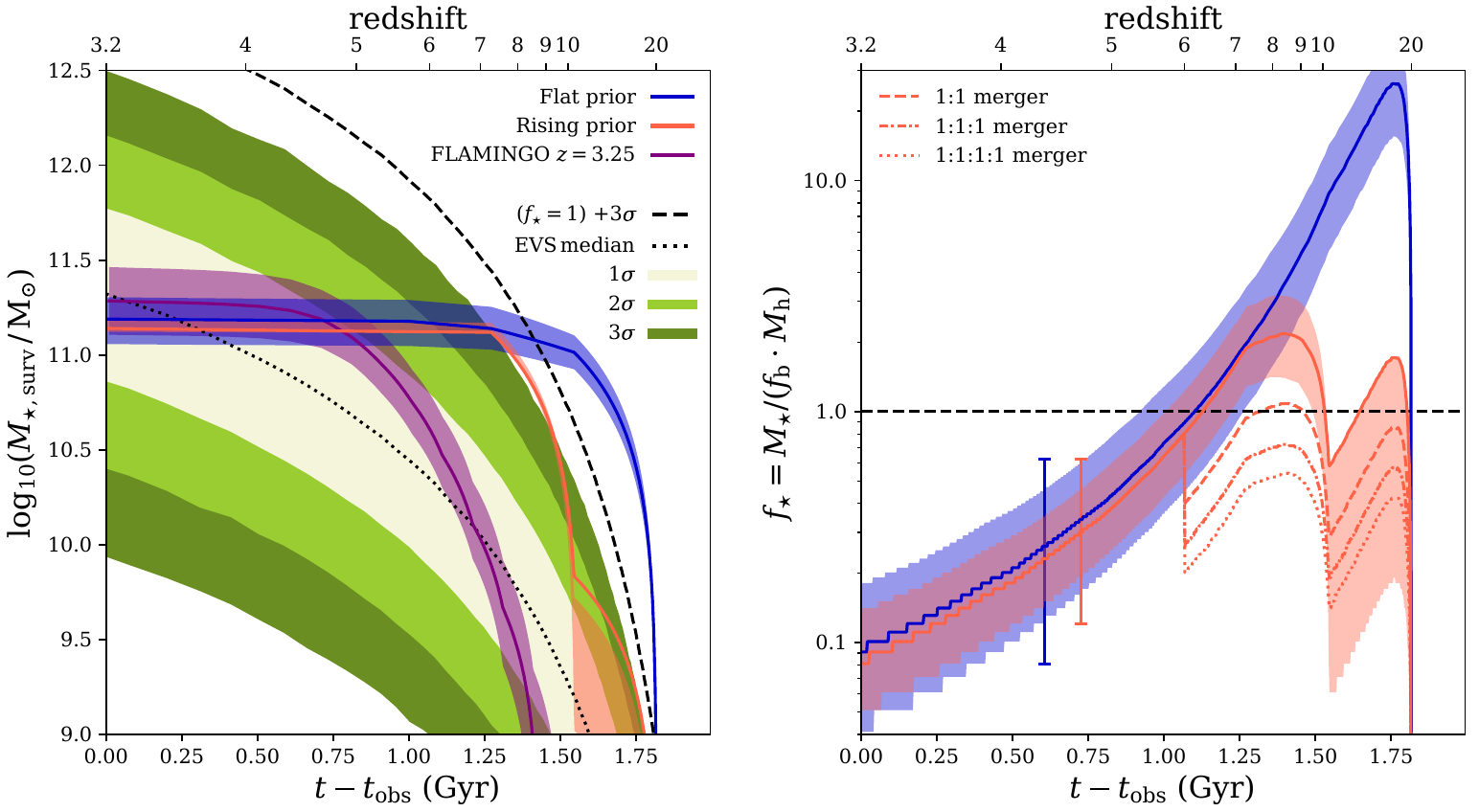}
       \caption{
       (Left) The cumulative MAP SFHs of the rising (orange) and flat (blue) continuity priors, with the 1$\sigma$ regions additionally shaded.  The solid purple line denotes the median SFH from the 100 oldest massive ($\mstar > 10^{11} \, \Msun$) galaxies with $\mathrm{SFR}<10~\Msun~\mathrm{yr}^{-1}$ at $z_{\mathrm{obs}}=3.25$ from the $1 \, \Gpc^3$ \flamingo simulation, with the sample 1$\sigma$ also shaded.  The EVS prediction from \citet{lovell23_evs} for the most massive galaxy at each redshift is also plotted with beige/green bands for the 1$\sigma$ to 3$\sigma$ confidence levels.  The dashed line shows the 3$\sigma$ upper-limit when assuming a maximal stellar fraction, $f_{\star} = 1 $.  The results show that the fiducial model is significantly discrepant with the predictions of the EVS approach.  Forcing a younger age using the rising prior lowers the disagreement to within $3\text{--}4\sigma$, but leads to much smaller uncertainty estimates from our Bayesian approach.  (Right) 
       The implied stellar fractions (sampled in 1$\%$ increments) for each of the models assuming the same EVS model and colour-coded identically.  These are allowed to go (unphysically) above 1 to demonstrate the disagreement. Regions of 1$\sigma$ consistency between the EVS and \prospector uncertainties are shaded accordingly.  Dashed, dash-dotted, and dotted lines show possible scenarios for a major merger at $z=6$, by halving, trisecting, or quartering the rising model's SFH, respectively.  These show that more reasonable values for $f_{\star}$ can be obtained by considering $\sim 3$ progenitors merging after $z \sim 6$. The effects of cosmic variance are hypothesised to inflate the 1$\sigma$ scatter of the EVS by up to a factor of $\sim 2$ \citep[see, e.g.,][]{chen23_cv, xiao24}. This would roughly double the presented errors for $z < 10$, which we indicate with two vertical error bars.
   }\label{fig.ageSFHs}
\end{figure*}

Our analysis of ZF-UDS-7329 shows that this is a massive ($M_{\star}\approx10^{11.4}~\Msun$), quiescent ($\mathrm{sSFR}\approx0.03~\Gyr^{-1}$) galaxy at $z=3.2$, consistent with previous studies of this object \citep{glazebrook24, carnall24}. While the spectrum is consistent with a very old stellar population (mass-weighted stellar age of 1.6 Gyr at an epoch when the Universe was only 2 Gyr old), we show that the spectrum is similarly well fit by a range of SFHs that have stellar ages of $1.3-1.8$ Gyr. Importantly, by introducing a new SFH prior that follows dark matter halo accretion and therefore is increasing with cosmic time, the best-fit SFH is consistent with direct, high-redshift observations of star-forming galaxies. 

In this section, we discuss the implications of ZF-UDS-7329 for the star-formation efficiency of early galaxies (Sections~\ref{subsec:agediscussion} and \ref{subsec:halomass_size}), the environment of ZF-UDS-7329 (Section~\ref{subsec:environment}), and highlight systematics related to the possible $\alpha$-element enhancement (Section~\ref{subsec:alphenc}).

\subsection{Early forming galaxies}
\label{subsec:agediscussion}

ZF-UDS-7329 -- given its early formation -- has been put forward as serious challenge to our current understanding of cosmology \citep{glazebrook24} and galaxy formation \citep{carnall24}. Our analysis shows that the JWST/NIRSpec+NIRCam data also support solutions that are consistent with the current paradigms of cosmology and galaxy formation. 

While the SFH of ZF-UDS-7329 indeed shows a formation redshift of $z>7$, the exact formation epoch is not well determined: the flat SFH prior gives a formation epoch of $z\approx11$, consistent with \citet{glazebrook24}, while the rising SFH prior leads to $z\approx8$ (Fig.~\ref{fig.priorSFHs}). While this redshift difference is significant and the Universe roughly doubles its age over this redshift range, the stellar age only changes from 1.6 to 1.4 Gyr, a difference of only 200 Myr, which leads to negligible changes in the NIRSpec PRISM spectrum (see Fig.~\ref{fig.ageMAPs}). While we largely break the age-metallicity degeneracy (Figs.~\ref{fig.posteriors} and \ref{fig.metallicity}), there is still the fundamental limitation that we can not self-consistently model $\alpha$-element enhancement, which can lead to systematics in our stellar age estimates (see Section~\ref{subsec:alphenc}). 

Importantly, by using the rising SFH prior, which is physically motivated, we obtain a SFH for ZF-UDS-7329 that is roughly consistent with the latest high-redshift observations of star-forming galaxies. Specifically, Fig.~\ref{fig.priorSFHs} shows SFRs from several spectroscopically confirmed galaxies, including GS-z14-0 and GS-z14-1 \citep[$\mstar \approx 10^8~\Msun$,][]{carniani24_z14}, GN-z11 \citep[$\mstar \approx 10^9~\Msun$,][]{tacchella23}, GLASS-z12 \citep[$\mstar \approx 10^9~\Msun$,][]{castellano24}, REBELS-04 \citep[$\mstar \approx 10^9~\Msun$,][]{bouwens22}, REBELS-25 \citep[$\mstar \approx 10^{10}~\Msun$,][]{bouwens22}, JADES-318593 \citep[$\mstar \approx 10^{11}~\Msun$][]{simmonds24}, RUBIES-UDS-z7 \citep[$\mstar \approx 10^{10}~\Msun$,][]{weibel24}, and proto-cluster SPT0311-58 \citep[$\mstar \approx 10^{11}~\Msun$,][]{arribas24}. In contrast, the flat SFH prior gives a SFR of $\sim600~\Msun~\yr^{-1}$ at $z\approx10-20$, which is nearly 2 orders of magnitude above those direct SFR measurements. On the other hand, the rising SFH prior leads to a SFR of $\sim40~\Msun~\yr^{-1}$ at $z\approx10-20$, which is much closer to these direct observations.
Another important consideration is the metallicity of higher-redshift galaxies. While ZF-UDS-7329 has solar metallicity and high $\alpha$-element abundance \citep[see also][]{carnall24}, galaxies at $z>10$ have invariably sub-solar gas metallicity. Given that gas metallicity probes the $\alpha$ element oxygen, the mismatch between the stellar metallicity in ZF-UDS-7329 and in the gas of these higher-redshift systems may be problematic. Scenarios where gas metallicity is diluted by inflows are possible, but would imply that we preferentially observe this class of systems, given that there are essentially no current observations of solar-metallicity systems at sufficiently high redshift \citetext{with the closest known examples being at too low mass and redshift, i.e. $\mstar \sim 10^{10}~\Msun$ at $z=4.6$, \citealp{deugenio25}; and $z=6.7$, \citealp{shapley24}}. We can envision a period of rapid metal enrichment between $z=10$ and the peak of the SFH, where most of the high-SFR sources are detected by ALMA; among these, only SPT0311-58 has a measurement of the gas metallicity, which is still only half solar \citep{arribas24}. Characterizing the gas metallicity of these systems would be a powerful consistency test for the scenario presented here. 

Still, even disregarding the uncertainties due to metallicity evolution, our inferred SFHs lies roughly a factor of 2 above those direct estimates. A possible way to alleviate this tension is by considering mergers: we are are measuring a SFH of a galaxy that assembled through both in-situ star formation and ex-situ mass accretion (mergers of  galaxies). It is therefore not surprising that the SFR we estimate at early times in ZF-UDS-7329 is higher than what observed in any single high-redshift galaxy, since the SFH of ZF-UDS-7329 includes both the SFR of the main progenitor and all of its accreted satellites. While we cannot measure the ex-situ mass fraction for this specific galaxy, we find that the extended morphology (relatively large size of $>1$ kpc) with its colour gradient is consistent with having undergone (multiple) mergers. Specifically, the colour gradient is consistent with inside-out quenching \citep{tacchella15, tacchella18_dust}. The high central stellar density could be formed through a gas compaction event \citep{springel05, robertson06, hopkins13_merger, zolotov15, tacchella16_MS}, which is triggered by mergers or counter-rotating streams \citep{dekel14_nugget, lapiner23}. This nuclear starburst will consume and expel most of the gas, leading to a quenching episode, soon after which the re-accreted and newly accreted gas with higher angular momentum will form mostly stars in an extended disc configuration \citep{tacchella16_profile}. In addition, this galaxy lies probably in an over-density (see Section~\ref{subsec:environment}), where mergers are prevalent, in particular at higher redshifts \citep{hopkins06a}. 

In order to quantitatively assess the implication of the SFH on the efficiency of star formation in the early Universe, we connect ZF-UDS-7329 to the underlying dark matter halo mass function to estimate the stellar fraction, which is defined as $f_{\star}=M_{\star}/(f_{\rm b}\times M_{\rm h})$, where $f_{\rm b}=0.16$ is the cosmic baryon fraction and $M_{\rm h}$ is the halo mass. The stellar fraction is also sometimes referred to as the integrated star-formation efficiency. For example, $f_{\star}=1$ corresponds to the case in which all baryons available to the galaxy have been converted into stars. It is challenging to estimate the halo mass $M_{\rm h}$ for a single galaxy because of cosmic variance, stochastic sampling of the halo mass and galaxy stellar mass functions, and the limited volumes of observational surveys and simulations. To overcome the problem of characterising a precise selection function for population studies, we use the extreme value statistics \citep[EVS; e.g.,][]{harrison11} approach introduced by \citet{lovell23_evs} and discussed in \citet{carnall24}, which provides a full probability distribution of halo masses given a survey volume and stellar mass of a galaxy. We adopt the \textsc{Evstats} package\footnote{\url{https://github.com/christopherlovell/evstats}} provided by \citet{lovell23_evs}. We assume a survey area of PRIMER from which ZF-UDS-7329 has been selected, which is a total of 378.2 arcmin$^2$ (effective comoving volume $\approx 1.27 \times 10^{-3}~\Gpc^3$ for $3 < z < 4$).

The left panel of Fig.~\ref{fig.ageSFHs} presents the stellar mass growth histories obtained from our SFHs for the flat SFH prior (blue line) and the rising SFH prior (orange line). We compare these growth histories with EVS-based estimates for the most-massive galaxy expected in our survey volume as a function of redshift. Specifically, the dotted line shows the fiducial model for $f_{\star}$, where $f_{\star}$ is a truncated log-normal following $f_{\star}=\ln(\mathcal{N}(\mu,\,\sigma^{2}))$, with $\mu=0.135$ and $\sigma=1$ across the interval $0 \leq f_{\star} \leq 1$, motivated by a variety of theoretical and observational constraints \citep{lovell23_evs}. The shaded regions indicate the 1$\sigma$, 2$\sigma$ and 3$\sigma$ contours of this fiducial EVS model. The dashed line is the 3$\sigma$ upper limit of the maximal case, in which $f_{\star}=1$ and all available baryons are converted into stars. 

Comparing the mass growth history of ZF-UDS-7329 with the EVS model estimates, we find that ZF-UDS-7329 is indeed consistent with being one of the most massive systems at the redshift of observations ($z=3.2$), but well within the fiducial model. However, at earlier times, because of the early formation and the associated high stellar mass, the growth history of both priors is in a $>3\sigma$ tension with the fiducial EVS model at $z=8-9$. Even more challenging, the flat SFH prior leads to a growth history that is in significant tension with the $f_{\star}=1$ model at $z>10$. However, since the growth history with the rising SFH prior shows most of the growth at $z=7-10$, this growth history lies always below the $3\sigma$ line of the $f_{\star}=1$ model (shown as a dashed line), implying consistency with a more reasonable stellar fraction at early cosmic times. 

The right panel of Fig.~\ref{fig.ageSFHs} shows the implied stellar fraction assuming that ZF-UDS-7329 is the most massive system in the survey volume at each redshift. We find a stellar fraction of $f_{\star}=0.08_{-0.03}^{+0.05}$ at $z=3.2$. For comparison, at $z=0$, the stellar fraction peaks at a halo mass of $\approx10^{12}~M_{\odot}$ with a value of $f_{\star}=0.2$, and there seems to be only little evolution in the stellar fraction towards $z=3$, with some studies finding values of $f_{\star}=0.1$ \citep{moster10, behroozi13b, behroozi18, rodriguez-puebla17}. We conclude that the stellar fraction at $z=3.2$ of ZF-UDS-7329 is consistent with lower-redshift estimates from empirical models and abundance matching approaches. 

At $z=7-12$ (right panel of Fig.~\ref{fig.ageSFHs}), we find a stellar fraction of $f_{\star}\approx1-2$ and $2-10$ for the rising and flat SFH prior, respectively. This shows that the flat SFH prior implies an unphysically high stellar fraction, while the rising prior is at maximal efficiency of star formation. Consistently, as discussed in Appendix~\ref{a.sfh}, starting the SFH of the galaxy at $z=10$ (instead of $z=20$) is able to reproduce the observational data similarly well and significantly reduces the star-formation activity at early ($z>10$) times, thereby reducing the need for extreme star-formation efficiencies at high redshifts. 

Importantly, mergers can significantly reduce the high star-formation efficiency. In the right panel of Fig.~\ref{fig.ageSFHs}, we also show the stellar fraction evolution by assuming that ZF-UDS-7329 underwent one, two or three major mergers (1:1, 1:1:1 and 1:1:1:1) at $z=6$, in which the progenitors have the same SFHs. We find that such a single major merger, double major merger and triple major merger can reduce the stellar fraction to $f_{\star}=1.0$, $f_{\star}=0.7$ and $f_{\star}=0.5$ at $z=8-9$. This highlights that mergers can further reduce the inferred stellar fraction at early times, further lowering the necessity for extreme galaxy formation physics.  

But are such mergers expected at $z>3$? Considering $\Lambda$CDM, \citet[][see their figure 3]{fakhouri10_dm} obtained a mean merger rate per halo per unit redshift that is nearly independent of $z$ and only weakly dependent on $M_h$ \citep[see also][]{genel10, rodriguez-gomez15}. Considering major mergers (mass ratio of $>1/3$), the mean merger rate is $dN_m/dz\approx0.6$ and $0.3$ for a halo mass of $M_h=10^{13}~M_{\odot}$ and $M_h=10^{11}~M_{\odot}$, respectively. Considering the halo mass evolution obtained by EVS model, we estimate a total of $4\pm1$ major mergers to take place from $z=12$ to $z=3.2$. This indicates that our scenarios outlined above of one, two or three major mergers lies within the expectation of $\Lambda$CDM.

The EVS model provides a simple description of how to relate the stellar masses of galaxies to their host dark matter halo masses. Specifically, within the EVS framework, we assume that ZF-UDS-7329 traces the most massive halo in the probed cosmic volume at each redshift. A caveat of this approach is that each redshift is treated independently, i.e. there is a lack of continuity in the model. Expanding on this, we compare our growth histories to the cosmological hydrodynamic simulation \flamingo \citep{schaye23, kugel23}. Specifically, we use the box with a volume of (1 Gpc)$^3$ and a baryonic particle mass of $1\times10^8~\Msun$. We select the 100 oldest galaxies with $M_{\star}>10^{11}~\Msun$ and $\mathrm{SFR}<20~\Msun~\yr^{-1}$ at the redshift of observation. We plot the median SFH of those galaxies in the left-panel of Fig.~\ref{fig.ageSFHs}. We find that the \flamingo galaxies' SFHs have a significantly lower formation redshift ($z \sim 5-6$) with respect to ZF-UDS-7329, even when adopting the rising prior.  Similar results are find in other theoretical and numerical models \citep[e.g.,][]{lagos24}. While part of this difference could be due to the numerical resolution, it might also point towards a higher star-formation efficiency at early cosmic times than what current numerical models are assuming. In particular, as put forward by \citet[][see also \citealt{li24_fbb, renzini23}]{dekel23}, the high densities and low metallicities at these early epochs could lead to a high star-formation efficiency in the most massive dark-matter haloes, because the free-fall time is shorter than $\sim1$ Myr, below the time for low-metallicity massive stars to develop winds and supernovae. In such a regime, the integrated star-formation efficiency lies in the range of $0.2-1.0$, consistent with our results presented here. A more detailed comparison, including a larger set of galaxies, is needed to further constrain the numerical models and the star-formation efficiency.

Finally, cosmic variance could be an important factor as well. In the above EVS-based discussion, we assume that ZF-UDS-7329 is the most massive system in the PRIMER survey. However, we do not know how representative this survey area is of the universe as a whole, which can lead to biased estimates of the abundance of rare objects such as ZF-UDS-7329. Cosmic variance may reach $\gtrsim100\%$ of the stellar mass density at the high-mass end \citep[see, e.g.,][]{chen23_cv, xiao24}. This would roughly double our 1$\sigma$ uncertainties for the implied $f_{\star}$ (indicated by the errorbar in Fig.~\ref{fig.ageSFHs}), further reducing the tension.

In summary, while ZF-UDS-7329 is an early forming galaxy, we estimate a stellar fraction (i.e., integrated star-formation efficiency) of $f_{\star} \approx 0.5-1.0$ at $z\approx7-12$, in particular when considering the impact of mergers and cosmic variance. While this stellar fraction at early times is higher than lower-redshift estimates, we find that ZF-UDS-7329 potentially evolves to a stellar fraction of 0.1 by $z=3.2$, which is consistent with the peak of the stellar-to-halo mass ratio at $z\approx0-3$. We therefore conclude that ZF-UDS-7329 formed very efficiently at high redshifts, but is consistent with direct, high-redshift observations and realistic galaxy formation physics.



\subsection{Dark matter halo mass from the galaxy size}
\label{subsec:halomass_size}

In addition to assess the integrated star-formation efficiency based on the most massive halo in the survey volume as done in the previous section, we now perform an independent crosscheck of the halo mass estimation by using the theoretically motivated and observationally confirmed relation between galaxy sizes and their virial radius. 

To first order, the physical origin of the galaxy size-mass relation and its evolution can be understood by considering that both dark matter and diffuse gas acquire angular momentum via tidal torques and mergers \citep{peebles69, white84, porciani02, vitvitska02}. The specific angular momentum is often written using the dimensionless spin parameter:
\begin{equation}
    \lambda = \frac{J |E|^{1/2}}{G M^{5/2}},
\end{equation}
where $J$ is the total angular momentum, $E$ is the total energy, $G$ is Newton's gravitational constant and $M$ is the total mass. In the classical picture, diffuse gas acquires about the same amount of specific angular momentum as the dark matter, and conserves most of this angular momentum as it cools, collapses, and forms stars. The \citet{mo98} model predicts 
\begin{equation}
    R_{\rm eff} = 1.678 \cdot R_d = 1.187\frac{j_d}{m_d}f_c^{-1/2}f_R \lambda R_{200},
\end{equation}
where $j_d$ and $m_d$ are fractions of baryon angular momentum and mass budget within halo in the central disc, $f_c$ is a function of halo concentration, and $f_R$ is a function that takes into account baryonic contraction of halo in response to halo formation. At $z=3$, the expectation is that $R_{\rm eff} \approx 0.01 R_{200}$, which is supported by observations based on abundance matching \citep{kravtsov13, somerville18, shibuya19}.

Our measured half-light size in F444W of $R_{\rm eff} = 1.27 \pm 0.02$ kpc translates to $R_{200}\approx 127 \pm 25$ kpc (assuming an uncertainty of 20\% for the spin parameter). At $z=3$, this translate to a dark matter halo mass of $M_{\rm h} = (4/3) \pi R_{200}^3 200\rho_{\rm crit}(z=3.2) = (8.6_{-4.2}^{+6.2})\times10^{12}~\Msun$, i.e. $\log(M_h/\Msun) = 12.9\pm0.3$. This implies a stellar fraction of $f_{\star}=M_{\star}/(f_b \cdot M_h)=0.16_{-0.07}^{+0.15}$, which is consistent within the uncertainties with our estimate discussed in the previous section.

\subsection{Environment of ZF-UDS-7329}
\label{subsec:environment}

The expectation in $\Lambda$CDM structure formation is that the first collapsed structures are highly clustered \citep{springel06}. Observationally, several examples of overdensities have been put forward. For example, JADES-GS-z14-0 -- the most distant galaxy known at $z = 14.32$ -- has indeed a neighbour (JADES-GS-z14-1 at $z = 13.9$) and it seems likely that these galaxies are at least mildly associated in an extended large-scale structure \citep{carniani24_z14}. GN-z11 has nine galaxies within $\sim5$ comoving Mpc transverse with photometric redshifts consistent with $z = 10.6$, which is consistent with GN-z11 being hosted by a massive dark-matter halo of $\approx8\times10^{10}~\Msun$ \citep{tacchella23}. Furthermore, several galaxy over-densities and proto-clusters have been identified in the Epoch of Reionzation \citep{higuchi19, helton24_overdensity, sun24}. Theoretically, semi-analytical models and numerical simulations show that that the cosmic SFR density at $z>5$ is dominated by galaxies in proto-clusters \citep{chiang17, lim24}. Recently, \citet{rennehan24} shows that galaxies such as ZF-UDS-7329 indeed have SFHs and formation epochs that are consistent with the most massive galaxies within protocluster cores of galaxy clusters at $z\sim2$.

As early-forming galaxy, we expect ZF-UDS-7329 to be part of an over-density, and possibly proto-cluster. We find that ZF-UDS-7329 has two companions at redshift $z=3.2$. These galaxies are ZF-UDS-7542 \citep[MSA ID 13079 from program \#2565;][]{schreiber18, nanayakkara24} and RUBIES-UDS-46261 from program \#4233\footnote{Obtained from the DJA \citep{heintz24_dja}.} \citep{de-graaff24_rubies}, and lie within a projected distances of 12 and 6.5~arcsec, respectively, corresponding to 90 and 50~kpc. While RUBIES-UDS-46261 is star forming, ZF-UDS-7542 is a massive, post-starburst galaxy \citep{nanayakkara24}, which suggests a connection with the massive quiescent nature of ZF-UDS-7329 itself. The presence and proximity of another massive quiescent galaxy in the vicinity of ZF-UDS-7329 suggests a link between the quiescent nature of these galaxies and their surrounding environment. Furthermore, it supports the idea that ZF-UDS-7329 might have undergone a merger-rich history. Other massive, quiescent galaxies have been reported to occupy overdensities \citep{de-graaff24}; however, the exact nature of the environment surrounding ZF-UDS-7329 requires further investigation (de~Graaff et al., in~prep.).

\subsection{Metallicity: \texorpdfstring{$\alpha$}{Alpha}-element enhancement}
\label{subsec:alphenc}

The best-fit metallicity values for ZF-UDS-7329 can only be possible for its stellar age if there is some deviation from a solar composition, because some of the solar chemical abundances are due to enrichment processes that act on long timescales \citep[e.g.,][]{maiolino19}, whereas the SFH of ZF-UDS-7329 is necessarily short. We may derive loose bounds on [$\mathrm{\alpha/Fe}$] using the simplified relation $[Z/\mathrm{H}] = [\mathrm{Fe/H}] + A[\mathrm{\alpha/Fe}]$ introduced in \citet{trager00}, where $A$ is a constant dependent on the detailed abundance pattern (= 0.93 if all $\mathrm{\alpha}$-elements are varied identically). Taking the C3K best-fit value $[Z/\mathrm{H}] = 0.11$ to be genuine, we find $[\mathrm{Fe/H}] \thickapprox -0.1$ to $-0.3$, for values of $[\mathrm{\alpha/Fe}] = 0.2$ to $0.5$. Such results would be concordant with studies of massive quiescent galaxies at $z \sim 0.7$ from the LEGA-C survey \citep{van-der-wel16, beverage21}. Furthermore, both chemical evolution models \citep{thomas05}, and analyses of early-type galaxies from the Sloan Digital Sky Survey \citep{de-la-rosa11}, have provided strong evidence for a correlation between short star-formation timescales and $[\mathrm{\alpha/Fe}]$. We adopt the relation of \citet{thomas05}, $[\mathrm{\alpha/Fe}] \thickapprox 1/5 - 1/6\log_{10}(\Delta t)$, which assumes a Gaussian SFH of FWHM $\Delta t$. Taking $\Delta t$ to be the time between $16 \%$ and $84 \%$ of the mass forming, we find $[\mathrm{\alpha/Fe}] \thickapprox $ 0.3 from our best-fit SFHs. Finally, \citet{carnall24} fit a higher-resolution version of the spectrum using the Absorption Line Fitter (\textsc{ALF}) code \citep{conroy12,conroy18} to obtain $[\mathrm{Mg/Fe}] = 0.42^{+0.19}_{-0.17} $, consistent with our estimations. Given our treatment has assumed a solar abundance pattern throughout, how do these results curtail our confidence in the inferred population parameters?

In Fig.~\ref{fig.metallicity} we display the spectra of synthetic $\alpha$-enhanced SSPs based on the MILES library from \citet{vazdekis15} at two separate ages, convolved to the same resolution and binning as our data and offset for clarity. The SEDs demonstrate that a 1.0 Gyr-old population with $[\mathrm{\alpha/Fe}] = 0.4$ can appear similar to a 1.5 Gyr-old one with a solar composition. This effect is most prominent at precisely the ages between $1 \text{--} 1.5 \, \Gyr$, with the TiO band at $7150 \, \AA$ becoming particularly sensitive to the degree of $\alpha$-enhancement. This highlights that considering $\alpha$-element enhancement is important to derive more accurate stellar ages for quiescent galaxies. It should be noted, however, that these $\alpha$-enhanced SSP templates do not self-consistently account for the effect of [$\alpha$/Fe] on the stellar isochrones, which may change the lifetimes on the stellar main sequence, and hence the colour of the spectra. Thanks to recent advancements \citep{park24}, this will however become possible to self-consistently model. 


\subsection{Possible TP-AGB contribution}
\label{subsec:tpagb}

The importance and contribution of thermally pulsing asymptotic giant branch (TP-AGB) stars has been debated for decades \citep{maraston05, conroy09a, kriek10}. The contribution peaks in a stellar age range of $0.2-2$ Gyr, but large uncertainties in the modelling of the TP-AGB phase existed due to its double-shell burning regime, leading to instabilities and ``thermal pulses'' on short timescales alongside a strong mass loss. Since we are fitting the spectrum until $\sim1~\mu$m, where molecular features are present, TP-ABG could contribute to or even dominate the emission. \citet{lu24} suggest the presence of a strong TP-AGB component in a galaxy at $z=1.08$, which seems to have a similar stellar age and metallicity as ZF-UDS-7329, though possible a longer formation period. Depending on the TP-AGB models used, the authors find that the inferred ages of the delayed-$\tau$ model SFH can change by several 100 Myr. While we cannot exclude that a similar effect on our inferred ages, we note that we do not fit the MILES model beyond $\sim0.7~\mu$m due to resolution and find consistent age constraints with the C3K models (Table~\ref{t.prosp}). This is probably the case because the $D_n4000$ strength together with the age of the Universe derive our stellar age constraint.

\section{Conclusions}
\label{conclusion}

Our analysis of the JWST data for ZF-UDS-7329 provides new insights into the formation and evolution of massive quiescent galaxies in the early Universe. ZF-UDS-7329, with a stellar mass of $M_{\star}\approx10^{11.4}~\Msun$ and a sSFR of $\approx0.03~\Gyr^{-1}$ at $z=3.2$, has a strong 4000\AA break, implying an extremely high formation redshift. There are only a few spectroscopically confirmed massive quiescent galaxies at $z \gtrsim 3$ with such an extreme SFH, with most appearing as post-starbursts \citep{deugenio20}, or having quenched at $z \sim 4$ \citep{nanayakkara24}. Therefore, ZF-UDS-7329 presents an interesting object for understanding the complexities of early galaxy formation, and has been used for challenging cosmology \citep{glazebrook24} and claiming extreme galaxy formation physics \citep{carnall24}.

We extended the analysis by \citet{glazebrook24} to include different SFH priors, stellar libraries (MILES versus C3K), metallicity, and IMF assumptions. We re-reduce the NIRSpec/PRISM spectrum with the NIRSpec GTO pipeline \citep[e.g.,][]{curtis-lake23, curti23_ero, curti24_Z, bunker24}. Our findings show that the spectrum of ZF-UDS-7329 is consistent with a range of SFHs, suggesting stellar ages between 1.3 and 1.8 Gyr. This range reflects uncertainties but also indicates the robustness of our methodology in constraining the age and formation history of this galaxy. We also discuss an enhanced [$\alpha$/Fe] abundance, which we argue could reduce the real age by a $\sim200$ Myr. Options for fitting SPS models with $\alpha$-enhanced libraries are currently limited, but upcoming models will hopefully shed light on this hypothesis. 

Importantly, by employing a physically motivated rising SFH prior, which tracks dark matter accretion histories but allows for significant deviation from this, we derived a formation history for ZF-UDS-7329 that aligns well with direct high-redshift observations, indicating SFRs and sSFRs more consistent with known galaxies at $z>10$ (see Fig.~\ref{fig.priorSFHs}). We find that ZF-UDS-7329 is an extremely early forming galaxy, consistent with a stellar fraction (i.e., integrated star-formation efficiency) of $f_{\star} \approx 0.5-1.0$ at $z\approx7-12$, in particular when considering the impact of mergers (Fig.~\ref{fig.ageSFHs}). We stress that we infer the stellar fraction from the EVS model, where we assume that ZF-UDS-7329 traces the most massive halo in the probed cosmic volume at each redshift. A caveat of this approach is that each redshift is treated independently, i.e. there is a lack of continuity in the model. Nevertheless, this stellar fraction at early times is higher than lower-redshift estimates, but we find that ZF-UDS-7329 potentially evolves to a stellar fraction of 0.1 by $z=3.2$, which is consistent with the peak of the stellar-to-halo mass ratio at $z\approx0-3$. We therefore conclude that ZF-UDS-7329 formed very efficiently at high redshifts, but does not necessitate unseen galaxies at redshift $z>10$.

The presence of a colour gradient, as observed in NIRCam imaging data, suggests a complex morphology with an older, quiescent bulge component and a younger disc component (Fig.~\ref{fig.uvj}). This supports the notion that ZF-UDS-7329 may have undergone mergers, contributing to its mass assembly and extended SFH.

Our analysis also highlights the significance of the environment in shaping galaxy evolution. ZF-UDS-7329's possible association with an over-density or proto-cluster environment at $z=3.2$ is consistent with theoretical predictions of early structure formation in the $\Lambda$CDM cosmology: massive, early forming galaxies are high-$\sigma$ peaks and therefore highly clustered. 

In summary, ZF-UDS-7329 represents an early-forming galaxy. Its efficient star formation at high redshifts, coupled with its significant stellar mass and complex morphology, provides a valuable benchmark for future studies of galaxy evolution in the early Universe. Specifically, upcoming high spectral resolution JWST/NIRSpec IFU observations of ZF-UDS-7329 (JWST Proposal Cycle 3, ID \#5069) will constrain its dynamical mass, chemical abundances, gas properties, and spatially resolved SFH, helping us to test IMF assumptions, to constrain star-formation timescales, and to look for supermassive black hole feedback induced quenching.




\section*{Acknowledgements}
We thank the referee for their constructive comments that helped to improve the paper.
We are grateful to Andrea Ferrara, Avishai Dekel, and Andrey Kravtsov for insightful discussions. 
We thank Joop Schaye, Matthieu Schaller, Roi Kugel and John Helly for creating the FLAMINGO project and providing us with the data.
This work has been conducted as part of a Part III Project at the Cavendish Laboratory. CT thanks for insightful feedback from Didier Queloz. ST acknowledges support by the Royal Society Research Grant G125142. FDE acknowledges support by the Science and Technology Facilities Council (STFC), by the ERC through Advanced Grant 695671 ``QUENCH'', and by the UKRI Frontier Research grant RISEandFALL. K.G. and T.N. acknowledge support from Australian Research Council Laureate Fellowship FL180100060.
This research was supported in part by grant NSF PHY-2309135 to the Kavli Institute for Theoretical Physics (KITP). 
The NIRCam cutouts presented herein were retrieved from the Dawn JWST Archive (DJA). DJA is an initiative of the Cosmic Dawn Center (DAWN), which is funded by the Danish National Research Foundation under grant DNRF140. These data were reduced using a combination of the pipeline \textsc{jwst} and \textsc{grizli}\footnote{\href{10.5281/zenodo.1146904}{10.5281/zenodo.1146904}}. NIRSpec spectra from the RUBIES survey were obtained from the DJA, and reduced using the pipeline \textsc{jwst} and \textsc{msaexp}\footnote{\href{https://github.com/gbrammer/msaexp}{https://doi.org/10.5281/zenodo.7299500}}.
This work used the DiRAC@Durham facility managed by the Institute for Computational Cosmology on behalf of the STFC DiRAC HPC Facility (\href{www.dirac.ac.uk}{www.dirac.ac.uk}). The equipment was funded by BEIS capital funding via STFC capital grants ST/K00042X/1, ST/P002293/1, ST/R002371/1 and ST/S002502/1, Durham University and STFC operations grant ST/R000832/1. DiRAC is part of the National e-Infrastructure.

\section*{DATA AVAILABILITY}

The data underlying this article will be shared on reasonable request to the corresponding author.

 


\bibliographystyle{config/mnras}

\begin{thebibliography}{}
\makeatletter
\relax
\def\mn@urlcharsother{\let\do\@makeother \do\$\do\&\do\#\do\^\do\_\do\%\do\~}
\def\mn@doi{\begingroup\mn@urlcharsother \@ifnextchar [ {\mn@doi@}
  {\mn@doi@[]}}
\def\mn@doi@[#1]#2{\def\@tempa{#1}\ifx\@tempa\@empty \href
  {http://dx.doi.org/#2} {doi:#2}\else \href {http://dx.doi.org/#2} {#1}\fi
  \endgroup}
\def\mn@eprint#1#2{\mn@eprint@#1:#2::\@nil}
\def\mn@eprint@arXiv#1{\href {http://arxiv.org/abs/#1} {{\tt arXiv:#1}}}
\def\mn@eprint@dblp#1{\href {http://dblp.uni-trier.de/rec/bibtex/#1.xml}
  {dblp:#1}}
\def\mn@eprint@#1:#2:#3:#4\@nil{\def\@tempa {#1}\def\@tempb {#2}\def\@tempc
  {#3}\ifx \@tempc \@empty \let \@tempc \@tempb \let \@tempb \@tempa \fi \ifx
  \@tempb \@empty \def\@tempb {arXiv}\fi \@ifundefined
  {mn@eprint@\@tempb}{\@tempb:\@tempc}{\expandafter \expandafter \csname
  mn@eprint@\@tempb\endcsname \expandafter{\@tempc}}}

\bibitem[\protect\citeauthoryear{{Adams} et~al.,}{{Adams}
  et~al.}{2024}]{adams24}
{Adams} N.~J.,  et~al., 2024, \mn@doi [\apj] {10.3847/1538-4357/ad2a7b}, \href
  {https://ui.adsabs.harvard.edu/abs/2024ApJ...965..169A} {965, 169}

\bibitem[\protect\citeauthoryear{{Arribas} et~al.,}{{Arribas}
  et~al.}{2024}]{arribas24}
{Arribas} S.,  et~al., 2024, \mn@doi [\aap] {10.1051/0004-6361/202348824},
  \href {https://ui.adsabs.harvard.edu/abs/2024A&A...688A.146A} {688, A146}

\bibitem[\protect\citeauthoryear{{Asplund}, {Grevesse}, {Sauval}  \&
  {Scott}}{{Asplund} et~al.}{2009}]{asplund09}
{Asplund} M.,  {Grevesse} N.,  {Sauval} A.~J.,   {Scott} P.,  2009, \mn@doi
  [\araa] {10.1146/annurev.astro.46.060407.145222}, \href
  {http://adsabs.harvard.edu/abs/2009ARA%26A..47..481A} {47, 481}

\bibitem[\protect\citeauthoryear{{Behroozi} \& {Silk}}{{Behroozi} \&
  {Silk}}{2018}]{behroozi18}
{Behroozi} P.,  {Silk} J.,  2018, \mn@doi [\mnras] {10.1093/mnras/sty945},
  \href {https://ui.adsabs.harvard.edu/abs/2018MNRAS.477.5382B} {477, 5382}

\bibitem[\protect\citeauthoryear{{Behroozi}, {Wechsler}  \&
  {Conroy}}{{Behroozi} et~al.}{2013}]{behroozi13b}
{Behroozi} P.~S.,  {Wechsler} R.~H.,   {Conroy} C.,  2013, \mn@doi [\apj]
  {10.1088/0004-637X/770/1/57}, \href
  {http://adsabs.harvard.edu/abs/2013ApJ...770...57B} {770, 57}

\bibitem[\protect\citeauthoryear{{Behroozi} et~al.,}{{Behroozi}
  et~al.}{2020}]{behroozi20}
{Behroozi} P.,  et~al., 2020, \mn@doi [\mnras] {10.1093/mnras/staa3164}, \href
  {https://ui.adsabs.harvard.edu/abs/2020MNRAS.499.5702B} {499, 5702}

\bibitem[\protect\citeauthoryear{{Belli}, {Newman}  \& {Ellis}}{{Belli}
  et~al.}{2019}]{belli19}
{Belli} S.,  {Newman} A.~B.,   {Ellis} R.~S.,  2019, \mn@doi [\apj]
  {10.3847/1538-4357/ab07af}, \href
  {https://ui.adsabs.harvard.edu/abs/2019ApJ...874...17B} {874, 17}

\bibitem[\protect\citeauthoryear{{Belli} et~al.,}{{Belli}
  et~al.}{2024}]{belli24}
{Belli} S.,  et~al., 2024, \mn@doi [\nat] {10.1038/s41586-024-07412-1}, \href
  {https://ui.adsabs.harvard.edu/abs/2024Natur.630...54B} {630, 54}

\bibitem[\protect\citeauthoryear{{Beverage}, {Kriek}, {Conroy}, {Bezanson},
  {Franx}  \& {van der Wel}}{{Beverage} et~al.}{2021}]{beverage21}
{Beverage} A.~G.,  {Kriek} M.,  {Conroy} C.,  {Bezanson} R.,  {Franx} M.,
  {van der Wel} A.,  2021, \mn@doi [\apjl] {10.3847/2041-8213/ac12cd}, \href
  {https://ui.adsabs.harvard.edu/abs/2021ApJ...917L...1B} {917, L1}

\bibitem[\protect\citeauthoryear{{Beverage} et~al.,}{{Beverage}
  et~al.}{2024}]{beverage24}
{Beverage} A.~G.,  et~al., 2024, \mn@doi [arXiv e-prints]
  {10.48550/arXiv.2407.02556}, \href
  {https://ui.adsabs.harvard.edu/abs/2024arXiv240702556B} {p. arXiv:2407.02556}

\bibitem[\protect\citeauthoryear{Bingham et~al.,}{Bingham
  et~al.}{2019}]{bingham19}
Bingham E.,  et~al., 2019, Journal of Machine Learning Research, 20, 1

\bibitem[\protect\citeauthoryear{{Bond}, {Cole}, {Efstathiou}  \&
  {Kaiser}}{{Bond} et~al.}{1991}]{bond91}
{Bond} J.~R.,  {Cole} S.,  {Efstathiou} G.,   {Kaiser} N.,  1991, \mn@doi
  [\apj] {10.1086/170520}, \href
  {http://adsabs.harvard.edu/abs/1991ApJ...379..440B} {379, 440}

\bibitem[\protect\citeauthoryear{{Bouwens} et~al.,}{{Bouwens}
  et~al.}{2022}]{bouwens22}
{Bouwens} R.~J.,  et~al., 2022, \mn@doi [\apj] {10.3847/1538-4357/ac5a4a},
  \href {https://ui.adsabs.harvard.edu/abs/2022ApJ...931..160B} {931, 160}

\bibitem[\protect\citeauthoryear{{Brammer} et~al.,}{{Brammer}
  et~al.}{2012}]{brammer12}
{Brammer} G.~B.,  et~al., 2012, \mn@doi [\apjs] {10.1088/0067-0049/200/2/13},
  \href {https://ui.adsabs.harvard.edu/abs/2012ApJS..200...13B} {200, 13}

\bibitem[\protect\citeauthoryear{{Bunker} et~al.,}{{Bunker}
  et~al.}{2024}]{bunker24}
{Bunker} A.~J.,  et~al., 2024, \mn@doi [\aap] {10.1051/0004-6361/202347094},
  \href {https://ui.adsabs.harvard.edu/abs/2024A&A...690A.288B} {690, A288}

\bibitem[\protect\citeauthoryear{{Byler}, {Dalcanton}, {Conroy}, {Johnson},
  {Choi}, {Dotter}  \& {Rosenfield}}{{Byler} et~al.}{2019}]{byler19}
{Byler} N.,  {Dalcanton} J.~J.,  {Conroy} C.,  {Johnson} B.~D.,  {Choi} J.,
  {Dotter} A.,   {Rosenfield} P.,  2019, \mn@doi [\aj]
  {10.3847/1538-3881/ab1b70}, \href
  {https://ui.adsabs.harvard.edu/abs/2019AJ....158....2B} {158, 2}

\bibitem[\protect\citeauthoryear{{Calzetti}, {Armus}, {Bohlin}, {Kinney},
  {Koornneef}  \& {Storchi-Bergmann}}{{Calzetti} et~al.}{2000}]{calzetti00}
{Calzetti} D.,  {Armus} L.,  {Bohlin} R.~C.,  {Kinney} A.~L.,  {Koornneef} J.,
   {Storchi-Bergmann} T.,  2000, \mn@doi [\apj] {10.1086/308692}, \href
  {http://adsabs.harvard.edu/abs/2000ApJ...533..682C} {533, 682}

\bibitem[\protect\citeauthoryear{{Cappellari}}{{Cappellari}}{2023}]{cappellari23}
{Cappellari} M.,  2023, \mn@doi [\mnras] {10.1093/mnras/stad2597}, \href
  {https://ui.adsabs.harvard.edu/abs/2023MNRAS.526.3273C} {526, 3273}

\bibitem[\protect\citeauthoryear{{Cappellari} et~al.,}{{Cappellari}
  et~al.}{2013}]{cappellari13d}
{Cappellari} M.,  et~al., 2013, \mn@doi [\mnras] {10.1093/mnras/stt562}, \href
  {https://ui.adsabs.harvard.edu/abs/2013MNRAS.432.1709C} {432, 1709}

\bibitem[\protect\citeauthoryear{{Carnall} et~al.,}{{Carnall}
  et~al.}{2023}]{carnall23}
{Carnall} A.~C.,  et~al., 2023, \mn@doi [\nat] {10.1038/s41586-023-06158-6},
  \href {https://ui.adsabs.harvard.edu/abs/2023Natur.619..716C} {619, 716}

\bibitem[\protect\citeauthoryear{{Carnall} et~al.,}{{Carnall}
  et~al.}{2024}]{carnall24}
{Carnall} A.~C.,  et~al., 2024, \mn@doi [\mnras] {10.1093/mnras/stae2092},
  \href {https://ui.adsabs.harvard.edu/abs/2024MNRAS.534..325C} {534, 325}

\bibitem[\protect\citeauthoryear{{Carniani} et~al.,}{{Carniani}
  et~al.}{2024a}]{carniani24_z14}
{Carniani} S.,  et~al., 2024a, \mn@doi [\nat] {10.1038/s41586-024-07860-9},
  \href {https://ui.adsabs.harvard.edu/abs/2024Natur.633..318C} {633, 318}

\bibitem[\protect\citeauthoryear{{Carniani} et~al.,}{{Carniani}
  et~al.}{2024b}]{carniani24_outflow}
{Carniani} S.,  et~al., 2024b, \mn@doi [\aap] {10.1051/0004-6361/202347230},
  \href {https://ui.adsabs.harvard.edu/abs/2024A&A...685A..99C} {685, A99}

\bibitem[\protect\citeauthoryear{{Castellano} et~al.,}{{Castellano}
  et~al.}{2022}]{castellano22}
{Castellano} M.,  et~al., 2022, \mn@doi [\apjl] {10.3847/2041-8213/ac94d0},
  \href {https://ui.adsabs.harvard.edu/abs/2022ApJ...938L..15C} {938, L15}

\bibitem[\protect\citeauthoryear{{Castellano} et~al.,}{{Castellano}
  et~al.}{2024}]{castellano24}
{Castellano} M.,  et~al., 2024, \mn@doi [\apj] {10.3847/1538-4357/ad5f88},
  \href {https://ui.adsabs.harvard.edu/abs/2024ApJ...972..143C} {972, 143}

\bibitem[\protect\citeauthoryear{{Charlot} \& {Fall}}{{Charlot} \&
  {Fall}}{2000}]{charlot00}
{Charlot} S.,  {Fall} S.~M.,  2000, \mn@doi [\apj] {10.1086/309250}, \href
  {https://ui.adsabs.harvard.edu/abs/2000ApJ...539..718C} {539, 718}

\bibitem[\protect\citeauthoryear{{Chen}, {Mo}  \& {Wang}}{{Chen}
  et~al.}{2023}]{chen23_cv}
{Chen} Y.,  {Mo} H.~J.,   {Wang} K.,  2023, \mn@doi [\mnras]
  {10.1093/mnras/stad2866}, \href
  {https://ui.adsabs.harvard.edu/abs/2023MNRAS.526.2542C} {526, 2542}

\bibitem[\protect\citeauthoryear{{Chiang}, {Overzier}, {Gebhardt}  \&
  {Henriques}}{{Chiang} et~al.}{2017}]{chiang17}
{Chiang} Y.-K.,  {Overzier} R.~A.,  {Gebhardt} K.,   {Henriques} B.,  2017,
  \mn@doi [\apjl] {10.3847/2041-8213/aa7e7b}, \href
  {http://adsabs.harvard.edu/abs/2017ApJ...844L..23C} {844, L23}

\bibitem[\protect\citeauthoryear{{Choi}, {Dotter}, {Conroy}, {Cantiello},
  {Paxton}  \& {Johnson}}{{Choi} et~al.}{2016}]{choi16}
{Choi} J.,  {Dotter} A.,  {Conroy} C.,  {Cantiello} M.,  {Paxton} B.,
  {Johnson} B.~D.,  2016, \mn@doi [\apj] {10.3847/0004-637X/823/2/102}, \href
  {http://adsabs.harvard.edu/abs/2016ApJ...823..102C} {823, 102}

\bibitem[\protect\citeauthoryear{{Conroy}}{{Conroy}}{2013}]{conroy13_rev}
{Conroy} C.,  2013, \mn@doi [\araa] {10.1146/annurev-astro-082812-141017},
  \href {http://adsabs.harvard.edu/abs/2013ARA%26A..51..393C} {51, 393}

\bibitem[\protect\citeauthoryear{{Conroy} \& {Gunn}}{{Conroy} \&
  {Gunn}}{2010}]{conroy10_code}
{Conroy} C.,  {Gunn} J.~E.,  2010, {FSPS: Flexible Stellar Population
  Synthesis}, Astrophysics Source Code Library, record ascl:1010.043

\bibitem[\protect\citeauthoryear{{Conroy} \& {van Dokkum}}{{Conroy} \& {van
  Dokkum}}{2012}]{conroy12}
{Conroy} C.,  {van Dokkum} P.~G.,  2012, \mn@doi [\apj]
  {10.1088/0004-637X/760/1/71}, \href
  {https://ui.adsabs.harvard.edu/abs/2012ApJ...760...71C} {760, 71}

\bibitem[\protect\citeauthoryear{{Conroy}, {Gunn}  \& {White}}{{Conroy}
  et~al.}{2009}]{conroy09a}
{Conroy} C.,  {Gunn} J.~E.,   {White} M.,  2009, \mn@doi [\apj]
  {10.1088/0004-637X/699/1/486}, \href
  {http://adsabs.harvard.edu/abs/2009ApJ...699..486C} {699, 486}

\bibitem[\protect\citeauthoryear{{Conroy}, {Villaume}, {van Dokkum}  \&
  {Lind}}{{Conroy} et~al.}{2018}]{conroy18}
{Conroy} C.,  {Villaume} A.,  {van Dokkum} P.~G.,   {Lind} K.,  2018, \mn@doi
  [\apj] {10.3847/1538-4357/aaab49}, \href
  {http://adsabs.harvard.edu/abs/2018ApJ...854..139C} {854, 139}

\bibitem[\protect\citeauthoryear{{Conroy}, {Naidu}, {Zaritsky}, {Bonaca},
  {Cargile}, {Johnson}  \& {Caldwell}}{{Conroy} et~al.}{2019}]{conroy19}
{Conroy} C.,  {Naidu} R.~P.,  {Zaritsky} D.,  {Bonaca} A.,  {Cargile} P.,
  {Johnson} B.~D.,   {Caldwell} N.,  2019, \mn@doi [\apj]
  {10.3847/1538-4357/ab5710}, \href
  {https://ui.adsabs.harvard.edu/abs/2019ApJ...887..237C} {887, 237}

\bibitem[\protect\citeauthoryear{{Cowley}, {Baugh}, {Cole}, {Frenk}  \&
  {Lacey}}{{Cowley} et~al.}{2018}]{cowley18}
{Cowley} W.~I.,  {Baugh} C.~M.,  {Cole} S.,  {Frenk} C.~S.,   {Lacey} C.~G.,
  2018, \mn@doi [\mnras] {10.1093/mnras/stx2897}, \href
  {http://adsabs.harvard.edu/abs/2018MNRAS.474.2352C} {474, 2352}

\bibitem[\protect\citeauthoryear{{Cueto}, {Hutter}, {Dayal}, {Gottl{\"o}ber},
  {Heintz}, {Mason}, {Trebitsch}  \& {Yepes}}{{Cueto} et~al.}{2024}]{cueto24}
{Cueto} E.~R.,  {Hutter} A.,  {Dayal} P.,  {Gottl{\"o}ber} S.,  {Heintz} K.~E.,
   {Mason} C.,  {Trebitsch} M.,   {Yepes} G.,  2024, \mn@doi [\aap]
  {10.1051/0004-6361/202349017}, \href
  {https://ui.adsabs.harvard.edu/abs/2024A&A...686A.138C} {686, A138}

\bibitem[\protect\citeauthoryear{{Curti} et~al.,}{{Curti}
  et~al.}{2023}]{curti23_ero}
{Curti} M.,  et~al., 2023, \mn@doi [\mnras] {10.1093/mnras/stac2737}, \href
  {https://ui.adsabs.harvard.edu/abs/2023MNRAS.518..425C} {518, 425}

\bibitem[\protect\citeauthoryear{{Curti} et~al.,}{{Curti}
  et~al.}{2024}]{curti24_Z}
{Curti} M.,  et~al., 2024, \mn@doi [\aap] {10.1051/0004-6361/202346698}, \href
  {https://ui.adsabs.harvard.edu/abs/2024A&A...684A..75C} {684, A75}

\bibitem[\protect\citeauthoryear{{Curtis-Lake} et~al.,}{{Curtis-Lake}
  et~al.}{2023}]{curtis-lake23}
{Curtis-Lake} E.,  et~al., 2023, \mn@doi [Nature Astronomy]
  {10.1038/s41550-023-01918-w}, \href
  {https://ui.adsabs.harvard.edu/abs/2023NatAs...7..622C} {7, 622}

\bibitem[\protect\citeauthoryear{{D'Eugenio} et~al.,}{{D'Eugenio}
  et~al.}{2020}]{deugenio20}
{D'Eugenio} C.,  et~al., 2020, \mn@doi [\apjl] {10.3847/2041-8213/ab7a96},
  \href {https://ui.adsabs.harvard.edu/abs/2020ApJ...892L...2D} {892, L2}

\bibitem[\protect\citeauthoryear{{D'Eugenio}, {Daddi}, {Liu}  \&
  {Gobat}}{{D'Eugenio} et~al.}{2023}]{deugenio23_cii}
{D'Eugenio} C.,  {Daddi} E.,  {Liu} D.,   {Gobat} R.,  2023, \mn@doi [\aap]
  {10.1051/0004-6361/202347233}, \href
  {https://ui.adsabs.harvard.edu/abs/2023A&A...678L...9D} {678, L9}

\bibitem[\protect\citeauthoryear{{D'Eugenio} et~al.,}{{D'Eugenio}
  et~al.}{2024a}]{deugenio24}
{D'Eugenio} F.,  et~al., 2024a, \mn@doi [arXiv e-prints]
  {10.48550/arXiv.2404.06531}, \href
  {https://ui.adsabs.harvard.edu/abs/2024arXiv240406531D} {p. arXiv:2404.06531}

\bibitem[\protect\citeauthoryear{{D'Eugenio} et~al.,}{{D'Eugenio}
  et~al.}{2024b}]{deugenio24_psb}
{D'Eugenio} F.,  et~al., 2024b, \mn@doi [Nature Astronomy]
  {10.1038/s41550-024-02345-1}, \href
  {https://ui.adsabs.harvard.edu/abs/2024NatAs...8.1443D} {8, 1443}

\bibitem[\protect\citeauthoryear{{D'Eugenio} et~al.,}{{D'Eugenio}
  et~al.}{2024c}]{deugenio24_C}
{D'Eugenio} F.,  et~al., 2024c, \mn@doi [\aap] {10.1051/0004-6361/202348636},
  \href {https://ui.adsabs.harvard.edu/abs/2024A&A...689A.152D} {689, A152}

\bibitem[\protect\citeauthoryear{{D'Eugenio} et~al.,}{{D'Eugenio}
  et~al.}{2025}]{deugenio25}
{D'Eugenio} F.,  et~al., 2025, \mn@doi [\mnras] {10.1093/mnras/stae2545}, \href
  {https://ui.adsabs.harvard.edu/abs/2025MNRAS.536...51D} {536, 51}

\bibitem[\protect\citeauthoryear{{Dabringhausen}, {Kroupa}  \&
  {Baumgardt}}{{Dabringhausen} et~al.}{2009}]{dabringhausen09}
{Dabringhausen} J.,  {Kroupa} P.,   {Baumgardt} H.,  2009, \mn@doi [\mnras]
  {10.1111/j.1365-2966.2009.14425.x}, \href
  {https://ui.adsabs.harvard.edu/abs/2009MNRAS.394.1529D} {394, 1529}

\bibitem[\protect\citeauthoryear{{Dav{\'e}}, {Angl{\'e}s-Alc{\'a}zar},
  {Narayanan}, {Li}, {Rafieferantsoa}  \& {Appleby}}{{Dav{\'e}}
  et~al.}{2019}]{dave19}
{Dav{\'e}} R.,  {Angl{\'e}s-Alc{\'a}zar} D.,  {Narayanan} D.,  {Li} Q.,
  {Rafieferantsoa} M.~H.,   {Appleby} S.,  2019, \mn@doi [\mnras]
  {10.1093/mnras/stz937}, \href
  {https://ui.adsabs.harvard.edu/abs/2019MNRAS.486.2827D} {486, 2827}

\bibitem[\protect\citeauthoryear{{Davies} et~al.,}{{Davies}
  et~al.}{2024}]{davies24}
{Davies} R.~L.,  et~al., 2024, \mn@doi [\mnras] {10.1093/mnras/stae327}, \href
  {https://ui.adsabs.harvard.edu/abs/2024MNRAS.528.4976D} {528, 4976}

\bibitem[\protect\citeauthoryear{{Dayal}, {Ferrara}, {Dunlop}  \&
  {Pacucci}}{{Dayal} et~al.}{2014}]{dayal14}
{Dayal} P.,  {Ferrara} A.,  {Dunlop} J.~S.,   {Pacucci} F.,  2014, \mn@doi
  [\mnras] {10.1093/mnras/stu1848}, \href
  {https://ui.adsabs.harvard.edu/abs/2014MNRAS.445.2545D} {445, 2545}

\bibitem[\protect\citeauthoryear{{Dekel} \& {Burkert}}{{Dekel} \&
  {Burkert}}{2014}]{dekel14_nugget}
{Dekel} A.,  {Burkert} A.,  2014, \mn@doi [\mnras] {10.1093/mnras/stt2331},
  \href {http://adsabs.harvard.edu/abs/2014MNRAS.438.1870D} {438, 1870}

\bibitem[\protect\citeauthoryear{{Dekel}, {Zolotov}, {Tweed}, {Cacciato},
  {Ceverino}  \& {Primack}}{{Dekel} et~al.}{2013}]{dekel13}
{Dekel} A.,  {Zolotov} A.,  {Tweed} D.,  {Cacciato} M.,  {Ceverino} D.,
  {Primack} J.~R.,  2013, \mn@doi [\mnras] {10.1093/mnras/stt1338}, \href
  {http://adsabs.harvard.edu/abs/2013MNRAS.435..999D} {435, 999}

\bibitem[\protect\citeauthoryear{{Dekel}, {Sarkar}, {Birnboim}, {Mandelker}  \&
  {Li}}{{Dekel} et~al.}{2023}]{dekel23}
{Dekel} A.,  {Sarkar} K.~C.,  {Birnboim} Y.,  {Mandelker} N.,   {Li} Z.,  2023,
  \mn@doi [\mnras] {10.1093/mnras/stad1557}, \href
  {https://ui.adsabs.harvard.edu/abs/2023MNRAS.523.3201D} {523, 3201}

\bibitem[\protect\citeauthoryear{{Ding}, {Silverman}, {Birrer}, {Treu}, {Tang},
  {Yang}  \& {Bottrell}}{{Ding} et~al.}{2022}]{ding22_code}
{Ding} X.,  {Silverman} J.,  {Birrer} S.,  {Treu} T.,  {Tang} S.,  {Yang} L.,
  {Bottrell} C.,  2022, {GaLight: 2D modeling of galaxy images}, Astrophysics
  Source Code Library, record ascl:2209.011 (\mn@eprint {ascl} {2209.011})

\bibitem[\protect\citeauthoryear{{Donnan} et~al.,}{{Donnan}
  et~al.}{2024}]{donnan24}
{Donnan} C.~T.,  et~al., 2024, \mn@doi [\mnras] {10.1093/mnras/stae2037}, \href
  {https://ui.adsabs.harvard.edu/abs/2024MNRAS.533.3222D} {533, 3222}

\bibitem[\protect\citeauthoryear{{Dotter}}{{Dotter}}{2016}]{dotter16}
{Dotter} A.,  2016, \mn@doi [\apjs] {10.3847/0067-0049/222/1/8}, \href
  {https://ui.adsabs.harvard.edu/abs/2016ApJS..222....8D} {222, 8}

\bibitem[\protect\citeauthoryear{{Endsley} et~al.,}{{Endsley}
  et~al.}{2024}]{endsley24}
{Endsley} R.,  et~al., 2024, \mn@doi [\mnras] {10.1093/mnras/stae1857}, \href
  {https://ui.adsabs.harvard.edu/abs/2024MNRAS.533.1111E} {533, 1111}

\bibitem[\protect\citeauthoryear{{Esdaile} et~al.,}{{Esdaile}
  et~al.}{2021}]{esdaile21}
{Esdaile} J.,  et~al., 2021, \mn@doi [\apjl] {10.3847/2041-8213/abe11e}, \href
  {https://ui.adsabs.harvard.edu/abs/2021ApJ...908L..35E} {908, L35}

\bibitem[\protect\citeauthoryear{{Fakhouri}, {Ma}  \&
  {Boylan-Kolchin}}{{Fakhouri} et~al.}{2010}]{fakhouri10_dm}
{Fakhouri} O.,  {Ma} C.-P.,   {Boylan-Kolchin} M.,  2010, \mn@doi [\mnras]
  {10.1111/j.1365-2966.2010.16859.x}, \href
  {http://adsabs.harvard.edu/abs/2010MNRAS.406.2267F} {406, 2267}

\bibitem[\protect\citeauthoryear{{Ferland} et~al.,}{{Ferland}
  et~al.}{2013}]{ferland13}
{Ferland} G.~J.,  et~al., 2013, \rmxaa, \href
  {https://ui.adsabs.harvard.edu/abs/2013RMxAA..49..137F} {49, 137}

\bibitem[\protect\citeauthoryear{{Ferruit} et~al.,}{{Ferruit}
  et~al.}{2022}]{ferruit22}
{Ferruit} P.,  et~al., 2022, \mn@doi [\aap] {10.1051/0004-6361/202142673},
  \href {https://ui.adsabs.harvard.edu/abs/2022A&A...661A..81F} {661, A81}

\bibitem[\protect\citeauthoryear{{Finkelstein} et~al.,}{{Finkelstein}
  et~al.}{2023}]{finkelstein23_ceers}
{Finkelstein} S.~L.,  et~al., 2023, \mn@doi [\apjl] {10.3847/2041-8213/acade4},
  \href {https://ui.adsabs.harvard.edu/abs/2023ApJ...946L..13F} {946, L13}

\bibitem[\protect\citeauthoryear{{Finkelstein} et~al.,}{{Finkelstein}
  et~al.}{2024}]{finkelstein24}
{Finkelstein} S.~L.,  et~al., 2024, \mn@doi [\apjl] {10.3847/2041-8213/ad4495},
  \href {https://ui.adsabs.harvard.edu/abs/2024ApJ...969L...2F} {969, L2}

\bibitem[\protect\citeauthoryear{{Fujimoto} et~al.,}{{Fujimoto}
  et~al.}{2023}]{fujimoto23_zspec}
{Fujimoto} S.,  et~al., 2023, \mn@doi [\apjl] {10.3847/2041-8213/acd2d9}, \href
  {https://ui.adsabs.harvard.edu/abs/2023ApJ...949L..25F} {949, L25}

\bibitem[\protect\citeauthoryear{{Geda}, {Crawford}, {Hunt}, {Bershady},
  {Tollerud}  \& {Randriamampandry}}{{Geda} et~al.}{2022}]{geda22}
{Geda} R.,  {Crawford} S.~M.,  {Hunt} L.,  {Bershady} M.,  {Tollerud} E.,
  {Randriamampandry} S.,  2022, \mn@doi [\aj] {10.3847/1538-3881/ac5908}, \href
  {https://ui.adsabs.harvard.edu/abs/2022AJ....163..202G} {163, 202}

\bibitem[\protect\citeauthoryear{{Gelli}, {Mason}  \& {Hayward}}{{Gelli}
  et~al.}{2024}]{gelli24}
{Gelli} V.,  {Mason} C.,   {Hayward} C.~C.,  2024, \mn@doi [\apj]
  {10.3847/1538-4357/ad7b36}, \href
  {https://ui.adsabs.harvard.edu/abs/2024ApJ...975..192G} {975, 192}

\bibitem[\protect\citeauthoryear{{Genel}, {Bouch{\'e}}, {Naab}, {Sternberg}  \&
  {Genzel}}{{Genel} et~al.}{2010}]{genel10}
{Genel} S.,  {Bouch{\'e}} N.,  {Naab} T.,  {Sternberg} A.,   {Genzel} R.,
  2010, \mn@doi [\apj] {10.1088/0004-637X/719/1/229}, \href
  {http://adsabs.harvard.edu/abs/2010ApJ...719..229G} {719, 229}

\bibitem[\protect\citeauthoryear{{Glazebrook} et~al.,}{{Glazebrook}
  et~al.}{2024}]{glazebrook24}
{Glazebrook} K.,  et~al., 2024, \mn@doi [\nat] {10.1038/s41586-024-07191-9},
  \href {https://ui.adsabs.harvard.edu/abs/2024Natur.628..277G} {628, 277}

\bibitem[\protect\citeauthoryear{{Greggio} \& {Renzini}}{{Greggio} \&
  {Renzini}}{2011}]{greggio11_book}
{Greggio} L.,  {Renzini} A.,  2011, {Stellar Populations. A User Guide from Low
  to High Redshift}

\bibitem[\protect\citeauthoryear{{Hainline} et~al.,}{{Hainline}
  et~al.}{2024}]{hainline24}
{Hainline} K.~N.,  et~al., 2024, \mn@doi [\apj] {10.3847/1538-4357/ad1ee4},
  \href {https://ui.adsabs.harvard.edu/abs/2024ApJ...964...71H} {964, 71}

\bibitem[\protect\citeauthoryear{{Harikane} et~al.,}{{Harikane}
  et~al.}{2023}]{harikane23_uvlf}
{Harikane} Y.,  et~al., 2023, \mn@doi [\apjs] {10.3847/1538-4365/acaaa9}, \href
  {https://ui.adsabs.harvard.edu/abs/2023ApJS..265....5H} {265, 5}

\bibitem[\protect\citeauthoryear{{Harrison} \& {Coles}}{{Harrison} \&
  {Coles}}{2011}]{harrison11}
{Harrison} I.,  {Coles} P.,  2011, \mn@doi [\mnras]
  {10.1111/j.1745-3933.2011.01134.x}, \href
  {https://ui.adsabs.harvard.edu/abs/2011MNRAS.418L..20H} {418, L20}

\bibitem[\protect\citeauthoryear{{Hegde}, {Wyatt}  \& {Furlanetto}}{{Hegde}
  et~al.}{2024}]{hegde24}
{Hegde} S.,  {Wyatt} M.~M.,   {Furlanetto} S.~R.,  2024, \mn@doi [\jcap]
  {10.1088/1475-7516/2024/08/025}, \href
  {https://ui.adsabs.harvard.edu/abs/2024JCAP...08..025H} {2024, 025}

\bibitem[\protect\citeauthoryear{{Heintz} et~al.,}{{Heintz}
  et~al.}{2024}]{heintz24_dja}
{Heintz} K.~E.,  et~al., 2024, \mn@doi [arXiv e-prints]
  {10.48550/arXiv.2404.02211}, \href
  {https://ui.adsabs.harvard.edu/abs/2024arXiv240402211H} {p. arXiv:2404.02211}

\bibitem[\protect\citeauthoryear{{Helton} et~al.,}{{Helton}
  et~al.}{2024a}]{helton24}
{Helton} J.~M.,  et~al., 2024a, \mn@doi [arXiv e-prints]
  {10.48550/arXiv.2405.18462}, \href
  {https://ui.adsabs.harvard.edu/abs/2024arXiv240518462H} {p. arXiv:2405.18462}

\bibitem[\protect\citeauthoryear{{Helton} et~al.,}{{Helton}
  et~al.}{2024b}]{helton24_overdensity}
{Helton} J.~M.,  et~al., 2024b, \mn@doi [\apj] {10.3847/1538-4357/ad0da7},
  \href {https://ui.adsabs.harvard.edu/abs/2024ApJ...962..124H} {962, 124}

\bibitem[\protect\citeauthoryear{{Higson}, {Handley}, {Hobson}  \&
  {Lasenby}}{{Higson} et~al.}{2018}]{higson18}
{Higson} E.,  {Handley} W.,  {Hobson} M.,   {Lasenby} A.,  2018, \mn@doi
  [Bayesian Analysis] {10.1214/17-BA1075}, \href
  {https://ui.adsabs.harvard.edu/abs/2018BayAn..13..873H} {13, 873}

\bibitem[\protect\citeauthoryear{{Higuchi} et~al.,}{{Higuchi}
  et~al.}{2019}]{higuchi19}
{Higuchi} R.,  et~al., 2019, \mn@doi [\apj] {10.3847/1538-4357/ab2192}, \href
  {https://ui.adsabs.harvard.edu/abs/2019ApJ...879...28H} {879, 28}

\bibitem[\protect\citeauthoryear{{Hopkins}, {Hernquist}, {Cox}, {Di Matteo},
  {Robertson}  \& {Springel}}{{Hopkins} et~al.}{2006}]{hopkins06a}
{Hopkins} P.~F.,  {Hernquist} L.,  {Cox} T.~J.,  {Di Matteo} T.,  {Robertson}
  B.,   {Springel} V.,  2006, \mn@doi [\apjs] {10.1086/499298}, \href
  {http://adsabs.harvard.edu/abs/2006ApJS..163....1H} {163, 1}

\bibitem[\protect\citeauthoryear{{Hopkins}, {Cox}, {Hernquist}, {Narayanan},
  {Hayward}  \& {Murray}}{{Hopkins} et~al.}{2013}]{hopkins13_merger}
{Hopkins} P.~F.,  {Cox} T.~J.,  {Hernquist} L.,  {Narayanan} D.,  {Hayward}
  C.~C.,   {Murray} N.,  2013, \mn@doi [\mnras] {10.1093/mnras/stt017}, \href
  {https://ui.adsabs.harvard.edu/abs/2013MNRAS.430.1901H} {430, 1901}

\bibitem[\protect\citeauthoryear{{Inayoshi}, {Harikane}, {Inoue}, {Li}  \&
  {Ho}}{{Inayoshi} et~al.}{2022}]{inayoshi22}
{Inayoshi} K.,  {Harikane} Y.,  {Inoue} A.~K.,  {Li} W.,   {Ho} L.~C.,  2022,
  \mn@doi [\apjl] {10.3847/2041-8213/ac9310}, \href
  {https://ui.adsabs.harvard.edu/abs/2022ApJ...938L..10I} {938, L10}

\bibitem[\protect\citeauthoryear{{Jakobsen} et~al.,}{{Jakobsen}
  et~al.}{2022}]{jakobsen22}
{Jakobsen} P.,  et~al., 2022, \mn@doi [\aap] {10.1051/0004-6361/202142663},
  \href {https://ui.adsabs.harvard.edu/abs/2022A&A...661A..80J} {661, A80}

\bibitem[\protect\citeauthoryear{{Jeong}, {Yi}, {Kyeong}, {Sarzi}, {Sung}  \&
  {Oh}}{{Jeong} et~al.}{2013}]{jeong13}
{Jeong} H.,  {Yi} S.~K.,  {Kyeong} J.,  {Sarzi} M.,  {Sung} E.-C.,   {Oh} K.,
  2013, \mn@doi [\apjs] {10.1088/0067-0049/208/1/7}, \href
  {https://ui.adsabs.harvard.edu/abs/2013ApJS..208....7J} {208, 7}

\bibitem[\protect\citeauthoryear{{Ji} et~al.,}{{Ji} et~al.}{2024}]{ji24}
{Ji} Z.,  et~al., 2024, \mn@doi [\apj] {10.3847/1538-4357/ad6e7f}, \href
  {https://ui.adsabs.harvard.edu/abs/2024ApJ...974..135J} {974, 135}

\bibitem[\protect\citeauthoryear{{Johnson}, {Leja}, {Conroy}  \&
  {Speagle}}{{Johnson} et~al.}{2019}]{johnson19}
{Johnson} B.~D.,  {Leja} J.~L.,  {Conroy} C.,   {Speagle} J.~S.,  2019,
  {Prospector: Stellar population inference from spectra and SEDs} (\mn@eprint
  {ascl} {1905.025})

\bibitem[\protect\citeauthoryear{{Johnson}, {Leja}, {Conroy}  \&
  {Speagle}}{{Johnson} et~al.}{2021}]{johnson21}
{Johnson} B.~D.,  {Leja} J.,  {Conroy} C.,   {Speagle} J.~S.,  2021, \mn@doi
  [\apjs] {10.3847/1538-4365/abef67}, \href
  {https://ui.adsabs.harvard.edu/abs/2021ApJS..254...22J} {254, 22}

\bibitem[\protect\citeauthoryear{{Kannan} et~al.,}{{Kannan}
  et~al.}{2023}]{kannan23}
{Kannan} R.,  et~al., 2023, \mn@doi [\mnras] {10.1093/mnras/stac3743}, \href
  {https://ui.adsabs.harvard.edu/abs/2023MNRAS.524.2594K} {524, 2594}

\bibitem[\protect\citeauthoryear{{Kocevski} et~al.,}{{Kocevski}
  et~al.}{2023}]{kocevski23}
{Kocevski} D.~D.,  et~al., 2023, \mn@doi [\apjl] {10.3847/2041-8213/acad00},
  \href {https://ui.adsabs.harvard.edu/abs/2023ApJ...946L..14K} {946, L14}

\bibitem[\protect\citeauthoryear{{Kocevski} et~al.,}{{Kocevski}
  et~al.}{2024}]{kocevski24}
{Kocevski} D.~D.,  et~al., 2024, \mn@doi [arXiv e-prints]
  {10.48550/arXiv.2404.03576}, \href
  {https://ui.adsabs.harvard.edu/abs/2024arXiv240403576K} {p. arXiv:2404.03576}

\bibitem[\protect\citeauthoryear{{Koposov} et~al.,}{{Koposov}
  et~al.}{2022}]{koposov22}
{Koposov} S.,  et~al., 2022, {joshspeagle/dynesty: v2.0.3},
  \mn@doi{10.5281/zenodo.7388523}

\bibitem[\protect\citeauthoryear{{Kravtsov}}{{Kravtsov}}{2013}]{kravtsov13}
{Kravtsov} A.~V.,  2013, \mn@doi [\apjl] {10.1088/2041-8205/764/2/L31}, \href
  {http://adsabs.harvard.edu/abs/2013ApJ...764L..31K} {764, L31}

\bibitem[\protect\citeauthoryear{{Kravtsov} \& {Belokurov}}{{Kravtsov} \&
  {Belokurov}}{2024}]{kravtsov24}
{Kravtsov} A.,  {Belokurov} V.,  2024, \mn@doi [arXiv e-prints]
  {10.48550/arXiv.2405.04578}, \href
  {https://ui.adsabs.harvard.edu/abs/2024arXiv240504578K} {p. arXiv:2405.04578}

\bibitem[\protect\citeauthoryear{{Kriek} \& {Conroy}}{{Kriek} \&
  {Conroy}}{2013}]{kriek13}
{Kriek} M.,  {Conroy} C.,  2013, \mn@doi [\apjl] {10.1088/2041-8205/775/1/L16},
  \href {http://adsabs.harvard.edu/abs/2013ApJ...775L..16K} {775, L16}

\bibitem[\protect\citeauthoryear{{Kriek} et~al.,}{{Kriek}
  et~al.}{2010}]{kriek10}
{Kriek} M.,  et~al., 2010, \mn@doi [\apjl] {10.1088/2041-8205/722/1/L64}, \href
  {https://ui.adsabs.harvard.edu/abs/2010ApJ...722L..64K} {722, L64}

\bibitem[\protect\citeauthoryear{{Kroupa}}{{Kroupa}}{2001}]{kroupa01}
{Kroupa} P.,  2001, \mn@doi [\mnras] {10.1046/j.1365-8711.2001.04022.x}, \href
  {http://adsabs.harvard.edu/abs/2001MNRAS.322..231K} {322, 231}

\bibitem[\protect\citeauthoryear{{Kugel} et~al.,}{{Kugel}
  et~al.}{2023}]{kugel23}
{Kugel} R.,  et~al., 2023, \mn@doi [\mnras] {10.1093/mnras/stad2540}, \href
  {https://ui.adsabs.harvard.edu/abs/2023MNRAS.526.6103K} {526, 6103}

\bibitem[\protect\citeauthoryear{{La Barbera}, {Ferreras}, {Vazdekis}, {de la
  Rosa}, {de Carvalho}, {Trevisan}, {Falc{\'o}n-Barroso}  \&
  {Ricciardelli}}{{La Barbera} et~al.}{2013}]{la-barbera13}
{La Barbera} F.,  {Ferreras} I.,  {Vazdekis} A.,  {de la Rosa} I.~G.,  {de
  Carvalho} R.~R.,  {Trevisan} M.,  {Falc{\'o}n-Barroso} J.,   {Ricciardelli}
  E.,  2013, \mn@doi [\mnras] {10.1093/mnras/stt943}, \href
  {https://ui.adsabs.harvard.edu/abs/2013MNRAS.433.3017L} {433, 3017}

\bibitem[\protect\citeauthoryear{{La Barbera}, {Vazdekis}, {Ferreras},
  {Pasquali}, {Allende Prieto}, {R{\"o}ck}, {Aguado}  \& {Peletier}}{{La
  Barbera} et~al.}{2017}]{la-barbera17}
{La Barbera} F.,  {Vazdekis} A.,  {Ferreras} I.,  {Pasquali} A.,  {Allende
  Prieto} C.,  {R{\"o}ck} B.,  {Aguado} D.~S.,   {Peletier} R.~F.,  2017,
  \mn@doi [\mnras] {10.1093/mnras/stw2407}, \href
  {https://ui.adsabs.harvard.edu/abs/2017MNRAS.464.3597L} {464, 3597}

\bibitem[\protect\citeauthoryear{{Lagos} et~al.,}{{Lagos}
  et~al.}{2024}]{lagos24}
{Lagos} C. d.~P.,  et~al., 2024, \mn@doi [arXiv e-prints]
  {10.48550/arXiv.2409.16916}, \href
  {https://ui.adsabs.harvard.edu/abs/2024arXiv240916916L} {p. arXiv:2409.16916}

\bibitem[\protect\citeauthoryear{{Lapiner} et~al.,}{{Lapiner}
  et~al.}{2023}]{lapiner23}
{Lapiner} S.,  et~al., 2023, \mn@doi [\mnras] {10.1093/mnras/stad1263}, \href
  {https://ui.adsabs.harvard.edu/abs/2023MNRAS.522.4515L} {522, 4515}

\bibitem[\protect\citeauthoryear{{Leja}, {Carnall}, {Johnson}, {Conroy}  \&
  {Speagle}}{{Leja} et~al.}{2019a}]{leja19_nonparm}
{Leja} J.,  {Carnall} A.~C.,  {Johnson} B.~D.,  {Conroy} C.,   {Speagle} J.~S.,
   2019a, \mn@doi [\apj] {10.3847/1538-4357/ab133c}, \href
  {https://ui.adsabs.harvard.edu/abs/2019ApJ...876....3L} {876, 3}

\bibitem[\protect\citeauthoryear{{Leja} et~al.,}{{Leja} et~al.}{2019b}]{leja19}
{Leja} J.,  et~al., 2019b, \mn@doi [\apj] {10.3847/1538-4357/ab1d5a}, \href
  {https://ui.adsabs.harvard.edu/abs/2019ApJ...877..140L} {877, 140}

\bibitem[\protect\citeauthoryear{{Leja}, {Tacchella}  \& {Conroy}}{{Leja}
  et~al.}{2019c}]{leja19_uvj}
{Leja} J.,  {Tacchella} S.,   {Conroy} C.,  2019c, \mn@doi [\apjl]
  {10.3847/2041-8213/ab2f8c}, \href
  {https://ui.adsabs.harvard.edu/abs/2019ApJ...880L...9L} {880, L9}

\bibitem[\protect\citeauthoryear{{Li}, {Dekel}, {Sarkar}, {Aung}, {Giavalisco},
  {Mandelker}  \& {Tacchella}}{{Li} et~al.}{2024}]{li24_fbb}
{Li} Z.,  {Dekel} A.,  {Sarkar} K.~C.,  {Aung} H.,  {Giavalisco} M.,
  {Mandelker} N.,   {Tacchella} S.,  2024, \mn@doi [\aap]
  {10.1051/0004-6361/202348727}, \href
  {https://ui.adsabs.harvard.edu/abs/2024A&A...690A.108L} {690, A108}

\bibitem[\protect\citeauthoryear{{Lim}, {Tacchella}, {Schaye}, {Schaller},
  {Helton}, {Kugel}  \& {Maiolino}}{{Lim} et~al.}{2024}]{lim24}
{Lim} S.,  {Tacchella} S.,  {Schaye} J.,  {Schaller} M.,  {Helton} J.~M.,
  {Kugel} R.,   {Maiolino} R.,  2024, \mn@doi [\mnras]
  {10.1093/mnras/stae1790}, \href
  {https://ui.adsabs.harvard.edu/abs/2024MNRAS.532.4551L} {532, 4551}

\bibitem[\protect\citeauthoryear{{Looser} et~al.,}{{Looser}
  et~al.}{2023}]{looser23_pop}
{Looser} T.~J.,  et~al., 2023, \mn@doi [arXiv e-prints]
  {10.48550/arXiv.2306.02470}, \href
  {https://ui.adsabs.harvard.edu/abs/2023arXiv230602470L} {p. arXiv:2306.02470}

\bibitem[\protect\citeauthoryear{{Looser} et~al.,}{{Looser}
  et~al.}{2024}]{looser24}
{Looser} T.~J.,  et~al., 2024, \mn@doi [\nat] {10.1038/s41586-024-07227-0},
  \href {https://ui.adsabs.harvard.edu/abs/2024Natur.629...53L} {629, 53}

\bibitem[\protect\citeauthoryear{{Lovell}, {Harrison}, {Harikane}, {Tacchella}
  \& {Wilkins}}{{Lovell} et~al.}{2023}]{lovell23_evs}
{Lovell} C.~C.,  {Harrison} I.,  {Harikane} Y.,  {Tacchella} S.,   {Wilkins}
  S.~M.,  2023, \mn@doi [\mnras] {10.1093/mnras/stac3224}, \href
  {https://ui.adsabs.harvard.edu/abs/2023MNRAS.518.2511L} {518, 2511}

\bibitem[\protect\citeauthoryear{{Lu} et~al.,}{{Lu} et~al.}{2024}]{lu24}
{Lu} S.,  et~al., 2024, \mn@doi [Nature Astronomy]
  {10.1038/s41550-024-02391-9}, \href
  {https://ui.adsabs.harvard.edu/abs/2024NatAs.tmp..252L} {}

\bibitem[\protect\citeauthoryear{{Maiolino} \& {Mannucci}}{{Maiolino} \&
  {Mannucci}}{2019}]{maiolino19}
{Maiolino} R.,  {Mannucci} F.,  2019, \mn@doi [\aapr]
  {10.1007/s00159-018-0112-2}, \href
  {http://adsabs.harvard.edu/abs/2019A%26ARv..27....3M} {27, 3}

\bibitem[\protect\citeauthoryear{{Maiolino} et~al.,}{{Maiolino}
  et~al.}{2024}]{maiolino24_gnz11}
{Maiolino} R.,  et~al., 2024, \mn@doi [\nat] {10.1038/s41586-024-07052-5},
  \href {https://ui.adsabs.harvard.edu/abs/2024Natur.627...59M} {627, 59}

\bibitem[\protect\citeauthoryear{{Maraston}}{{Maraston}}{2005}]{maraston05}
{Maraston} C.,  2005, \mn@doi [\mnras] {10.1111/j.1365-2966.2005.09270.x},
  \href {http://adsabs.harvard.edu/abs/2005MNRAS.362..799M} {362, 799}

\bibitem[\protect\citeauthoryear{{Mason}, {Trenti}  \& {Treu}}{{Mason}
  et~al.}{2015}]{mason15}
{Mason} C.~A.,  {Trenti} M.,   {Treu} T.,  2015, \mn@doi [\apj]
  {10.1088/0004-637X/813/1/21}, \href
  {http://adsabs.harvard.edu/abs/2015ApJ...813...21M} {813, 21}

\bibitem[\protect\citeauthoryear{{Mason}, {Trenti}  \& {Treu}}{{Mason}
  et~al.}{2023}]{mason23}
{Mason} C.~A.,  {Trenti} M.,   {Treu} T.,  2023, \mn@doi [\mnras]
  {10.1093/mnras/stad035}, \href
  {https://ui.adsabs.harvard.edu/abs/2023MNRAS.521..497M} {521, 497}

\bibitem[\protect\citeauthoryear{{Mirocha} \& {Furlanetto}}{{Mirocha} \&
  {Furlanetto}}{2023}]{mirocha23}
{Mirocha} J.,  {Furlanetto} S.~R.,  2023, \mn@doi [\mnras]
  {10.1093/mnras/stac3578}, \href
  {https://ui.adsabs.harvard.edu/abs/2023MNRAS.519..843M} {519, 843}

\bibitem[\protect\citeauthoryear{{Mo}, {Mao}  \& {White}}{{Mo}
  et~al.}{1998}]{mo98}
{Mo} H.~J.,  {Mao} S.,   {White} S.~D.~M.,  1998, \mn@doi [\mnras]
  {10.1046/j.1365-8711.1998.01227.x}, \href
  {http://adsabs.harvard.edu/abs/1998MNRAS.295..319M} {295, 319}

\bibitem[\protect\citeauthoryear{{Momcheva} et~al.,}{{Momcheva}
  et~al.}{2016}]{momcheva16}
{Momcheva} I.~G.,  et~al., 2016, \mn@doi [\apjs] {10.3847/0067-0049/225/2/27},
  \href {http://adsabs.harvard.edu/abs/2016ApJS..225...27M} {225, 27}

\bibitem[\protect\citeauthoryear{{Morishita} et~al.,}{{Morishita}
  et~al.}{2024}]{morishita24}
{Morishita} T.,  et~al., 2024, \mn@doi [\apj] {10.3847/1538-4357/ad1404}, \href
  {https://ui.adsabs.harvard.edu/abs/2024ApJ...963....9M} {963, 9}

\bibitem[\protect\citeauthoryear{{Moster}, {Somerville}, {Maulbetsch}, {van den
  Bosch}, {Macci{\`o}}, {Naab}  \& {Oser}}{{Moster} et~al.}{2010}]{moster10}
{Moster} B.~P.,  {Somerville} R.~S.,  {Maulbetsch} C.,  {van den Bosch} F.~C.,
  {Macci{\`o}} A.~V.,  {Naab} T.,   {Oser} L.,  2010, \mn@doi [\apj]
  {10.1088/0004-637X/710/2/903}, \href
  {http://adsabs.harvard.edu/abs/2010ApJ...710..903M} {710, 903}

\bibitem[\protect\citeauthoryear{{Naidu} et~al.,}{{Naidu}
  et~al.}{2022}]{naidu22_highz}
{Naidu} R.~P.,  et~al., 2022, \mn@doi [\apjl] {10.3847/2041-8213/ac9b22}, \href
  {https://ui.adsabs.harvard.edu/abs/2022ApJ...940L..14N} {940, L14}

\bibitem[\protect\citeauthoryear{{Nanayakkara}, {Esdaile}, {Glazebrook},
  {Espejo Salcedo}, {Durre}  \& {Jacobs}}{{Nanayakkara}
  et~al.}{2022}]{nanayakkara22}
{Nanayakkara} T.,  {Esdaile} J.,  {Glazebrook} K.,  {Espejo Salcedo} J.~M.,
  {Durre} M.,   {Jacobs} C.,  2022, \mn@doi [\pasa] {10.1017/pasa.2021.61},
  \href {https://ui.adsabs.harvard.edu/abs/2022PASA...39....2N} {39, e002}

\bibitem[\protect\citeauthoryear{{Nanayakkara} et~al.,}{{Nanayakkara}
  et~al.}{2024}]{nanayakkara24}
{Nanayakkara} T.,  et~al., 2024, \mn@doi [Scientific Reports]
  {10.1038/s41598-024-52585-4}, \href
  {https://ui.adsabs.harvard.edu/abs/2024NatSR..14.3724N} {14, 3724}

\bibitem[\protect\citeauthoryear{{Narayanan} \& {Dav{\'e}}}{{Narayanan} \&
  {Dav{\'e}}}{2013}]{narayanan13}
{Narayanan} D.,  {Dav{\'e}} R.,  2013, \mn@doi [\mnras]
  {10.1093/mnras/stt1548}, \href
  {https://ui.adsabs.harvard.edu/abs/2013MNRAS.436.2892N} {436, 2892}

\bibitem[\protect\citeauthoryear{{Neistein} \& {Dekel}}{{Neistein} \&
  {Dekel}}{2008}]{neistein08}
{Neistein} E.,  {Dekel} A.,  2008, \mn@doi [\mnras]
  {10.1111/j.1365-2966.2008.13525.x}, \href
  {http://adsabs.harvard.edu/abs/2008MNRAS.388.1792N} {388, 1792}

\bibitem[\protect\citeauthoryear{{Noll}, {Burgarella}, {Giovannoli}, {Buat},
  {Marcillac}  \& {Mu{\~n}oz-Mateos}}{{Noll} et~al.}{2009}]{noll09}
{Noll} S.,  {Burgarella} D.,  {Giovannoli} E.,  {Buat} V.,  {Marcillac} D.,
  {Mu{\~n}oz-Mateos} J.~C.,  2009, \mn@doi [\aap]
  {10.1051/0004-6361/200912497}, \href
  {http://adsabs.harvard.edu/abs/2009A%26A...507.1793N} {507, 1793}

\bibitem[\protect\citeauthoryear{{Papovich} et~al.,}{{Papovich}
  et~al.}{2023}]{papovich23}
{Papovich} C.,  et~al., 2023, \mn@doi [\apjl] {10.3847/2041-8213/acc948}, \href
  {https://ui.adsabs.harvard.edu/abs/2023ApJ...949L..18P} {949, L18}

\bibitem[\protect\citeauthoryear{{Park}, {Conroy}, {Johnson}, {Leja}, {Dotter}
  \& {Cargile}}{{Park} et~al.}{2024a}]{park24}
{Park} M.,  {Conroy} C.,  {Johnson} B.~D.,  {Leja} J.,  {Dotter} A.,
  {Cargile} P.~A.,  2024a, \mn@doi [arXiv e-prints]
  {10.48550/arXiv.2410.21375}, \href
  {https://ui.adsabs.harvard.edu/abs/2024arXiv241021375P} {p. arXiv:2410.21375}

\bibitem[\protect\citeauthoryear{{Park} et~al.,}{{Park}
  et~al.}{2024b}]{park24_quench}
{Park} M.,  et~al., 2024b, \mn@doi [\apj] {10.3847/1538-4357/ad7e15}, \href
  {https://ui.adsabs.harvard.edu/abs/2024ApJ...976...72P} {976, 72}

\bibitem[\protect\citeauthoryear{{Pasha} \& {Miller}}{{Pasha} \&
  {Miller}}{2023}]{pasha23}
{Pasha} I.,  {Miller} T.~B.,  2023, \mn@doi [The Journal of Open Source
  Software] {10.21105/joss.05703}, \href
  {https://ui.adsabs.harvard.edu/abs/2023JOSS....8.5703P} {8, 5703}

\bibitem[\protect\citeauthoryear{{Peebles}}{{Peebles}}{1969}]{peebles69}
{Peebles} P.~J.~E.,  1969, \mn@doi [\apj] {10.1086/149876}, \href
  {http://adsabs.harvard.edu/abs/1969ApJ...155..393P} {155, 393}

\bibitem[\protect\citeauthoryear{{Perrin}, {Sivaramakrishnan}, {Lajoie},
  {Elliott}, {Pueyo}, {Ravindranath}  \& {Albert}}{{Perrin}
  et~al.}{2014}]{webbpsf}
{Perrin} M.~D.,  {Sivaramakrishnan} A.,  {Lajoie} C.-P.,  {Elliott} E.,
  {Pueyo} L.,  {Ravindranath} S.,   {Albert} L.,  2014, in {Oschmann}
  Jacobus~M. J.,  {Clampin} M.,  {Fazio} G.~G.,   {MacEwen} H.~A.,  eds,
  Society of Photo-Optical Instrumentation Engineers (SPIE) Conference Series
  Vol. 9143, Space Telescopes and Instrumentation 2014: Optical, Infrared, and
  Millimeter Wave. p. 91433X, \mn@doi{10.1117/12.2056689}

\bibitem[\protect\citeauthoryear{{Phan}, {Pradhan}  \& {Jankowiak}}{{Phan}
  et~al.}{2019}]{phan19}
{Phan} D.,  {Pradhan} N.,   {Jankowiak} M.,  2019, \mn@doi [arXiv e-prints]
  {10.48550/arXiv.1912.11554}, \href
  {https://ui.adsabs.harvard.edu/abs/2019arXiv191211554P} {p. arXiv:1912.11554}

\bibitem[\protect\citeauthoryear{{Planck Collaboration} et~al.,}{{Planck
  Collaboration} et~al.}{2020}]{planck-collaboration20}
{Planck Collaboration} et~al., 2020, \mn@doi [\aap]
  {10.1051/0004-6361/201833910}, \href
  {https://ui.adsabs.harvard.edu/abs/2020A&A...641A...6P} {641, A6}

\bibitem[\protect\citeauthoryear{{Porciani}, {Dekel}  \& {Hoffman}}{{Porciani}
  et~al.}{2002}]{porciani02}
{Porciani} C.,  {Dekel} A.,   {Hoffman} Y.,  2002, \mn@doi [\mnras]
  {10.1046/j.1365-8711.2002.05305.x}, \href
  {https://ui.adsabs.harvard.edu/abs/2002MNRAS.332..325P} {332, 325}

\bibitem[\protect\citeauthoryear{{Press} \& {Schechter}}{{Press} \&
  {Schechter}}{1974}]{press74}
{Press} W.~H.,  {Schechter} P.,  1974, \mn@doi [\apj] {10.1086/152650}, \href
  {http://adsabs.harvard.edu/abs/1974ApJ...187..425P} {187, 425}

\bibitem[\protect\citeauthoryear{{Rauscher} et~al.,}{{Rauscher}
  et~al.}{2007}]{rauscher07}
{Rauscher} B.~J.,  et~al., 2007, \mn@doi [\pasp] {10.1086/520887}, \href
  {https://ui.adsabs.harvard.edu/abs/2007PASP..119..768R} {119, 768}

\bibitem[\protect\citeauthoryear{{Rennehan}}{{Rennehan}}{2024}]{rennehan24}
{Rennehan} D.,  2024, \mn@doi [\apj] {10.3847/1538-4357/ad793d}, \href
  {https://ui.adsabs.harvard.edu/abs/2024ApJ...975..114R} {975, 114}

\bibitem[\protect\citeauthoryear{{Renzini}}{{Renzini}}{2023}]{renzini23}
{Renzini} A.,  2023, \mn@doi [\mnras] {10.1093/mnrasl/slad091}, \href
  {https://ui.adsabs.harvard.edu/abs/2023MNRAS.525L.117R} {525, L117}

\bibitem[\protect\citeauthoryear{{Robertson}, {Bullock}, {Cox}, {Di Matteo},
  {Hernquist}, {Springel}  \& {Yoshida}}{{Robertson}
  et~al.}{2006}]{robertson06}
{Robertson} B.,  {Bullock} J.~S.,  {Cox} T.~J.,  {Di Matteo} T.,  {Hernquist}
  L.,  {Springel} V.,   {Yoshida} N.,  2006, \mn@doi [\apj] {10.1086/504412},
  \href {http://adsabs.harvard.edu/abs/2006ApJ...645..986R} {645, 986}

\bibitem[\protect\citeauthoryear{{Robertson} et~al.,}{{Robertson}
  et~al.}{2023}]{robertson23}
{Robertson} B.~E.,  et~al., 2023, \mn@doi [Nature Astronomy]
  {10.1038/s41550-023-01921-1}, \href
  {https://ui.adsabs.harvard.edu/abs/2023NatAs...7..611R} {7, 611}

\bibitem[\protect\citeauthoryear{{Robertson} et~al.,}{{Robertson}
  et~al.}{2024}]{robertson24}
{Robertson} B.,  et~al., 2024, \mn@doi [\apj] {10.3847/1538-4357/ad463d}, \href
  {https://ui.adsabs.harvard.edu/abs/2024ApJ...970...31R} {970, 31}

\bibitem[\protect\citeauthoryear{{Rodriguez-Gomez} et~al.,}{{Rodriguez-Gomez}
  et~al.}{2015}]{rodriguez-gomez15}
{Rodriguez-Gomez} V.,  et~al., 2015, \mn@doi [\mnras] {10.1093/mnras/stv264},
  \href {http://adsabs.harvard.edu/abs/2015MNRAS.449...49R} {449, 49}

\bibitem[\protect\citeauthoryear{{Rodr{\'{\i}}guez-Puebla}, {Primack},
  {Avila-Reese}  \& {Faber}}{{Rodr{\'{\i}}guez-Puebla}
  et~al.}{2017}]{rodriguez-puebla17}
{Rodr{\'{\i}}guez-Puebla} A.,  {Primack} J.~R.,  {Avila-Reese} V.,   {Faber}
  S.~M.,  2017, \mn@doi [\mnras] {10.1093/mnras/stx1172}, \href
  {http://adsabs.harvard.edu/abs/2017MNRAS.470..651R} {470, 651}

\bibitem[\protect\citeauthoryear{{Sabti}, {Mu{\~n}oz}  \&
  {Kamionkowski}}{{Sabti} et~al.}{2024}]{sabti24}
{Sabti} N.,  {Mu{\~n}oz} J.~B.,   {Kamionkowski} M.,  2024, \mn@doi [\prl]
  {10.1103/PhysRevLett.132.061002}, \href
  {https://ui.adsabs.harvard.edu/abs/2024PhRvL.132f1002S} {132, 061002}

\bibitem[\protect\citeauthoryear{{S{\'a}nchez-Bl{\'a}zquez}
  et~al.,}{{S{\'a}nchez-Bl{\'a}zquez} et~al.}{2006}]{sanchez-blazquez06}
{S{\'a}nchez-Bl{\'a}zquez} P.,  et~al., 2006, \mn@doi [\mnras]
  {10.1111/j.1365-2966.2006.10699.x}, \href
  {https://ui.adsabs.harvard.edu/abs/2006MNRAS.371..703S} {371, 703}

\bibitem[\protect\citeauthoryear{{Schaye} et~al.,}{{Schaye}
  et~al.}{2023}]{schaye23}
{Schaye} J.,  et~al., 2023, \mn@doi [\mnras] {10.1093/mnras/stad2419}, \href
  {https://ui.adsabs.harvard.edu/abs/2023MNRAS.526.4978S} {526, 4978}

\bibitem[\protect\citeauthoryear{{Scholtz} et~al.,}{{Scholtz}
  et~al.}{2024}]{scholtz24}
{Scholtz} J.,  et~al., 2024, \mn@doi [arXiv e-prints]
  {10.48550/arXiv.2405.19401}, \href
  {https://ui.adsabs.harvard.edu/abs/2024arXiv240519401S} {p. arXiv:2405.19401}

\bibitem[\protect\citeauthoryear{{Schreiber} et~al.,}{{Schreiber}
  et~al.}{2018}]{schreiber18}
{Schreiber} C.,  et~al., 2018, \mn@doi [\aap] {10.1051/0004-6361/201833070},
  \href {https://ui.adsabs.harvard.edu/abs/2018A&A...618A..85S} {618, A85}

\bibitem[\protect\citeauthoryear{{Setton} et~al.,}{{Setton}
  et~al.}{2024}]{setton24}
{Setton} D.~J.,  et~al., 2024, \mn@doi [\apj] {10.3847/1538-4357/ad6a18}, \href
  {https://ui.adsabs.harvard.edu/abs/2024ApJ...974..145S} {974, 145}

\bibitem[\protect\citeauthoryear{{Shapley} et~al.,}{{Shapley}
  et~al.}{2024}]{shapley24}
{Shapley} A.~E.,  et~al., 2024, \mn@doi [arXiv e-prints]
  {10.48550/arXiv.2410.00110}, \href
  {https://ui.adsabs.harvard.edu/abs/2024arXiv241000110S} {p. arXiv:2410.00110}

\bibitem[\protect\citeauthoryear{{Shen} et~al.,}{{Shen} et~al.}{2020}]{shen20}
{Shen} X.,  et~al., 2020, \mn@doi [\mnras] {10.1093/mnras/staa1423}, \href
  {https://ui.adsabs.harvard.edu/abs/2020MNRAS.495.4747S} {495, 4747}

\bibitem[\protect\citeauthoryear{{Shen}, {Vogelsberger}, {Boylan-Kolchin},
  {Tacchella}  \& {Kannan}}{{Shen} et~al.}{2023}]{shen23}
{Shen} X.,  {Vogelsberger} M.,  {Boylan-Kolchin} M.,  {Tacchella} S.,
  {Kannan} R.,  2023, \mn@doi [\mnras] {10.1093/mnras/stad2508}, \href
  {https://ui.adsabs.harvard.edu/abs/2023MNRAS.525.3254S} {525, 3254}

\bibitem[\protect\citeauthoryear{{Shen}, {Vogelsberger}, {Boylan-Kolchin},
  {Tacchella}  \& {Naidu}}{{Shen} et~al.}{2024}]{shen24_ede}
{Shen} X.,  {Vogelsberger} M.,  {Boylan-Kolchin} M.,  {Tacchella} S.,   {Naidu}
  R.~P.,  2024, \mn@doi [\mnras] {10.1093/mnras/stae1932}, \href
  {https://ui.adsabs.harvard.edu/abs/2024MNRAS.533.3923S} {533, 3923}

\bibitem[\protect\citeauthoryear{{Shibuya}, {Ouchi}, {Harikane}  \&
  {Nakajima}}{{Shibuya} et~al.}{2019}]{shibuya19}
{Shibuya} T.,  {Ouchi} M.,  {Harikane} Y.,   {Nakajima} K.,  2019, \mn@doi
  [\apj] {10.3847/1538-4357/aaf64b}, \href
  {https://ui.adsabs.harvard.edu/abs/2019ApJ...871..164S} {871, 164}

\bibitem[\protect\citeauthoryear{{Simmonds} et~al.,}{{Simmonds}
  et~al.}{2024a}]{simmonds24_bursty}
{Simmonds} C.,  et~al., 2024a, \mn@doi [\mnras] {10.1093/mnras/stad3605}, \href
  {https://ui.adsabs.harvard.edu/abs/2024MNRAS.527.6139S} {527, 6139}

\bibitem[\protect\citeauthoryear{{Simmonds} et~al.,}{{Simmonds}
  et~al.}{2024b}]{simmonds24}
{Simmonds} C.,  et~al., 2024b, \mn@doi [\mnras] {10.1093/mnras/stae2537}, \href
  {https://ui.adsabs.harvard.edu/abs/2024MNRAS.535.2998S} {535, 2998}

\bibitem[\protect\citeauthoryear{{Skelton} et~al.,}{{Skelton}
  et~al.}{2014}]{skelton14}
{Skelton} R.~E.,  et~al., 2014, \mn@doi [\apjs] {10.1088/0067-0049/214/2/24},
  \href {http://adsabs.harvard.edu/abs/2014ApJS..214...24S} {214, 24}

\bibitem[\protect\citeauthoryear{{Somerville} et~al.,}{{Somerville}
  et~al.}{2018}]{somerville18}
{Somerville} R.~S.,  et~al., 2018, \mn@doi [\mnras] {10.1093/mnras/stx2040},
  \href {https://ui.adsabs.harvard.edu/abs/2018MNRAS.473.2714S} {473, 2714}

\bibitem[\protect\citeauthoryear{{Speagle}}{{Speagle}}{2020}]{speagle20}
{Speagle} J.~S.,  2020, \mn@doi [\mnras] {10.1093/mnras/staa278}, \href
  {https://ui.adsabs.harvard.edu/abs/2020MNRAS.493.3132S} {493, 3132}

\bibitem[\protect\citeauthoryear{{Spiniello}, {Trager}  \&
  {Koopmans}}{{Spiniello} et~al.}{2015}]{spiniello15}
{Spiniello} C.,  {Trager} S.~C.,   {Koopmans} L.~V.~E.,  2015, \mn@doi [\apj]
  {10.1088/0004-637X/803/2/87}, \href
  {https://ui.adsabs.harvard.edu/abs/2015ApJ...803...87S} {803, 87}

\bibitem[\protect\citeauthoryear{{Springel} \& {Hernquist}}{{Springel} \&
  {Hernquist}}{2005}]{springel05}
{Springel} V.,  {Hernquist} L.,  2005, \mn@doi [\apjl] {10.1086/429486}, \href
  {http://adsabs.harvard.edu/abs/2005ApJ...622L...9S} {622, L9}

\bibitem[\protect\citeauthoryear{{Springel}, {Frenk}  \& {White}}{{Springel}
  et~al.}{2006}]{springel06}
{Springel} V.,  {Frenk} C.~S.,   {White} S. D.~M.,  2006, \mn@doi [\nat]
  {10.1038/nature04805}, \href
  {https://ui.adsabs.harvard.edu/abs/2006Natur.440.1137S} {440, 1137}

\bibitem[\protect\citeauthoryear{{Strait} et~al.,}{{Strait}
  et~al.}{2023}]{strait23}
{Strait} V.,  et~al., 2023, \mn@doi [\apjl] {10.3847/2041-8213/acd457}, \href
  {https://ui.adsabs.harvard.edu/abs/2023ApJ...949L..23S} {949, L23}

\bibitem[\protect\citeauthoryear{{Sun} \& {Furlanetto}}{{Sun} \&
  {Furlanetto}}{2016}]{sun16}
{Sun} G.,  {Furlanetto} S.~R.,  2016, \mn@doi [\mnras] {10.1093/mnras/stw980},
  \href {http://adsabs.harvard.edu/abs/2016MNRAS.460..417S} {460, 417}

\bibitem[\protect\citeauthoryear{{Sun}, {Faucher-Gigu{\`e}re}, {Hayward},
  {Shen}, {Wetzel}  \& {Cochrane}}{{Sun} et~al.}{2023}]{sun23_bursty}
{Sun} G.,  {Faucher-Gigu{\`e}re} C.-A.,  {Hayward} C.~C.,  {Shen} X.,  {Wetzel}
  A.,   {Cochrane} R.~K.,  2023, \mn@doi [\apjl] {10.3847/2041-8213/acf85a},
  \href {https://ui.adsabs.harvard.edu/abs/2023ApJ...955L..35S} {955, L35}

\bibitem[\protect\citeauthoryear{{Sun} et~al.,}{{Sun} et~al.}{2024}]{sun24}
{Sun} F.,  et~al., 2024, \mn@doi [\apj] {10.3847/1538-4357/ad07e3}, \href
  {https://ui.adsabs.harvard.edu/abs/2024ApJ...961...69S} {961, 69}

\bibitem[\protect\citeauthoryear{{Tacchella}, {Trenti}  \&
  {Carollo}}{{Tacchella} et~al.}{2013}]{tacchella13}
{Tacchella} S.,  {Trenti} M.,   {Carollo} C.~M.,  2013, \mn@doi [\apjl]
  {10.1088/2041-8205/768/2/L37}, \href
  {http://adsabs.harvard.edu/abs/2013ApJ...768L..37T} {768, L37}

\bibitem[\protect\citeauthoryear{{Tacchella} et~al.,}{{Tacchella}
  et~al.}{2015}]{tacchella15}
{Tacchella} S.,  et~al., 2015, \mn@doi [\apj] {10.1088/0004-637X/802/2/101},
  \href {http://adsabs.harvard.edu/abs/2015ApJ...802..101T} {802, 101}

\bibitem[\protect\citeauthoryear{{Tacchella}, {Dekel}, {Carollo}, {Ceverino},
  {DeGraf}, {Lapiner}, {Mandelker}  \& {Primack Joel}}{{Tacchella}
  et~al.}{2016a}]{tacchella16_MS}
{Tacchella} S.,  {Dekel} A.,  {Carollo} C.~M.,  {Ceverino} D.,  {DeGraf} C.,
  {Lapiner} S.,  {Mandelker} N.,   {Primack Joel} R.,  2016a, \mn@doi [\mnras]
  {10.1093/mnras/stw131}, \href
  {http://adsabs.harvard.edu/abs/2016MNRAS.457.2790T} {457, 2790}

\bibitem[\protect\citeauthoryear{{Tacchella}, {Dekel}, {Carollo}, {Ceverino},
  {DeGraf}, {Lapiner}, {Mandelker}  \& {Primack}}{{Tacchella}
  et~al.}{2016b}]{tacchella16_profile}
{Tacchella} S.,  {Dekel} A.,  {Carollo} C.~M.,  {Ceverino} D.,  {DeGraf} C.,
  {Lapiner} S.,  {Mandelker} N.,   {Primack} J.~R.,  2016b, \mn@doi [\mnras]
  {10.1093/mnras/stw303}, \href
  {http://adsabs.harvard.edu/abs/2016MNRAS.458..242T} {458, 242}

\bibitem[\protect\citeauthoryear{{Tacchella} et~al.,}{{Tacchella}
  et~al.}{2018a}]{tacchella18_dust}
{Tacchella} S.,  et~al., 2018a, \mn@doi [\apj] {10.3847/1538-4357/aabf8b},
  \href {https://ui.adsabs.harvard.edu/abs/2018ApJ...859...56T} {859, 56}

\bibitem[\protect\citeauthoryear{{Tacchella}, {Bose}, {Conroy}, {Eisenstein}
  \& {Johnson}}{{Tacchella} et~al.}{2018b}]{tacchella18}
{Tacchella} S.,  {Bose} S.,  {Conroy} C.,  {Eisenstein} D.~J.,   {Johnson}
  B.~D.,  2018b, \mn@doi [\apj] {10.3847/1538-4357/aae8e0}, \href
  {http://adsabs.harvard.edu/abs/2018ApJ...868...92T} {868, 92}

\bibitem[\protect\citeauthoryear{{Tacchella} et~al.,}{{Tacchella}
  et~al.}{2022a}]{tacchella22_quench}
{Tacchella} S.,  et~al., 2022a, \mn@doi [\apj] {10.3847/1538-4357/ac449b},
  \href {https://ui.adsabs.harvard.edu/abs/2022ApJ...926..134T} {926, 134}

\bibitem[\protect\citeauthoryear{{Tacchella} et~al.,}{{Tacchella}
  et~al.}{2022b}]{tacchella22_highz}
{Tacchella} S.,  et~al., 2022b, \mn@doi [\apj] {10.3847/1538-4357/ac4cad},
  \href {https://ui.adsabs.harvard.edu/abs/2022ApJ...927..170T} {927, 170}

\bibitem[\protect\citeauthoryear{{Tacchella} et~al.,}{{Tacchella}
  et~al.}{2023a}]{tacchella23_metal}
{Tacchella} S.,  et~al., 2023a, \mn@doi [\mnras] {10.1093/mnras/stad1408},
  \href {https://ui.adsabs.harvard.edu/abs/2023MNRAS.522.6236T} {522, 6236}

\bibitem[\protect\citeauthoryear{{Tacchella} et~al.,}{{Tacchella}
  et~al.}{2023b}]{tacchella23}
{Tacchella} S.,  et~al., 2023b, \mn@doi [\apj] {10.3847/1538-4357/acdbc6},
  \href {https://ui.adsabs.harvard.edu/abs/2023ApJ...952...74T} {952, 74}

\bibitem[\protect\citeauthoryear{{Thomas}, {Maraston}, {Bender}  \& {Mendes de
  Oliveira}}{{Thomas} et~al.}{2005}]{thomas05}
{Thomas} D.,  {Maraston} C.,  {Bender} R.,   {Mendes de Oliveira} C.,  2005,
  \mn@doi [\apj] {10.1086/426932}, \href
  {https://ui.adsabs.harvard.edu/abs/2005ApJ...621..673T} {621, 673}

\bibitem[\protect\citeauthoryear{{Trager}, {Faber}, {Worthey}  \&
  {Gonz{\'a}lez}}{{Trager} et~al.}{2000}]{trager00}
{Trager} S.~C.,  {Faber} S.~M.,  {Worthey} G.,   {Gonz{\'a}lez} J.~J.,  2000,
  \mn@doi [\aj] {10.1086/301299}, \href
  {https://ui.adsabs.harvard.edu/abs/2000AJ....119.1645T} {119, 1645}

\bibitem[\protect\citeauthoryear{{Trinca}, {Schneider}, {Valiante}, {Graziani},
  {Ferrotti}, {Omukai}  \& {Chon}}{{Trinca} et~al.}{2024}]{trinca24}
{Trinca} A.,  {Schneider} R.,  {Valiante} R.,  {Graziani} L.,  {Ferrotti} A.,
  {Omukai} K.,   {Chon} S.,  2024, \mn@doi [\mnras] {10.1093/mnras/stae651},
  \href {https://ui.adsabs.harvard.edu/abs/2024MNRAS.529.3563T} {529, 3563}

\bibitem[\protect\citeauthoryear{{Valentino} et~al.,}{{Valentino}
  et~al.}{2023}]{valentino23}
{Valentino} F.,  et~al., 2023, \mn@doi [\apj] {10.3847/1538-4357/acbefa}, \href
  {https://ui.adsabs.harvard.edu/abs/2023ApJ...947...20V} {947, 20}

\bibitem[\protect\citeauthoryear{{Vazdekis} et~al.,}{{Vazdekis}
  et~al.}{2015}]{vazdekis15}
{Vazdekis} A.,  et~al., 2015, \mn@doi [\mnras] {10.1093/mnras/stv151}, \href
  {http://adsabs.harvard.edu/abs/2015MNRAS.449.1177V} {449, 1177}

\bibitem[\protect\citeauthoryear{{Ventura}, {Qin}, {Balu}  \&
  {Wyithe}}{{Ventura} et~al.}{2024}]{ventura24}
{Ventura} E.~M.,  {Qin} Y.,  {Balu} S.,   {Wyithe} J. S.~B.,  2024, \mn@doi
  [\mnras] {10.1093/mnras/stae567}, \href
  {https://ui.adsabs.harvard.edu/abs/2024MNRAS.529..628V} {529, 628}

\bibitem[\protect\citeauthoryear{{Vitvitska}, {Klypin}, {Kravtsov}, {Wechsler},
  {Primack}  \& {Bullock}}{{Vitvitska} et~al.}{2002}]{vitvitska02}
{Vitvitska} M.,  {Klypin} A.~A.,  {Kravtsov} A.~V.,  {Wechsler} R.~H.,
  {Primack} J.~R.,   {Bullock} J.~S.,  2002, \mn@doi [\apj] {10.1086/344361},
  \href {https://ui.adsabs.harvard.edu/abs/2002ApJ...581..799V} {581, 799}

\bibitem[\protect\citeauthoryear{{Vogelsberger} et~al.,}{{Vogelsberger}
  et~al.}{2020}]{vogelsberger20}
{Vogelsberger} M.,  et~al., 2020, \mn@doi [\mnras] {10.1093/mnras/staa137},
  \href {https://ui.adsabs.harvard.edu/abs/2020MNRAS.492.5167V} {492, 5167}

\bibitem[\protect\citeauthoryear{{Wan}, {Tacchella}, {Johnson}, {Iyer},
  {Speagle}  \& {Maiolino}}{{Wan} et~al.}{2024}]{wan24}
{Wan} J.~T.,  {Tacchella} S.,  {Johnson} B.~D.,  {Iyer} K.~G.,  {Speagle}
  J.~S.,   {Maiolino} R.,  2024, \mn@doi [\mnras] {10.1093/mnras/stae1734},
  \href {https://ui.adsabs.harvard.edu/abs/2024MNRAS.532.4002W} {532, 4002}

\bibitem[\protect\citeauthoryear{{Wang} et~al.,}{{Wang}
  et~al.}{2023a}]{wang23_prospectorbeta}
{Wang} B.,  et~al., 2023a, \mn@doi [\apjl] {10.3847/2041-8213/acba99}, \href
  {https://ui.adsabs.harvard.edu/abs/2023ApJ...944L..58W} {944, L58}

\bibitem[\protect\citeauthoryear{{Wang} et~al.,}{{Wang} et~al.}{2023b}]{wang23}
{Wang} B.,  et~al., 2023b, \mn@doi [\apjl] {10.3847/2041-8213/acfe07}, \href
  {https://ui.adsabs.harvard.edu/abs/2023ApJ...957L..34W} {957, L34}

\bibitem[\protect\citeauthoryear{{Wang} et~al.,}{{Wang}
  et~al.}{2024}]{wang24_sp}
{Wang} B.,  et~al., 2024, \mn@doi [\apjl] {10.3847/2041-8213/ad55f7}, \href
  {https://ui.adsabs.harvard.edu/abs/2024ApJ...969L..13W} {969, L13}

\bibitem[\protect\citeauthoryear{{Wechsler}, {Bullock}, {Primack}, {Kravtsov}
  \& {Dekel}}{{Wechsler} et~al.}{2002}]{wechsler02}
{Wechsler} R.~H.,  {Bullock} J.~S.,  {Primack} J.~R.,  {Kravtsov} A.~V.,
  {Dekel} A.,  2002, \mn@doi [\apj] {10.1086/338765}, \href
  {http://adsabs.harvard.edu/abs/2002ApJ...568...52W} {568, 52}

\bibitem[\protect\citeauthoryear{{Weibel} et~al.,}{{Weibel}
  et~al.}{2024}]{weibel24}
{Weibel} A.,  et~al., 2024, \mn@doi [arXiv e-prints]
  {10.48550/arXiv.2409.03829}, \href
  {https://ui.adsabs.harvard.edu/abs/2024arXiv240903829W} {p. arXiv:2409.03829}

\bibitem[\protect\citeauthoryear{{White}}{{White}}{1984}]{white84}
{White} S.~D.~M.,  1984, \mn@doi [\apj] {10.1086/162573}, \href
  {http://adsabs.harvard.edu/abs/1984ApJ...286...38W} {286, 38}

\bibitem[\protect\citeauthoryear{{Whitler}, {Stark}, {Endsley}, {Leja},
  {Charlot}  \& {Chevallard}}{{Whitler} et~al.}{2023}]{whitler23_sfh}
{Whitler} L.,  {Stark} D.~P.,  {Endsley} R.,  {Leja} J.,  {Charlot} S.,
  {Chevallard} J.,  2023, \mn@doi [\mnras] {10.1093/mnras/stad004}, \href
  {https://ui.adsabs.harvard.edu/abs/2023MNRAS.519.5859W} {519, 5859}

\bibitem[\protect\citeauthoryear{{Wilkins} et~al.,}{{Wilkins}
  et~al.}{2023}]{wilkins23_frontier}
{Wilkins} S.~M.,  et~al., 2023, \mn@doi [\mnras] {10.1093/mnras/stac3280},
  \href {https://ui.adsabs.harvard.edu/abs/2023MNRAS.519.3118W} {519, 3118}

\bibitem[\protect\citeauthoryear{{Williams}, {Quadri}, {Franx}, {van Dokkum}
  \& {Labb{\'e}}}{{Williams} et~al.}{2009}]{williams09}
{Williams} R.~J.,  {Quadri} R.~F.,  {Franx} M.,  {van Dokkum} P.,   {Labb{\'e}}
  I.,  2009, \mn@doi [\apj] {10.1088/0004-637X/691/2/1879}, \href
  {http://adsabs.harvard.edu/abs/2009ApJ...691.1879W} {691, 1879}

\bibitem[\protect\citeauthoryear{{Williams} et~al.,}{{Williams}
  et~al.}{2024}]{williams24}
{Williams} C.~C.,  et~al., 2024, \mn@doi [\apj] {10.3847/1538-4357/ad3f17},
  \href {https://ui.adsabs.harvard.edu/abs/2024ApJ...968...34W} {968, 34}

\bibitem[\protect\citeauthoryear{{Witten} et~al.,}{{Witten}
  et~al.}{2024}]{witten24}
{Witten} C.,  et~al., 2024, \mn@doi [arXiv e-prints]
  {10.48550/arXiv.2407.07937}, \href
  {https://ui.adsabs.harvard.edu/abs/2024arXiv240707937W} {p. arXiv:2407.07937}

\bibitem[\protect\citeauthoryear{{Xiao} et~al.,}{{Xiao} et~al.}{2024}]{xiao24}
{Xiao} M.,  et~al., 2024, \mn@doi [\nat] {10.1038/s41586-024-08094-5}, \href
  {https://ui.adsabs.harvard.edu/abs/2024Natur.635..311X} {635, 311}

\bibitem[\protect\citeauthoryear{{Yung}, {Somerville}, {Finkelstein}, {Popping}
   \& {Dav{\'e}}}{{Yung} et~al.}{2019}]{yung19_uvlf}
{Yung} L.~Y.~A.,  {Somerville} R.~S.,  {Finkelstein} S.~L.,  {Popping} G.,
  {Dav{\'e}} R.,  2019, \mn@doi [\mnras] {10.1093/mnras/sty3241}, \href
  {http://adsabs.harvard.edu/abs/2019MNRAS.483.2983Y} {483, 2983}

\bibitem[\protect\citeauthoryear{{Yung}, {Somerville}, {Finkelstein}, {Wilkins}
   \& {Gardner}}{{Yung} et~al.}{2024}]{yung24}
{Yung} L.~Y.~A.,  {Somerville} R.~S.,  {Finkelstein} S.~L.,  {Wilkins} S.~M.,
  {Gardner} J.~P.,  2024, \mn@doi [\mnras] {10.1093/mnras/stad3484}, \href
  {https://ui.adsabs.harvard.edu/abs/2024MNRAS.527.5929Y} {527, 5929}

\bibitem[\protect\citeauthoryear{{Zavala} et~al.,}{{Zavala}
  et~al.}{2024}]{zavala24}
{Zavala} J.~A.,  et~al., 2024, \mn@doi [arXiv e-prints]
  {10.48550/arXiv.2403.10491}, \href
  {https://ui.adsabs.harvard.edu/abs/2024arXiv240310491Z} {p. arXiv:2403.10491}

\bibitem[\protect\citeauthoryear{{Zhang}, {Romano}, {Ivison}, {Papadopoulos}
  \& {Matteucci}}{{Zhang} et~al.}{2018}]{zhang18}
{Zhang} Z.-Y.,  {Romano} D.,  {Ivison} R.~J.,  {Papadopoulos} P.~P.,
  {Matteucci} F.,  2018, \mn@doi [\nat] {10.1038/s41586-018-0196-x}, \href
  {http://adsabs.harvard.edu/abs/2018Natur.558..260Z} {558, 260}

\bibitem[\protect\citeauthoryear{{Zolotov} et~al.,}{{Zolotov}
  et~al.}{2015}]{zolotov15}
{Zolotov} A.,  et~al., 2015, \mn@doi [\mnras] {10.1093/mnras/stv740}, \href
  {http://adsabs.harvard.edu/abs/2015MNRAS.450.2327Z} {450, 2327}

\bibitem[\protect\citeauthoryear{{de Graaff} et~al.,}{{de Graaff}
  et~al.}{2024a}]{de-graaff24}
{de Graaff} A.,  et~al., 2024a, \mn@doi [arXiv e-prints]
  {10.48550/arXiv.2404.05683}, \href
  {https://ui.adsabs.harvard.edu/abs/2024arXiv240405683D} {p. arXiv:2404.05683}

\bibitem[\protect\citeauthoryear{{de Graaff} et~al.,}{{de Graaff}
  et~al.}{2024b}]{de-graaff24_rubies}
{de Graaff} A.,  et~al., 2024b, \mn@doi [arXiv e-prints]
  {10.48550/arXiv.2409.05948}, \href
  {https://ui.adsabs.harvard.edu/abs/2024arXiv240905948D} {p. arXiv:2409.05948}

\bibitem[\protect\citeauthoryear{{de Graaff} et~al.,}{{de Graaff}
  et~al.}{2024c}]{de-graaff24_kin}
{de Graaff} A.,  et~al., 2024c, \mn@doi [\aap] {10.1051/0004-6361/202347755},
  \href {https://ui.adsabs.harvard.edu/abs/2024A&A...684A..87D} {684, A87}

\bibitem[\protect\citeauthoryear{{de La Rosa}, {La Barbera}, {Ferreras}  \& {de
  Carvalho}}{{de La Rosa} et~al.}{2011}]{de-la-rosa11}
{de La Rosa} I.~G.,  {La Barbera} F.,  {Ferreras} I.,   {de Carvalho} R.~R.,
  2011, \mn@doi [\mnras] {10.1111/j.1745-3933.2011.01146.x}, \href
  {https://ui.adsabs.harvard.edu/abs/2011MNRAS.418L..74D} {418, L74}

\bibitem[\protect\citeauthoryear{{van der Wel} et~al.,}{{van der Wel}
  et~al.}{2016}]{van-der-wel16}
{van der Wel} A.,  et~al., 2016, \mn@doi [\apjs] {10.3847/0067-0049/223/2/29},
  \href {https://ui.adsabs.harvard.edu/abs/2016ApJS..223...29V} {223, 29}

\makeatother
\end{thebibliography}

\appendix

\section{NIRSpec reduction comparison}\label{a.red}

\begin{figure*}
   \includegraphics[width=1.0\textwidth]{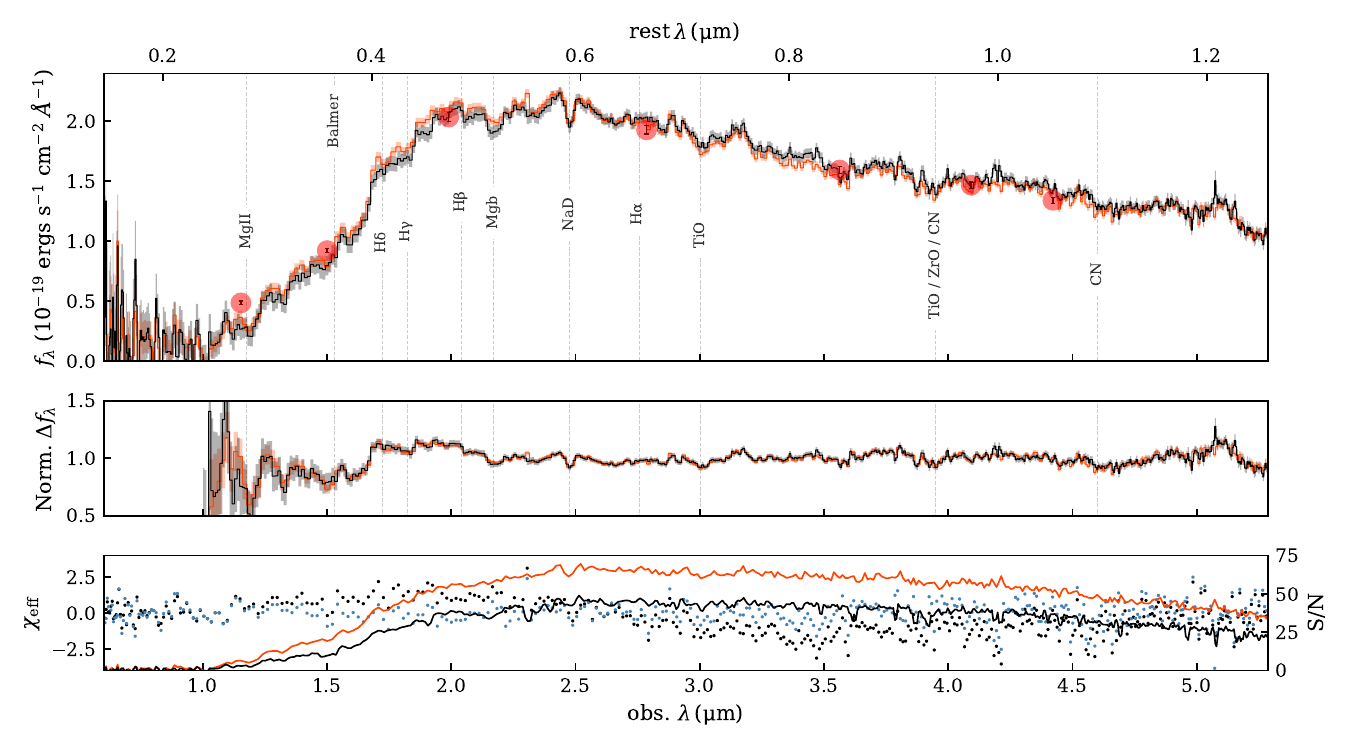}
   \caption{(Top) Spectrum of ZF-UDS-7329 reduced by \citet{glazebrook24} in orange and the pipeline from the NIRSpec-GTO team \citep[see, e.g.,][]{curtis-lake23, carniani24_outflow, curti24_Z} in black.  The shaded regions correspond to the estimated $1\sigma$ statistical errors, and the NIRCam photometry are error bars highlighted by red circles.  Vertical dashed lines indicate prominent absorption features. (Middle) Normalised spectral flux densities after removing the continuum shape through fitting each with an order-8 polynomial.  Both datasets have very similar equivalent widths for all the features.  (Bottom)  Residuals between $f^{\mathrm{K}}_{\lambda}$ and $f^{\mathrm{C}}_{\lambda}$ quantified by $\chi_{\mathrm{eff}}$ with (blue) and without (black) the continuum shape removed. SNR of each dataset are also plotted as lines colour-coded as before.  It should be noted that the finer binning used in $f^{\mathrm{C}}_{\lambda}$ necessitated a lower SNR.
   }\label{fig.compare}
\end{figure*}

We show the reduced spectrum adopted in this paper compared to that of \citet{glazebrook24} in Fig.~\ref{fig.compare}.  

\section{PPXF spectral fitting}\label{ppxfm}

\subsection{Methodology}\label{ppxfmodel}

The full spectrum is fitted using the Penalised Pixel-Fitting (\ppxf) method \citep{cappellari23} with a library of SSP templates that are a combination of the synthetic C3K model atmospheres \citep{conroy19} and MIST isochrones \citep{choi16} with solar abundance patterns and a Salpeter IMF.  In particular, the scheme minimises the function $\chi^2 + R \, \beta$, where $\chi^2$ are the weighted model residuals and $\beta$ is a function quantifying the smoothness of the weights (i.e. their second derivative) with adjustable coefficient $R$.  SSPs older than the age of the Universe at $z=3.2$ were excluded, leaving a logarithmic grid of 86 ages from $10^{5.0}$ to $ 10^{9.2}$ yr, across 12 metallicities, $\log_{10}(Z_\mathrm{SSP}/\Zsun) = -2.5$ to $0.5$.  While noise could bias the fit to younger ages, it is reasoned that allowing older SSPs would not be suitable given the extreme SFH of this galaxy.

Although the spectrum has not been directly calibrated, it agrees reasonably well with the photometric uncertainties, so a multiplicative polynomial is not applied to the fit.  This is to ensure the dust attenuation, which is included using the \citet{calzetti00} curve, could be properly constrained, as our current pipeline does not fit for photometry.  The velocity dispersions of the kinematic components are allowed to vary from 1 to 1000 km/s, but it should be noted that the PRISM disperser resolution is far too low to constrain this quantity.  The SSP weights are additionally assessed using bootstrapping of the initial \ppxf best-fit;  Here the fitting is repeated 100 times, with artificial noise from random samples of the original best-fit residuals added to the spectrum at each iteration.  In all cases second-order regularisation is used with the \ppxf regularisation parameter set to a (modest) value of $R=5$.

\subsection{Results}\label{ppxffiducial}

\begin{figure}
   \includegraphics[width=\columnwidth]{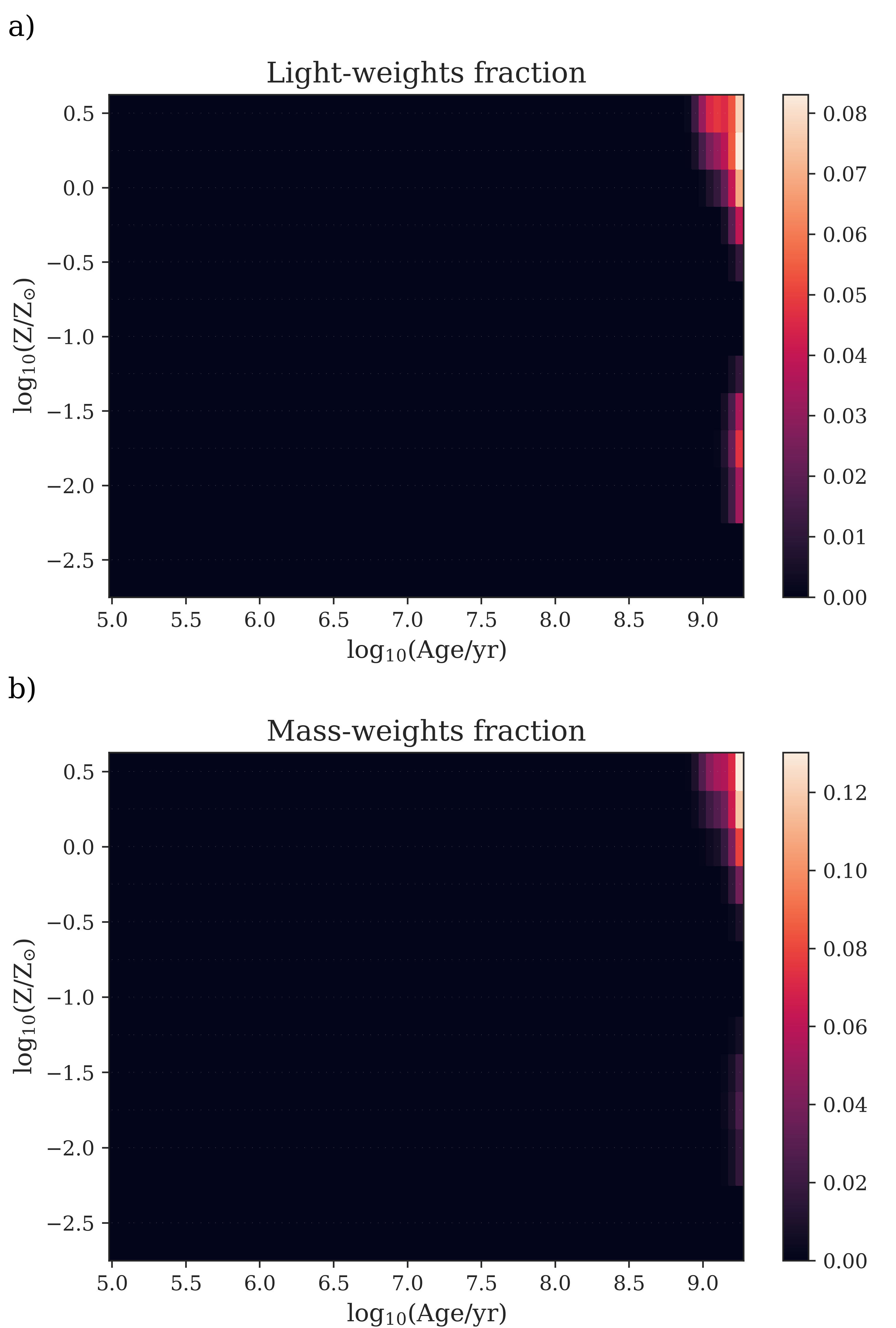}
   \caption{(a) Light weights inferred by \ppxf for the SSP metallicity-age grid. (b) \ppxf best-fit SFH of ZF-UDS-7329 given by the mass-weights of the SSPs.  $> 90\%$ of the mass is formed $> 1\, \Gyr$ prior to observation.
   }\label{fig.ppxf}
\end{figure}

The inferred SSP light-weights and mass-weights after the bootstrapping procedure are shown in Fig.~\ref{fig.ppxf}.  We find the vast majority of stellar mass is formed over 1 \Gyr prior to observation, with both the median mass-weighted and light-weighted age $\simeq 1.6$ Gyr.  The effective dust extinction is found as $
\tau_\mathrm{V} = 0.35$.  While most of the inferred population is super-solar, there is a marked sub-solar solution, accounting for $\sim 10 \%$ of the total flux, that vanishes at greater regularisation ($R \sim 20$).  The mean mass-weighted metallicity of [$Z$/H] = 0.07 is consistent with that obtained from full-spectrum fitting with \prospector (see next section) and implies some degree of template mismatch to be consistent with such an old age.  We find a high stellar mass, $\log_{10}{(\mstar/\Msun)} = 11.38$, $\sim$ 0.2 dex greater than with \prospector, which is expected given the usage of the bottom-heavier Salpeter IMF.  The best-fit spectrum and residuals are plotted in Fig.~\ref{fig.MAPs}.

\section{SFH prior comparison}\label{c.sfhprior}

\begin{figure*}
   \includegraphics[width=1.0\textwidth]{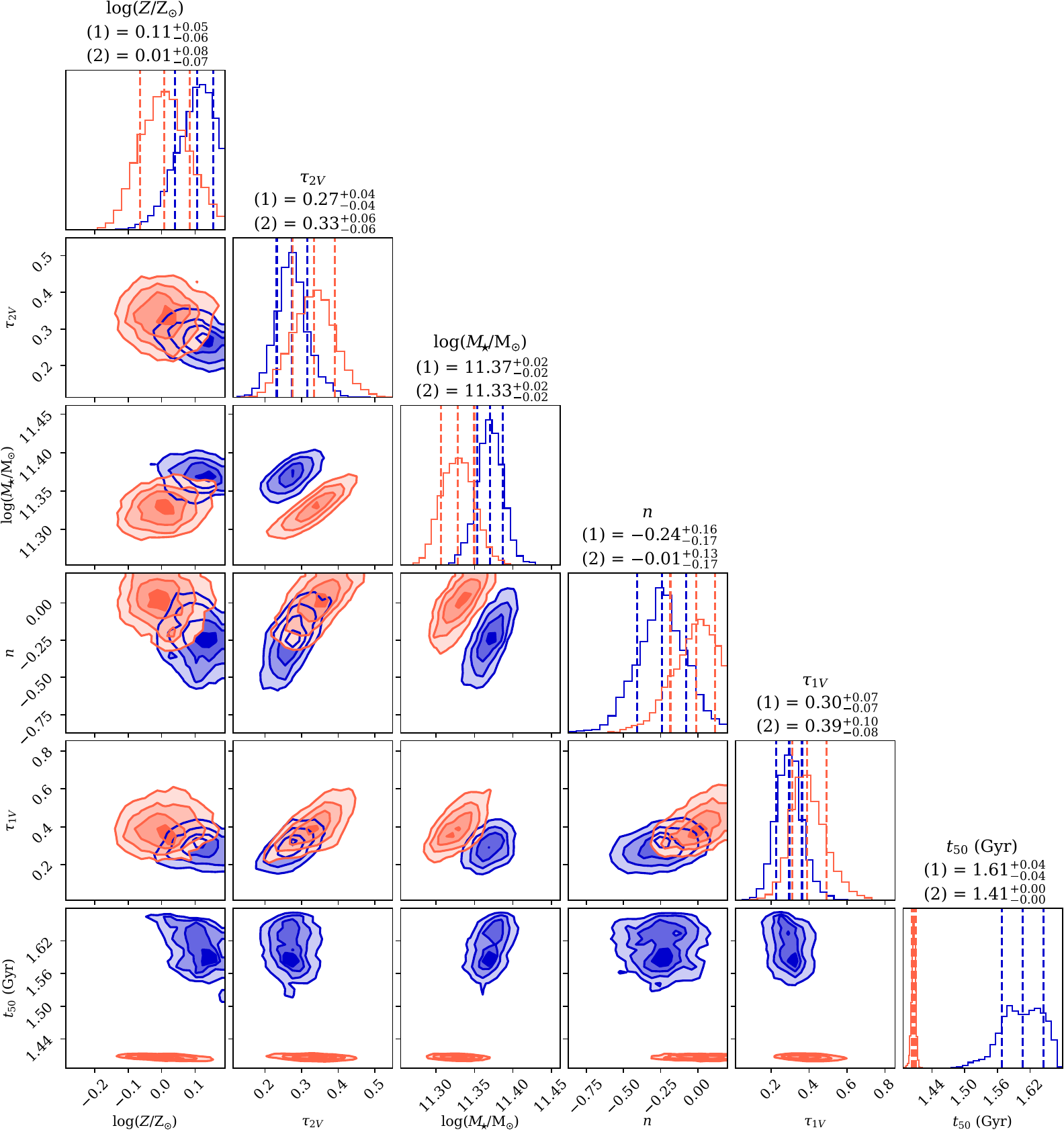}
   \caption{Marginalised posterior distributions for six key parameters of the C3K \prospector models with a flat continuity prior (blue), or rising continuity prior (orange), with median values labelled with (1), (2), respectively.  The rising prior favours a younger, more dusty solution.
   }\label{fig.priorposteriors}
\end{figure*}

We show that the results from the \prospector models with a flat versus "rising" continuity prior are largely consistent.  The posteriors for 6 key parameters are presented in Fig.~\ref{fig.priorposteriors}; the tight constraint on the median age arises from the rising model placing almost all the stellar mass in a single time bin.

\section{Mock SSP tests}\label{app.ssp}

We quantify the effect of changing the SFH prior on the derived stellar ages by fitting mock SSP spectra.  Two SSPs are generated, one with age $= 1.5~\Gyr$, and one with age $= 1.9~\Gyr$.  The other parameters are identically chosen as $\log_{10}{(\mstar/\Msun)} = 11.3$, [$Z$/H]$=0.01$, $\tau_{1,V}=\tau_{2,V}=0.3$, $n=-0.1$, $z=3.2$, and no nebular emission, to roughly mimic the results from our best fits of ZF-UDS-7329. The spectra are binned to the same resolution as our data, and convolved with our estimated LSF. In addition, we add Gaussian noise according to the same (fractional) noise vector.  

\begin{figure*}
   \includegraphics[width=1.0\textwidth]{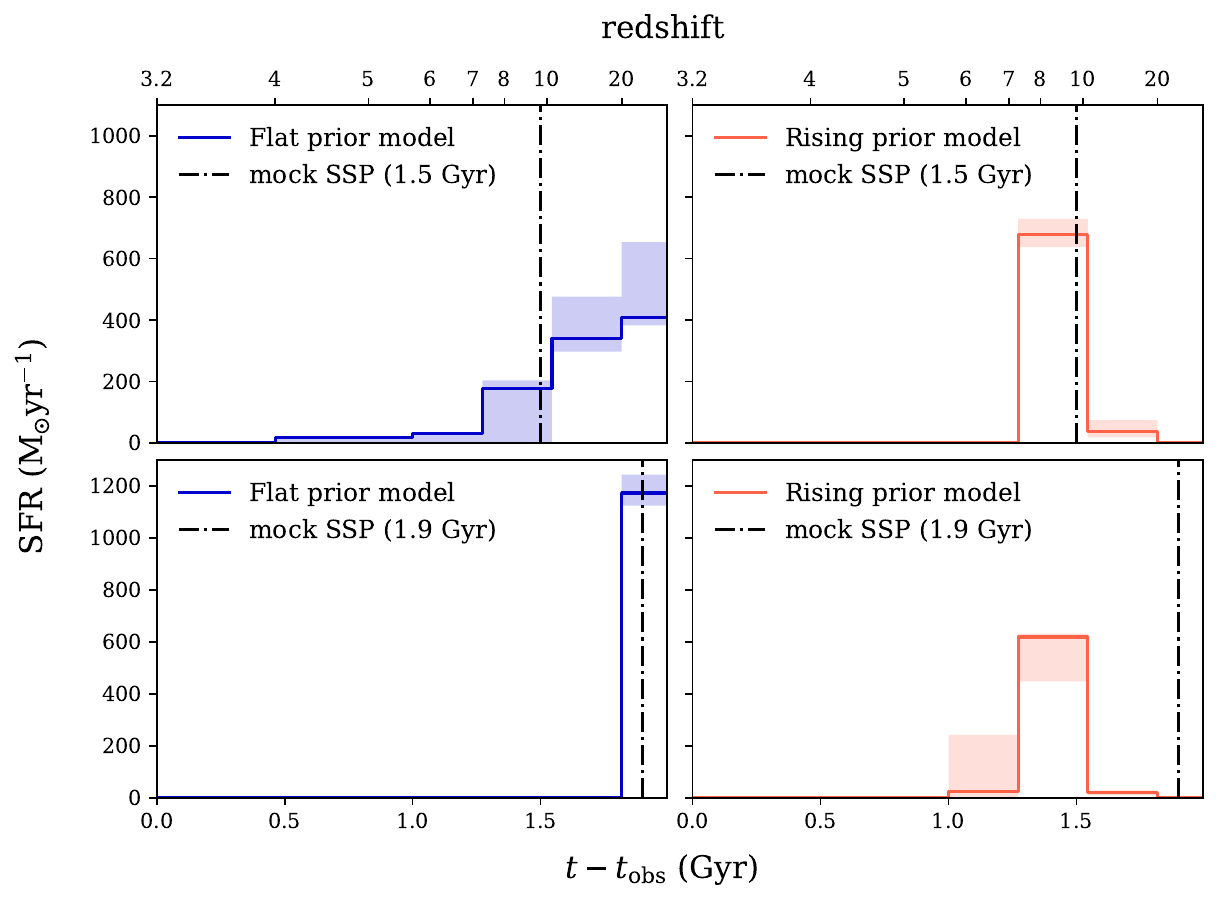}
   \caption{The effect of the SFH prior on derived stellar ages $\gtrsim 1.3~\Gyr$:  Best-fit SFHs for the flat continuity prior (left column) or rising continuity prior (right column) models when fitted to a mock SSP spectrum of age $= 1.5~\Gyr$ (top row) or $1.9~\Gyr$ (bottom row), indicated with a vertical dot-dashed line.  The MAP is shown as a solid line while the $16\textsuperscript{th}\text{--}84\textsuperscript{th}$ percentile region of the posterior is shaded. The flat prior manages to recover the age of the older SSP, but overestimates the age of the younger one by $\sim 250~\Myr$.  In contrast, the rising prior always forces a younger age $\lesssim 1.4~\Gyr$ for a population of this size.  Hence, it underestimates the age of the younger SSP by $\sim 100~\Myr$ and the older one by $\sim 500~\Myr$.  This demonstrates the difficulty of constraining stellar ages on an accuracy level of 10\% using non-parametric SFHs. 
   }\label{fig.mockssp}
\end{figure*}

Fig.~\ref{fig.mockssp} shows the best-fit SFHs produced by our models.  For the younger SSP, the flat prior overestimates the age by $\sim 250~\Myr$, while the rising prior underestimates it by $\sim 100~\Myr$.  Both posteriors on the median age are $\gtrsim 3\sigma$ inconsistent with the true value.  This is compounded by the use of time binning, for which the placement of bin edges will also influence the results.  Hence, the posterior distribution on the age should not be taken at face value for these non-parametric models.  For the older SSP, the flat prior now correctly recovers the age, but the rising one gives a MAP SFH virtually unchanged from the previous test, hence underestimating the age by $\sim 500~\Myr$. 
In all cases, the other stellar parameters are recovered within $1\text{--}2 \sigma$, showing the uniqueness of this problem.

\section{SFH cut-offs}\label{a.sfh}

\begin{figure}
   \includegraphics[width=\columnwidth]{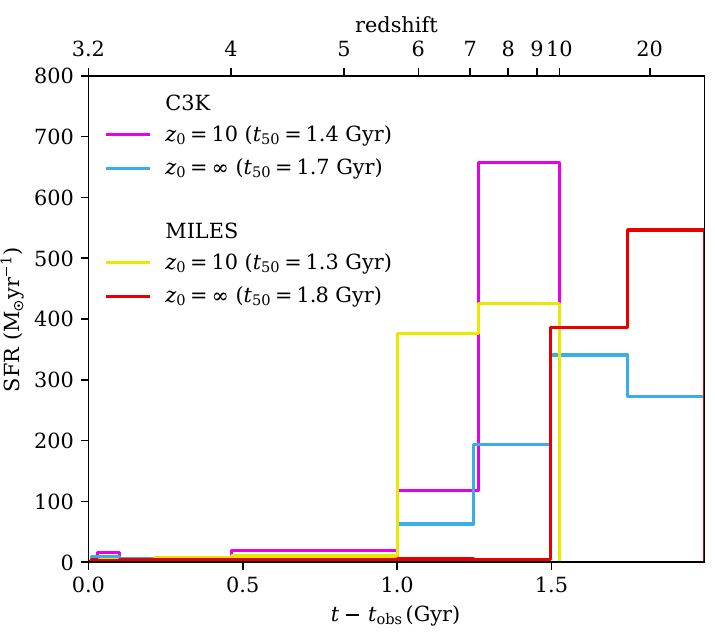}
   \caption{MAP SFHs for \prospector models with no star-formation allowed before $z_0=10$, or from the age of the universe (look-back time $\thickapprox 2 \, \Gyr$).  The median mass-weighted ages are bracketed in the legend.
   }\label{fig.appsfh_ages}
\end{figure}

\begin{figure*}
   \includegraphics[width=1.0\textwidth]{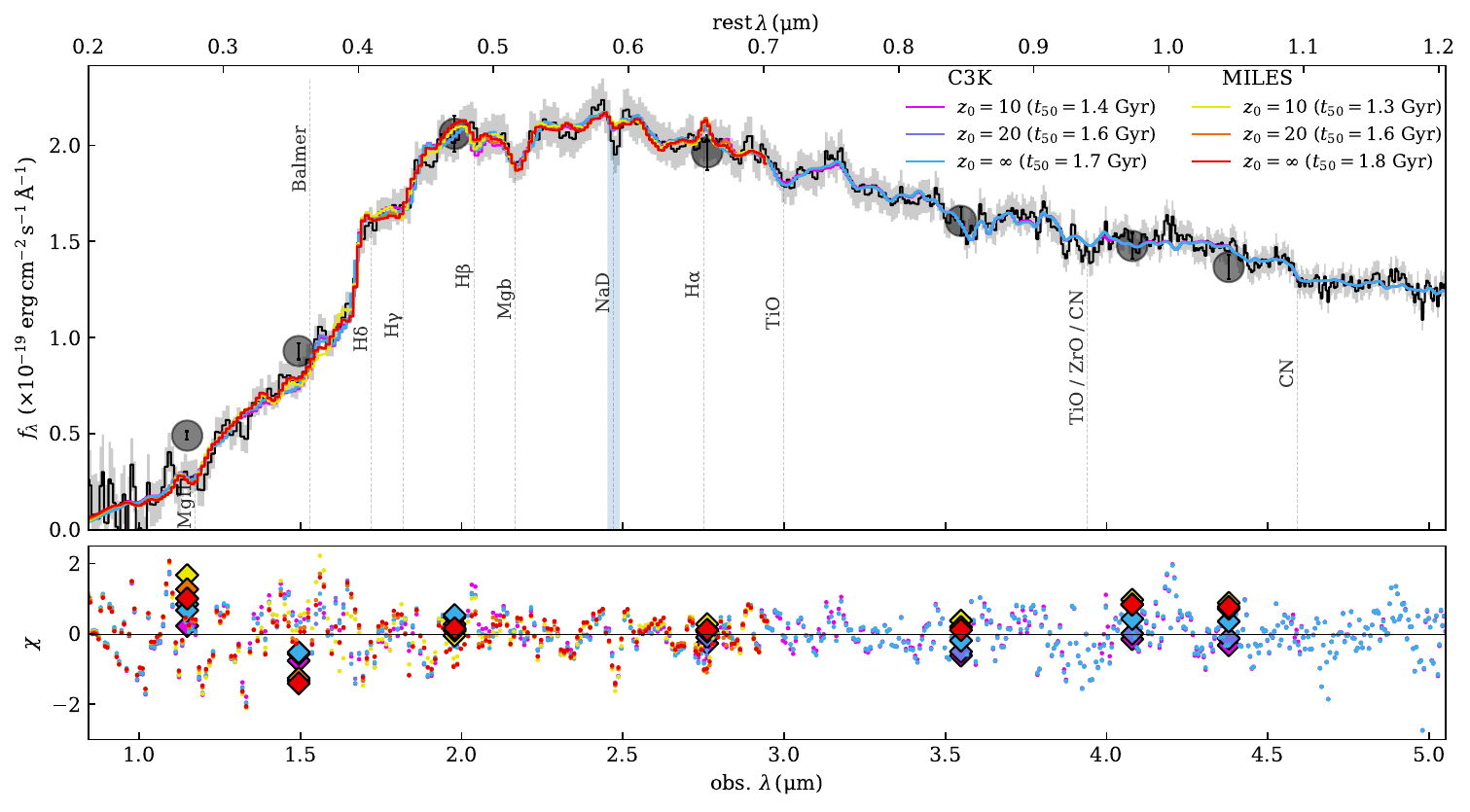}
   \caption{(Top) MAP best-fit spectra for \prospector models with no SF allowed before $z = 20$, $z = 10$, or the age of the Universe.  Corresponding median mass-weighted ages are bracketed in the legend.  All other lines and shaded regions are identical to Fig.~\ref{fig.MAPs}. (Bottom) Residuals for each spectrum as circles, colour-coded identically.  Photometric residuals, with the exception of F115W, which lies at $\chi \sim 8$, are displayed as diamonds.  While the youngest models do not reproduce the flux near the Balmer break as well, they generally remain within the estimated $1\sigma$ statistical errors.  
   }\label{fig.ageMAPs}
\end{figure*}

We present the best-fit spectra for all our \prospector models with no star formation allowed past $z_0 = 10, 20, \infty$ in Fig.~\ref{fig.ageMAPs}, with the corresponding SFHs in Fig.~\ref{fig.appsfh_ages}.
These demonstrate that forcing a $\sim 300 \, \Myr$ younger poor fit does not significantly disagree with the data, with most values remaining within the $5 \%$ error margin.  From oldest to youngest age, we find reduced $\chi^2$ values (calculated with $j_{\mathrm{spec}}=1$) of 0.43, 0.44, 0.49 for the MILES models and 0.37, 0.37, 0.38 for the C3K models (which cover a larger spectral range).  We also note that the inferred noise inflation factor, $j_{\mathrm{spec}}$, is relatively unchanged in these tests. 

\section{IMF tests}\label{b.imf}

\begin{figure*}
   \includegraphics[width=1.0\textwidth]{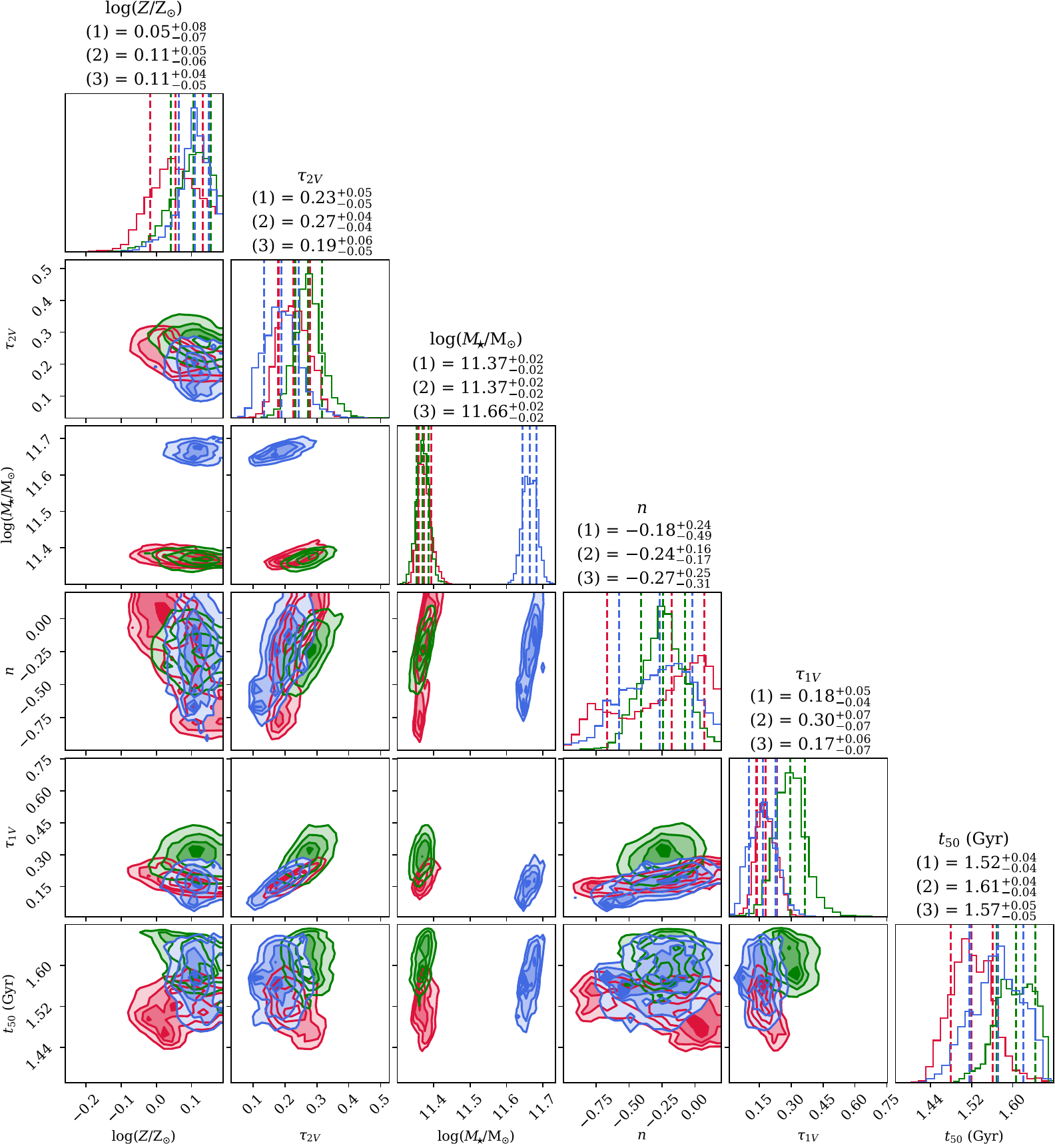}
   \caption{Marginalised posterior distributions for six key parameters of the \prospector models with a top-heavy (red), Kroupa (green), and bottom-heavy (blue) IMF, with median values labelled with (1), (2), and (3), respectively.  Other than the formed stellar mass, which is $\sim 0.3$ dex higher with the top-heavy IMF, the inferred values are all consistent.
   }\label{fig.imfposteriors}
\end{figure*}

Top-heavy and bottom-heavy IMFs were implemented in the C3K \prospector model as described in \ref{imfmethod}.  The posteriors for six key parameters are shown in Fig.~\ref{fig.imfposteriors} alongside the fiducial values, which used a \citet{kroupa01} IMF.  The results demonstrate that while a top-heavy IMF gives a $\sim 0.1$ dex reduction in $\msurv$, it is unable to reduce the total \textit{formed} mass, $\mstar$. All other population parameters are largely unchanged by the IMF variation.  

\label{lastpage}
\end{document}